\documentclass{elsart}
\usepackage{graphicx,bm}
\usepackage{amssymb}

\begin{document}

\begin{frontmatter}
\title{Atom-molecule coherence in Bose gases}

\author{R.A. Duine\corauthref{cor1}}
\ead{duine@phys.uu.nl} \ead[url]{http://www.phys.uu.nl/\~{}duine}
\corauth[cor1]{Corresponding author} and
\author{H.T.C. Stoof}
\address{Institute for Theoretical Physics,
         University of Utrecht, Leuvenlaan 4,
         3584 CE Utrecht, The Netherlands}

\begin{abstract}
In an atomic gas near a Feshbach resonance, the energy of two
colliding atoms is close to the energy of a bound state, i.e., a
molecular state, in a closed channel that is coupled to the
incoming open channel. Due to the different spin arrangements of
the atoms in the open channel and the atoms in the molecular
state, the energy difference between the bound state and the
two-atom continuum threshold is experimentally accessible by means
of the Zeeman interaction of the atomic spins with a magnetic
field. As a result, it is in principle possible to vary the
scattering length to any value by tuning the magnetic field. This
level of experimental control has opened the road for many
beautiful experiments, which recently led to the demonstration of
coherence between atoms and molecules. This is achieved by
observing coherent oscillations between atoms and molecules,
analogous to coherent Rabi oscillations that occur in ordinary
two-level systems. We review the many-body theory that describes
coherence between atoms and molecules in terms of an effective
quantum field theory for Feshbach-resonant interactions. The most
important feature of this effective quantum field theory is that
it incorporates the two-atom physics of the Feshbach resonance
exactly, which turns out to be necessary to fully explain
experiments with Bose-Einstein condensed atomic gases.
\end{abstract}

\end{frontmatter}

\newpage

\def\bx{{\bf x}}
\def\bk{{\bf k}}
\def\bq{{\bf q}}
\def\bK{{\bf K}}
\def\br{{\bf r}}
\def\half{\frac{1}{2}}
\def\args{(\bx,t)}
\def\phiup{\phi_{\uparrow}}
\def\phidup{\phi^*_{\uparrow}}
\def\phidown{\phi_{\downarrow}}
\def\phiddown{\phi^*_{\downarrow}}
\def\phim{\phi_{\rm m}}
\def\phimd{\phi^*_{\rm m}}
\def\phia{\phi_{\rm a}}
\def\phiad{\phi_{\rm a}^*}
\def\psim{\hat \psi_{\rm m}}
\def\psimd{\hat \psi_{\rm m}^{\dagger}}
\def\psia{\hat \psi_{\rm a}}
\def\psiad{\hat\psi_{\rm a}^{\dagger}}
\def\chia{\hat\chi_{\rm a}}
\def\chiad{\hat\chi_{\rm a}^{\dagger}}
\def\chim{\hat\chi_{\rm m}}
\def\chimd{\hat\chi_{\rm m}^{\dagger}}

  \section{Introduction} \label{sec:introduction}

Following the first experimental realization of Bose-Einstein
condensation \cite{anderson1995}, a great deal of experimental and
theoretical progress has been made in the field of ultracold
atomic gases
\cite{giorgini1999,leggett2001,pethickandsmithbook,pitaevskiiandstringaribook}.
One particular reason for this progress is the unprecedented
experimental control over the atomic gases of interest. This
experimental control over the ultracold magnetically-trapped
alkali gases, has recently culminated in the demonstration of
experimentally adjustable interactions between the atoms
\cite{inouye1998}. This is achieved by means of a so-called
Feshbach resonance \cite{feshbach1962}.

Feshbach resonances were introduced in nuclear physics to describe
the narrow resonances observed in the total cross section for a
neutron scattering of a nucleus \cite{levinandfeshbachbook}. These
very narrow resonances are the result of the formation of a
long-lived compound nucleus during the scattering process, with a
binding energy close to that of the incoming neutron. The defining
feature of a Feshbach resonance is that the bound state
responsible for the resonance exists in another part of the
quantum-mechanical Hilbert space than the part associated with the
incoming particles. In the simplest case, these two parts of the
Hilbert space are referred to as the closed and open channel,
respectively.

\begin{figure}
\begin{center}
\includegraphics{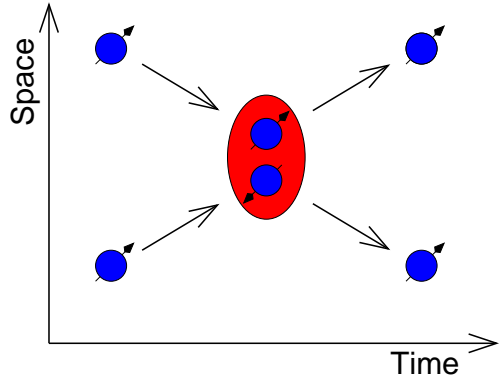}
\caption{\label{fig:collision}
 Illustration of a Feshbach-resonant atomic collision. Two atoms,
  with a hyperfine state indicated by the arrow, collide and form
  a long-lived molecule with a different spin arrangement, which
  ultimately decays again into two atoms.
   }
\end{center}
\end{figure}

\begin{figure}
\begin{center}
\includegraphics{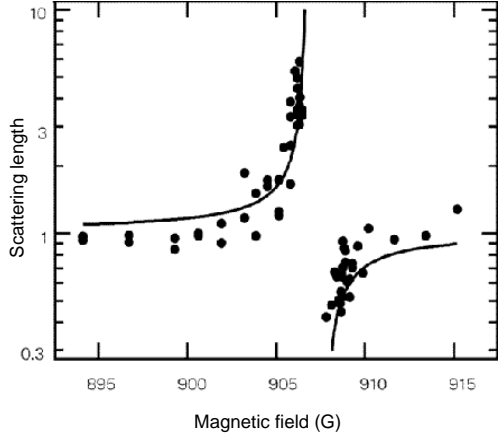}
\caption{\label{fig:inouye}
   The scattering length as a function of magnetic field as
  measured by Inouye {\it et al.} \cite{inouye1998}. The
  scattering length is normalized such that it is equal to one far
  off resonance.
   }
\end{center}
\end{figure}

Following these ideas from nuclear physics, Stwalley
\cite{stwalley1976} and Tiesinga {\it et al.} \cite{tiesinga1993}
considered Feshbach resonances in ultracold doubly spin-polarized
alkali gases. Due to the low temperatures of these gases, their
effective interatomic interactions are to a large extent
completely determined by the $s$-wave scattering length. Analogous
to the formation of a compound nucleus in neutron scattering, two
atoms can form a long-lived bound state, i.e., a diatomic
molecule, during an $s$-wave collision. This process is
illustrated in Fig.~\ref{fig:collision}. The two incoming atoms in
the open channel have a different hyperfine state than the bound
state in the closed channel and the coupling between the open and
closed channel is provided by the exchange interaction. As a
result of this difference in the hyperfine state, the two channels
have a different Zeeman shift in a magnetic field. Therefore, the
energy difference between the closed-channel bound state and the
two-atom continuum threshold, the so-called detuning, is
experimentally adjustable by tuning the magnetic field. This
implies that the $s$-wave scattering length, and hence the
magnitude and sign of the interatomic interactions, is also
adjustable to any desirable value. In Fig.~\ref{fig:inouye} the
scattering length, as measured by Inouye {\it et al.}
\cite{inouye1998}, is shown as a function of the magnetic field.
The position of the resonance in the magnetic field is at $B_0
\simeq 907$ (G)auss in this case. Following this first
experimental observation of Feshbach resonances in $^{23}$Na
\cite{inouye1998}, they have now been observed in various bosonic
atomic species
\cite{courteille1998,roberts1998,vuletic1999,marte2002,strecker2002},
as well as a number of fermionic isotopes
\cite{dieckmann2002,ohara2002,regal2003,bourdel2003}.

With this experimental degree of freedom it is possible to study
very interesting new regimes in the many-body physics of ultracold
atomic gases. The first experimental application was the detailed
study of the collapse of a condensate with attractive
interactions, corresponding to negative scattering lengths. In
general a collapse occurs when the attractive interactions
overcome the stabilizing kinetic energy of the condensate atoms in
the trap. Since the typical interaction energy is proportional to
the density, there is a certain maximum number of atoms above
which the condensate is unstable
\cite{ruprecht1995,shuryak1996,stoof1997,houbiers1996,bergeman1997}.
In the first observations of the condensate collapse by Bradley
\textit{et al.} \cite{bradley1995}, a condensate of doubly
spin-polarized $^7$Li atoms was used. In these experiments the
atoms have a fixed negative scattering length which for the
experimental trap parameters lead to a maximum number of
condensate atoms that was so small that nondestructive imaging of
the condensate was impossible. Moreover, thermal fluctuations due
to a large thermal component made the initiation of the collapse a
stochastic process \cite{duine2002}, thus preventing also a series
of destructive measurements of a single collapse event
\cite{sackett1998}. A statistical analysis has nevertheless
resulted in important information about the collapse process
\cite{sackett1999}. Very recently, it was even possible to
overcome these complications \cite{gerton2000}.

In addition to the experiment with $^7$Li, experiments with
$^{85}$Rb have been carried out \cite{cornish2000}. In particular,
Roberts \textit{et al.} \cite{roberts2001} also studied the
stability criterion for the condensate, and Donley \textit{et al.}
\cite{donley2001} studied the dynamics of a single collapse event
in great detail. Both of these experiments make use of a Feshbach
resonance to achieve a well-defined initial condition for each
destructive measurement. It turns out that during a collapse a
significant fraction of atoms is expelled from the condensate.
Moreover, one observes a burst of hot atoms with an energy of
about $150$ nK. Several mean-field analyses of the collapse, which
model the atom loss phenomenologically by a three-body
recombination rate constant
\cite{kagan1998,ueda1999,eleftheriou2000,adhikari2002,saito2002,santos2002,bao2003},
as well as an approach that considers elastic condensate
collisions \cite{duine2001,duine2003b}, and an approach that takes
into account the formation of molecules \cite{milstein2003}, have
offered a great deal of theoretical insight. Nevertheless, the
physical mechanism responsible for the explosion of atoms out of
the condensate and the formation of the noncondensed component is
to a great extent still not understood at present.

A second experimental application of a Feshbach resonance in a
Bose-Einstein condensed gas is the observation of a bright soliton
train by Strecker {\it et al.} \cite{strecker2002}. In this
experiment, one starts with a large one-dimensional Bose-Einstein
condensate of $^7$Li atoms with positive scattering length near a
Feshbach resonance. The scattering length is then abruptly changed
to a negative value. Due to its one-dimensional nature the
condensate does not collapse, but instead forms a train of on
average four bright solitary waves that repel each other. The
formation of these bright solitons is the result of phase
fluctuations \cite{khawaja2002}, which are in this case important
due to the low dimensionality
\cite{mullin1997,ho1999,petrov2000a,petrov2000b,petrov2001,andersen2002}.
The repulsion between the bright solitons is a result of their
relative phase difference of about $\pi$. In a similar experiment
Khaykovich {\it et al.} \cite{khaykovich2002} have observed the
formation of a single bright soliton.

A third experimental application are the experiments with trapped
gases of fermionic atoms, where the objective is to cool the gas
down to temperatures where the so-called BCS transition, i.e., the
Bose-Einstein condensation of Cooper pairs, may be observed. The
BCS transition temperature increases if the scattering length is
more negative \cite{stoof1996}, and hence a Feshbach resonance can
possibly be used to make the transition experimentally less
difficult to achieve. This possibility has inspired the study of
many-body effects in fermionic gases near a Feshbach resonance
\cite{holland2001,kokkelmans2002a,milstein2002,ohashi2002a,ohashi2002b,ohashi2003,combescot2003},
as well as fluctuation effects on the critical temperature
\cite{combescot1999,heiselberg2000}. One of the most interesting
features of a fermionic gas near a Feshbach resonance is the
crossover between a condensate of Cooper pairs and a condensate of
molecules, the so-called BCS-BEC crossover that was recently
studied by Ohashi and Griffin
\cite{ohashi2002a,ohashi2002b,ohashi2003} on the basis of the
Nozi\`eres-Schmitt-Rink formalism \cite{nozieres1985}. As a first
step towards this crossover, Regal {\it et al.} \cite{regal2003b}
were recently able to convert a fraction of the atoms in a gas of
fermionic atoms in the normal state into diatomic molecules, by
sweeping the magnetic field across a Feshbach resonance. Following
this observation, Strecker {\it et al.} observed the formation of
long-lived $^6$Li$_2$ molecules \cite{strecker2003}, and Xu {\it
et al.} observed $^{23}$Na molecules \cite{xu2003}. Very recently,
even the formation of Bose-Einstein condensates of molecules has
been observed by Jochim {\it et al.} \cite{jochim2003}, Greiner
{\it et al.} \cite{greiner2003}, and by Zwierlein {\it et al.}
\cite{zwierlein2003}. As another application of Feshbach
resonances in fermionic gases we mention here also the theoretical
proposal by Falco {\it et al.} to observe a new manifestation of
the Kondo effect in these systems \cite{falco2003}.

The experimental application on which we focus in this paper is
the observation of coherent atom-molecule oscillations
\cite{donley2002}. These experiments are inspired by the
theoretical proposal of Drummond {\it et al.} \cite{drummond1998}
and Timmermans {\it et al.} \cite{timmermans1999a} to describe the
Feshbach-resonant part of the interactions between the atoms in a
Bose-Einstein condensate by a coupling of the atomic condensate to
a molecular condensate. For this physical picture to be valid,
there has to be a well-defined phase between the wave function
that describes the atoms in the atomic condensate, and its
molecular counterpart. An equivalent statement is that there is
coherence between the atoms and the molecules. Since the energy
difference between the atoms and the molecular state is
experimentally tunable by adjusting the magnetic field, it is,
with this physical picture in mind, natural to perform a Rabi
experiment by means of one pulse in the magnetic field towards
resonance, and to perform a Ramsey experiment consisting of two
short pulses in the magnetic field. If the physical picture is
correct we expect to observe oscillations in the remaining number
of condensate atoms in both cases.

In the first experiment along these lines, Claussen {\it et al.}
\cite{claussen2002} started from a Bose-Einstein condensate of
$^{85}$Rb atoms without a visible thermal cloud and tuned the
magnetic field such that the atoms were effectively
noninteracting. With this atomic species this is possible, because
the off-resonant background scattering length is negative, which
can be compensated for by making the resonant part of the
scattering length positive. Next, one applied a trapezoidal pulse
in the magnetic field, directed towards resonance. As a function
of the duration of the pulse one observed that the number of atoms
first decreases but after some time increases again. This increase
can not be explained by a ``conventional'' loss process, such as
dipolar relaxation or three-body recombination, since the
magnitude of the loss is in these cases given by a rate constant
times the square and the cube of the density, respectively. As a
result, the loss always increases with longer times. A theoretical
description of this experiment is complicated by the fact that the
experiment is at long times close to the resonance where little is
known about the magnetic-field dependence of these rate constants.
Although the magnetic-field dependence has been calculated for a
shape resonance
\cite{moerdijk1996,fedichev1996,esry1999,braaten2001}, it is not
immediately obvious that the results carry over to the
multichannel situation of a Feshbach resonance. Moreover, precise
experimental data is unavailable \cite{roberts2000}. Therefore a
satisfying quantitative description is still lacking, although two
attempts have been made \cite{duine2003b,mackie2002}.

After these experiments, the same group performed an experiment
consisting of two short pulses in the magnetic field towards
resonance, separated by a longer evolution time \cite{donley2002}.
As a function of this evolution time an oscillation in the number
of condensate atoms was observed. Over the investigated range of
magnetic field during the evolution time, the frequency of this
oscillation agreed exactly with the molecular binding energy found
from a two-atom coupled-channels calculation
\cite{kokkelmans2002b}, indicating coherence between atoms and
molecules. Very recently, Claussen {\it et al.} have performed a
similar series of measurements over a larger range of magnetic
fields \cite{claussen2003}. It was found that close to resonance
the frequency of the oscillation deviates from the vacuum
molecular binding energy as a result of many-body effects
\cite{duine2003c,duine2003d}.

As already mentioned, the first theories for Feshbach-resonant
interactions introduce the physical picture of an interacting
atomic Bose-Einstein condensate coupled to a noninteracting
molecular condensate
\cite{drummond1998,timmermans1999a,timmermans1999b}. The first
description of the Ramsey experiments by Donley {\it et al.}
\cite{donley2002} was achieved within the Hartree-Fock-Bogoliubov
mean-field theory \cite{kokkelmans2002b,mackie2002,kohler2002}.

It turns out that, for a complete understanding of the
experiments, it is necessary to exactly incorporate the two-atom
physics into the theory. Although the above-mentioned theories
have provided a first understanding of the physics of a Bose gas
near a Feshbach resonance, these many-body theories do not contain
the two-atom collision properties exactly. To incorporate the
two-atom physics exactly, it is from a diagrammatic point of view
required to sum all the ladder Feynman diagrams of the microscopic
theory. By means of this procedure, we have recently derived an
effective quantum field theory describing the many-body properties
of an atomic gas near a Feshbach resonance \cite{duine2003a}. It
is the aim of this paper to review and extend this effective
atom-molecule theory and its applications
\cite{duine2003a,duine2003c,duine2003d}. Moreover, along the way
we discuss some of the differences and similarities between our
theory and a number of other theories for Feshbach-resonant
interactions in atomic Bose gases
\cite{drummond1998,timmermans1999a,timmermans1999b,kokkelmans2002b,mackie2002,kohler2002,kheruntsyan1998a,calsamiglia2001,mackie2002b,kohler2002b,kohler2002c}.

With this objective in mind, this paper is organized as follows.
In Section~\ref{sec:scattering} we review two-atom scattering
theory. In particular, we emphasize the relation between the
scattering amplitude of a potential and its bound states. Both the
single-channel case, as well as the multichannel case that can
give rise to Feshbach resonances, are discussed. This introductory
section introduces many important concepts in a simple setting,
and hence clarifies much of the physics that is discussed in later
sections. In Section~\ref{sec:many-body} we present in detail the
derivation of an effective quantum field theory applicable for
studying many-body properties of the system, starting from the
microscopic atomic hamiltonian for a Feshbach resonance. This
effective field theory consists of an atomic quantum field that is
coupled to a molecular quantum field responsible for the Feshbach
resonance. It is used in Section~\ref{sec:normalstate} to study
the normal state of the gas. In particular, we show here that the
two-atom scattering properties as well as the molecular binding
energy are correctly incorporated into the theory. Moreover, we
also discuss many-body effects on the molecular binding energy.
Section~\ref{sec:meanfieldbec} is devoted to the discussion of the
Bose-Einstein condensed phase of the gas. We derive the mean-field
theory resulting from our quantum field theory. We also discuss
the differences and similarities between this mean-field theory
and in particular the mean-field theories that were recently
proposed by Kokkelmans and Holland \cite{kokkelmans2002b}, Mackie
{\it et al.} \cite{mackie2002}, and K\"ohler {\it et al.}
\cite{kohler2002}. In Section~\ref{sec:oscillations} our
mean-field theory is applied to the two-pulse experiments
\cite{donley2002,claussen2003}. It is the perfect agreement
between theory and experiment obtained in this section that
ultimately justifies the {\it ab initio} approach to Bose gases
near a Feshbach resonance reviewed in this paper. We end in
Section~\ref{sec:concl} with our conclusions.

  \section{Scattering and bound states} \label{sec:scattering} In
this section we give a review of quantum-mechanical scattering
theory. We focus on the relation between the scattering amplitude
of a potential and its bound states \cite{sakuraibook,bjbook}. In
the first part we consider single-channel scattering and focus on
the example of the square well. In the second part we consider the
situation of two coupled channels, which can give rise to a
Feshbach resonance.

\subsection{Single-channel scattering: an example}
\label{subsec:singlechannel} 
\begin{figure}
\begin{center}
\includegraphics{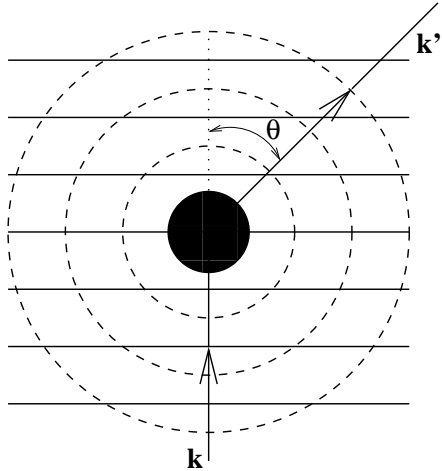}
\caption{\label{fig:scattering}
   Schematic representation of two-atom scattering in the
  center-of-mass reference frame. The atoms are initially
  in a plane-wave state with relative
  momentum $\hbar \bk$, and scatter into the spherical wave with
  relative momentum $\hbar \bk'$. Due to energy conservation we
  have that $k=k'$. The angle between $\bk$ and $\bk'$ is denoted
  by $\theta$. The region where the interaction takes place is
  indicated by the black circle.
   }
\end{center}
\end{figure}

We consider the situation of two
atoms of mass $m$ that interact via the potential $V(\br)$ that
vanishes for large distances between the atoms. The motion of the
atoms separates into the trivial center-of-mass motion and the
relative motion, described by the wave function $\psi (\br)$ where
$\br \equiv \bx_1-\bx_2$, and $\bx_1$ and $\bx_2$ are the
coordinates of the two atoms, respectively. This wave function is
determined by the time-independent Schr\"odinger equation
\begin{equation}
\label{eq:tindepse}
  \left[ -\frac{\hbar^2 {\bf \nabla}^2}{m} + V(\br) \right] \psi (\br) = E \psi
  (\br)~,
\end{equation}
with $E$ the energy of the atoms in the center-of-mass system.
Solutions of the Schr\"odinger equation with negative energy
correspond to bound states of the potential, i.e., to molecular
states. To describe atom-atom scattering we have to look for
solutions with positive energy $E=2 \epsilon_{\bk}$, with
$\epsilon_\bk \equiv \hbar^2 \bk^2/2m$ the kinetic energy of a
single atom with momentum $\hbar \bk$. Since any realistic
interatomic interaction potential vanishes rapidly as the distance
between the atoms becomes large, we know that the solution for $r
\to \infty$ of Eq.~(\ref{eq:tindepse}) is given by a superposition
of incoming and outgoing plane waves. More precisely, the
scattering wave function is given by an incoming plane wave and an
outgoing spherical wave and reads
\begin{eqnarray}
\label{eq:scatwf}
  \psi (\br)  \sim e^{i\bk\cdot\br} + f(\bk',\bk) \frac{e^{ik' r}}{r}~,
\end{eqnarray}
where the function $f(\bk',\bk)$ is known as the scattering
amplitude. The interatomic interaction potential depends only on
the distance between the atoms and hence the scattering amplitude
depends only on the angle $\theta$ between $\bk$ and $\bk' \equiv
k' \hat \br$, and the magnitude $k$. Because of energy
conservation we have that $k'=k$. The situation is shown
schematically in Fig.~\ref{fig:scattering}.

Following the partial-wave method we expand the scattering
amplitude in Legendre polynomials $P_{l} (x)$ according to
\begin{equation}
  f (\bk',\bk) = \sum_{l=0}^{\infty} f_l (k) P_l (\cos \theta)~.
\end{equation}
The wave function is expanded in a similar manner as
\begin{equation}
  \psi (r,\theta) = \sum_{l=0}^{\infty} R_l (k,r) P_l (\cos \theta)~,
\end{equation}
with  $R_l(k,r)=u_l(k,r)/r$ the radial wave function and
$u_l(k,r)$ determined by the radial Schr\"odinger equation
\begin{equation}
\label{eq:serad}
  \left[
  \frac{d^2}{dr^2}-\frac{l(l+1)}{r^2}-\frac{mV(r)}{\hbar^2}+k^2
  \right] u_l (k,r) = 0~.
\end{equation}
By expanding also the incident plane wave in partial waves according to
\begin{equation}
\label{eq:plwave}
  e^{i\bk \cdot \br} = \sum_{l=0}^\infty \frac{(2l+1)i^l}{kr} \sin
  \left(kr-\frac{l\pi}{2} \right) P_l (\cos \theta)~,
\end{equation}
we can show that to obey the boundary condition in
Eq.~(\ref{eq:scatwf}), the partial-wave amplitudes $f_l (k)$ have
to be of the form
\begin{equation}
\label{eq:partialwaveampl}
  f_l (k) = \frac{2l+1}{2ik} \left( e^{2i\delta_l (k)}-1\right)~,
\end{equation}
where $\delta_l (k)$ is the so-called phase shift of the $l$-th partial
wave.

For the ultracold alkali atoms, we are allowed to consider only
$s$-wave $(l=0)$ scattering, since the colliding atoms have too
low energies to penetrate the centrifugal barrier in the effective
hamiltonian in Eq.~(\ref{eq:serad}). Moreover, as we see later on,
the low-energy effective interactions between the atoms are fully
determined by the $s$-wave scattering length, defined by
\begin{equation}
\label{eq:defa}
  a = - \lim_{k\downarrow0} \frac{\delta_0 (k)}{k}~.
\end{equation}
From Eq.~(\ref{eq:partialwaveampl}) we find that the $s$-wave
scattering amplitude is given by
\begin{equation}
\label{eq:swavescattampl}
  f_0 (k) = \frac{1}{k \cot \delta_0 (k) -ik}~.
\end{equation}
As explained above, we take only the $s$-wave contribution into
account, which gives for the scattering amplitude at zero-momentum
\begin{equation}
\label{eq:swavescattamplzerok}
 f ({\bf 0},{\bf 0}) \simeq -a~.
\end{equation}

To illustrate the physical meaning of the $s$-wave scattering
length, we now calculate it explicitly for the simple case that
the interaction potential is a square well. We thus take the
interaction potential of the form
\begin{eqnarray}
\label{eq:squarewellpot}
  V(r) = \left\{
         \begin{array}{ll}
     V_0 & \mbox{if $r<R$}; \\
     0 & \mbox{if $r>R$},
     \end{array}
         \right.
\end{eqnarray}
with $R>0$. With this potential, the general solution of
Eq.~(\ref{eq:serad}) for $l=0$ is given by
\begin{eqnarray}
\label{eq:usquarewell}
  \begin{array}{ll}
  u^< (r) = A e^{i k^< r} + B e^{-i k^< r}, & \mbox{for $r<R$}; \\
  u^> (r) = C e^{i k r} + D e^{-i k r},  & \mbox{for $r>R$},
  \end{array}
\end{eqnarray}
with $k^< = \sqrt{k^2-mV_0/\hbar^2}$. Since the wave function
$\psi(r)$ has to obey the Schr\"odinger equation at the origin we
have to demand that the function $u^{<} (r)$ vanishes at this
point. This leads to the boundary condition $B=-A$. By comparing
the explicit form of the wave function $u^> (r)$ with the $s$-wave
component of the general scattering wave function for $r \to
\infty$, we find that
\begin{equation}
\label{eq:phaseshiftcd}
   e^{2 i \delta_0 (k)}=-\frac{C}{D}~.
\end{equation}
Hence, we determine the phase shift by demanding that the wave functions
for $r<R$ and $r>R$ join smoothly. This leads to the equations
\begin{eqnarray}
  A \left( e^{i k^< R} - e^{-i k^< R} \right) &=& -e^{2i\delta_0(k)} e^{i k R} + e^{-i k R},
  \nonumber \\
  A \left( k^< e^{i k^< R} + k^< e^{-i k^< R} \right)&=&  -e^{2 i \delta_0 (k)}k e^{i k R} - k
  e^{-ikR}~,
\end{eqnarray}
where we have chosen the normalization such that $D=1$. Multiplication of
the above equations with $e^{-i\delta_0 (k)}$ and dividing the result leads
to
\begin{equation}
  k \tan (k^< R) = k^< \tan (\delta_0 (k) + k R)~,
\end{equation}
from which it follows that
\begin{equation}
\label{eq:phaseshiftsqw}
  \delta_0 (k) = -kR + \tan^{-1} \left[ \frac{k}{k^<} \tan (k^< R) \right]~.
\end{equation}
Note that for a repulsive hard-core potential we have that $V_0
\to \infty$ and therefore, with the use of the definition in
Eq.~(\ref{eq:defa}), that the scattering length $a=R$. This
immediately gives a physical picture for a positive $s$-wave
scattering length: at low energy and momenta the details of the
potential are unimportant and we are allowed to model the
potential with an effective hard-core potential of radius $a$. For
a fully repulsive potential the scattering length is always
positive. For a potential with attractive parts the scattering
length can be both negative and positive, corresponding to
attractive and repulsive effective interactions, respectively.

\begin{figure}
\begin{center}
\includegraphics{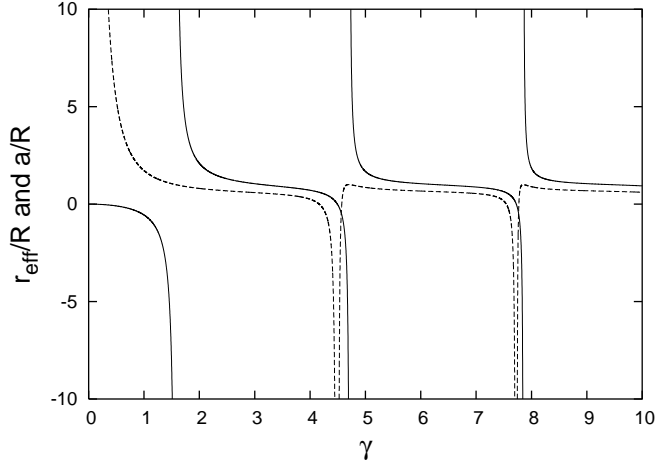}
\caption{\label{fig:asqw}
   Scattering length (solid line) and effective range (dashed line)
  for an attractive square well in units of the range of
  the potential, as a function of the
  dimensionless parameter $\gamma = R \sqrt{m|V_0|/\hbar^2}$.
   }
\end{center}
\end{figure}

This is seen by explicitly calculating the scattering length for
our example in the case that $V_0<0$.  As its definition in
Eq.~(\ref{eq:defa}) shows, the scattering length is determined by
the linear dependence of the phase shift on the magnitude of the
relative momentum $\hbar k$ of the scattering atoms for small
momentum. Generally, the phase shift can be expanded  according to
\cite{sakuraibook,bjbook,stoof1988}
\begin{equation}
\label{eq:deltaexpansion}
  k \cot (\delta_0 (k)) = - \frac{1}{a} + \frac{1}{2} r_{\rm eff} k^2 +
  \cdots~,
\end{equation}
from which the scattering length is determined by
\begin{equation}
\label{eq:asqw}
  a = R \left( 1 - \frac{\tan \gamma}{\gamma}\right)~,
\end{equation}
with $\gamma=R\sqrt{m|V_0|/\hbar^2}$ a dimensionless constant. The
parameter $r_{\rm eff}$ is the so-called effective range and is,
in our example of the square-well potential, given by
\begin{equation}
  r_{\rm eff} = R \left[1+ \frac{3 \tan \gamma- \gamma (3+\gamma^2)}
  {3 \gamma (\gamma-\tan \gamma)^2} \right]~.
\end{equation}
In Fig.~\ref{fig:asqw} the scattering length is shown as a
function of $\gamma$ by the solid line. Clearly, the scattering
length can be both negative and positive, and becomes equal to
zero at values of $\gamma$ such that $\gamma=\tan \gamma$. In the
same figure, the effective range is shown by the dashed line. Note
that the effective range diverges if the scattering length becomes
equal to zero. This is because the expansion in
Eq.~(\ref{eq:deltaexpansion}) is ill-defined for $a=0$. At values
of $\gamma=(n+1/2)\pi$ with $n$ a positive integer the scattering
length diverges and changes sign. This behaviour is called a
potential or shape resonance and in fact occurs each time the
potential is just deep enough to support a new bound state.
Therefore, for large and positive scattering length the square
well has a bound state with an energy just below the continuum
threshold. It turns out that there is an important relationship
between the energy of this bound state and the scattering length.

To find this relation we have to determine the bound-state energy
by solving the  Sch\"odinger equation for negative energy
$V_0<E<0$. This leads to solutions
\begin{eqnarray}
  \begin{array}{ll}
  u^< (r) =A \left (e^{i k^< r} - e^{-i k^< r}\right),& \mbox{for $r<R$};
  \\
  u^{>} (r) = B e^{- \kappa r},& \mbox{for $r>R$},
  \end{array}
\end{eqnarray}
with $k^<=\sqrt{m(E-V_0)/\hbar^2}$ and
$\kappa=\sqrt{m|E|/\hbar^2}$. Demanding again that these solutions
join smoothly at $r=R$, we find the equation for the bound-state
energy
\begin{equation}
\label{eq:boundstatee}
  \sqrt{\frac{m}{\hbar^2}|E_{\rm m}|}=-\sqrt{\frac{m}{\hbar^2}(E_{\rm m}-V_0)}
  \cot \left( \sqrt{\frac{m}{\hbar^2}(E_{\rm m}-V_0)}\right)~.
\end{equation}
We can show that for values of $\gamma$ such that
$(n-1/2)\pi<\gamma<(n+1/2) \pi$ this equation has $n$ solutions
for $V_0<E_{\rm m}<0$ \cite{bjbook}.

For small binding energy $|E_{\rm m}| \ll |V_0|$  we have from the
equation for the bound-state energy that
\begin{equation}
 \sqrt{\frac{m}{\hbar^2}|E_{\rm m}|} \simeq -\gamma \cot \gamma/R \simeq 1/a~,
\end{equation}
where we made use of the fact that $\gamma$ has to be close to the
resonant values $(n+1/2) \pi$ in this case. This leads to the
desired relation between the energy of the molecular state and the
scattering length given by
\begin{equation}
\label{eq:easqw}
 E_{\rm m}=-\frac{\hbar^2}{ma^2}~.
\end{equation}
This result does not depend on the specific details of the
potential and it turns out to be quite general. Any potential with
a large positive scattering length has a bound state just below
the continuum threshold with energy given by Eq.~(\ref{eq:easqw}).
Moreover, the relation will turn out to hold also in the
multichannel case of a Feshbach resonance as we will see in
Section~\ref{subsec:examplefb}. Before discussing this situation,
we first turn to some concepts of scattering theory which are of
importance for the remainder of this paper.

\subsection{Single-channel scattering: formal treatment}
\label{subsec:formal}

Let us give a more formal treatment of the scattering theory
described above. In a basis-independent formulation the
Schr\"odinger equation we have solved reads
\begin{equation}
\label{eq:sebasisindep}
  \left[ \hat H_0 + \hat V \right] | \psi \rangle = E | \psi \rangle~,
\end{equation}
with $\hat H_0 = \hat p^2/m$ the relative kinetic energy operator
for the atoms. To describe scattering, we have to look for
solutions which asymptotically represent an incoming plane wave,
and an outgoing spherical wave. In the absence of the potential
$\hat V$ there is no scattering, and hence we demand that the
solution of Eq.~(\ref{eq:sebasisindep}) reduces to a plane wave in
the limit of vanishing potential. The formal solution that obeys
this condition is given by
\begin{equation}
\label{eq:formalsolscatt}
  | \psi^{(+)}_\bk \rangle = | \bk \rangle + \frac{1}{E^+-\hat H_0}\hat
  V | \psi^{(+)}_\bk\rangle~,
\end{equation}
where $|\bk\rangle$ represents the incoming plane wave and we
recall that $E=2\epsilon_\bk$ is the kinetic energy of the atoms.
This energy is made slightly complex by the usual limiting
procedure $E^+\equiv  \lim_{\eta \downarrow 0} E+ i \eta$.
Moreover, we have for the scattering amplitude that
\begin{equation}
\label{eq:scattamplidentification}
  f (\bk',\bk) = -  \frac{m}{4 \pi \hbar^2}
    \langle \bk' | \hat V | \psi^{(+)}_\bk \rangle~.
\end{equation}

To determine the scattering amplitude directly, we introduce the
two-body T(ransition) matrix by means of
\begin{equation}
\label{eq:deftmatrix}
  \hat V | \psi^{(+)}_\bk \rangle = \hat T^{\rm 2B} (E^+) | \bk
  \rangle~.
\end{equation}
Multiplying the formal solution in Eq.~(\ref{eq:formalsolscatt})
by $\hat V$ we have that
\begin{equation}
 \hat T^{\rm 2B} (E^+)|\bk\rangle=\hat V|\bk\rangle+\hat V \frac{1}{E^+-\hat
 H_0}\hat T^{\rm 2B} (E^+) |\bk \rangle~.
\end{equation}
Since this equation holds for an arbitrary plane wave
$|\bk\rangle$ and because these plane waves form a complete set of
states we have the following operator equation for the two-body
T-matrix
\begin{equation}
\label{eq:lippmannschwinger1}
  \hat T^{\rm 2B} (z) = \hat V + \hat V \frac{1}{z-\hat H_0} \hat T^{\rm
  2B} (z)~.
\end{equation}
This equation is called the Lippmann-Schwinger equation and from
its solution we are able to determine the scattering properties of
the potential $\hat V$. To see this we first note that from the
definition of the T-matrix in Eq.~(\ref{eq:deftmatrix}), together
with Eq.~(\ref{eq:scattamplidentification}), it follows
immediately that
\begin{equation}
\label{eq:t2bandf}
  f(\bk',\bk) =  - \frac{m}{4 \pi \hbar^2}
    \langle \bk' | \hat T^{\rm 2B} (2 \epsilon_\bk^+)|\bk \rangle~.
\end{equation}
Therefore, we indeed see that the two-body T-matrix completely
determines the scattering amplitude. The Lippmann-Schwinger
equation for the two-body T-matrix can be solved in perturbation
theory in the potential. This results in the so-called Born series
given by
\begin{eqnarray}
\label{eq:born2b}
   \hat T^{\rm 2B} (z) = \hat V + \hat V
\hat G_0(z) \hat V + \hat V
 \hat G_0 (z) \hat V
  \hat G_0 (z) \hat V + \cdots~,
\end{eqnarray}
where
\begin{equation}
  \hat G_0 (z) = \frac{1}{z- \hat H_0}~,
\end{equation}
is the noninteracting propagator of the atoms. By using, instead
of the true interatomic interaction potential, a pseudopotential
of the form
\begin{equation}
\label{eq:pseudopot}
  V (\bx-\bx') = \frac{4 \pi a \hbar^2}{m} \delta (\bx-\bx')~,
\end{equation}
the first term in the Born series immediately yields the correct
result for the scattering amplitude at low energies and momenta,
given in Eq.~(\ref{eq:swavescattamplzerok}). Such a
pseudopotential should therefore not be used to calculate
higher-order terms in the Born series, but should be used only in
first-order perturbation theory.

The poles of the T-matrix in the complex-energy plane correspond
to bound states of the potential. To see this we note that the
formal solution of the Lippmann-Schwinger equation is given by
\begin{equation}
\label{eq:lippmannschwformalsol}
  \hat T^{\rm 2B} (z) = \hat V + \hat V \frac{1}{z-\hat H} \hat
  V~.
\end{equation}
After insertion of the complete set of eigenstates $|
\psi_{\alpha} \rangle$ of $\hat H = \hat H_0 +\hat V$ we have
\begin{equation}
\label{eq:t2banalytic}
 \hat T^{\rm 2B} (z) = \hat V + \sum_\alpha \hat V \frac{| \psi_\alpha
 \rangle \langle \psi_\alpha |}{z-\epsilon_\alpha} \hat
  V~,
\end{equation}
where the summation over $\alpha$ is discrete for the bound-state
energies $\epsilon_\alpha <0$, and represents an integration for
positive energies that correspond to scattering solutions of the
Schr\"odinger equation, so explicitly we have that
\begin{eqnarray}
   \hat T^{\rm 2B} (z) = \hat V + \sum_\kappa \hat V \frac{|
   \psi_\kappa
 \rangle \langle \psi_\kappa |}{z-\epsilon_\kappa} \hat
  V~ + \int \frac{d \bk}{(2\pi)^3} \hat V  \frac{|
   \psi^{(+)}_\bk
 \rangle \langle \psi^{(+)}_\bk |}{z-2\epsilon_\bk} \hat V.
\end{eqnarray}
From this equation we clearly see that the two-body T-matrix has
poles in the complex-energy plane, corresponding to the bound
states of the potential. In addition, the T-matrix contains a
branch cut on the positive real axis due to the continuum of
scattering states.

As an example, we note that for $s$-wave scattering the T-matrix
$T^{\rm 2B} (2 \epsilon_\bk^+)  \equiv \langle \bk' | \hat T^{\rm
2B} (2 \epsilon_\bk^+) | \bk \rangle$ is independent of the angle
between $\bk'$ and $\bk$. From the relation between the T-matrix
and the scattering amplitude, and the expression for the latter in
terms of the phase shift, we have for low positive  energies
\begin{eqnarray}
\label{eq:tmatrixexplicit}
  T^{\rm 2B} (E^+)
  &=& -\frac{4 \pi \hbar^2}{m}
     \frac{1}{\sqrt{\frac{mE}{\hbar^2}}
     \cot \left(\delta \left(\sqrt{\frac{mE}{\hbar^2}}\right)\right)
     - i \sqrt{\frac{mE}{\hbar^2}}}
  \nonumber \\
  &\simeq& \frac{4 \pi a \hbar^2}{m} \left[ \frac{1}{1+ia\sqrt{\frac{mE}{\hbar^2}}
  -\frac{ a r_{\rm eff}
  mE}{2 \hbar^2}} \right]~,
\end{eqnarray}
where we made use of the expansion in
Eq.~(\ref{eq:deltaexpansion}). From this result we deduce by
analytic continuation that
\begin{equation}
\label{eq:tmatrixcomplexplane}
  T^{\rm 2B} (z) \simeq
  \frac{4 \pi a \hbar^2}{m} \left[ \frac{1}{1-a\sqrt{-\frac{mz}{\hbar^2}}
  -\frac{ a r_{\rm eff}
  mz}{2 \hbar^2}} \right]~.
\end{equation}
Clearly, for large and positive scattering length the T-matrix has
a pole at negative energy $E_{\rm m}~=~-\hbar^2/ma^2$, in complete
agreement with our previous discussions.

Summarizing, we have found that the scattering length of an
attractive potential well can have any value and depends strongly
on the energy of the weakliest bound state in the potential. In
principle therefore, if we have experimental access to the energy
difference of this bound state and the continuum threshold we are
able to experimentally alter the scattering length and thereby the
effective interactions of the atoms. In the single-channel case
this is basically impossible to achieve. In a multichannel system,
however, the energy difference is experimentally accessible, which
makes the low-energy effective interactions between the atoms
tunable. In the next section we discuss this situation.

\subsection{Example of a Feshbach resonance}
\label{subsec:examplefb} We consider now the situation of
atom-atom scattering where the atoms have two internal states
\cite{other}. These states correspond, roughly speaking, to the
eigenstates of the spin operator ${\bf S}$ of the valence electron
of the alkali atoms. The effective interaction potential between
the atoms depends on the state of the valence electrons of the
colliding atoms. If these form a singlet the electrons are in
principle allowed to be on top of each other. For a triplet this
is forbidden. Hence, the singlet potential is generally much
deeper than the triplet potential.

Of course, in reality the atom also has a nucleus with spin ${\bf
I}$ which interacts with the spin of the electron via the
hyperfine interaction
\begin{equation}
\label{eq:hyperfineint}
  V_{\rm hf} = \frac{a_{\rm hf}}{\hbar^2} {\bf I} \cdot {\bf S},
\end{equation}
with $a_{\rm hf}$ the hyperfine constant. The hyperfine
interaction couples the singlet and triplet states. Moreover, in
the presence of a magnetic field the different internal states of
the atoms have a different Zeeman shift. In an experiment with
magnetically-trapped gases, the energy difference between these
states is therefore experimentally accessible. Putting these
results together, we can write down the Sch\"odinger equation that
models the above physics
\begin{eqnarray}
\label{eq:feshbachham}
  \left(
    \begin{array}{cc}
      -\frac{\hbar^2 {\bf \nabla}^2}{m}+ V_{\rm T} (\br)\!-\!E& V_{\rm hf}\\
      V_{\rm hf} & -\frac{\hbar^2 {\bf \nabla}^2}{m}+ \Delta \mu B +V_{\rm S} (\br)\!-\!E
    \end{array}
  \right)
  \left( \!
   \begin{array}{c}
     \psi_{\rm T} (\br)\\
     \psi_{\rm S} (\br)
     \end{array}
  \! \right)=0~.
\end{eqnarray}
Here, $V_{{\rm T}} (\br)$ and $V_{\rm S} (\br)$ are the
interaction potentials of atoms with internal state $| {\rm T}
\rangle$ and $| {\rm S} \rangle$, respectively, and $\Delta \mu B$
is their difference in Zeeman energy due to the interaction with
the magnetic field $B$, with $\Delta \mu$ the difference in
magnetic moment. In agreement with the above remarks, $|{\rm
T}\rangle$ is referred to as the triplet channel, whereas $|{\rm
S}\rangle$ is referred to as the singlet channel. The potentials
$V_{{\rm T}} (\br)$ and $V_{{\rm S}} (\br)$ are the triplet and
singlet interaction potentials, respectively.

\begin{figure}
\begin{center}
\includegraphics[width=7cm]{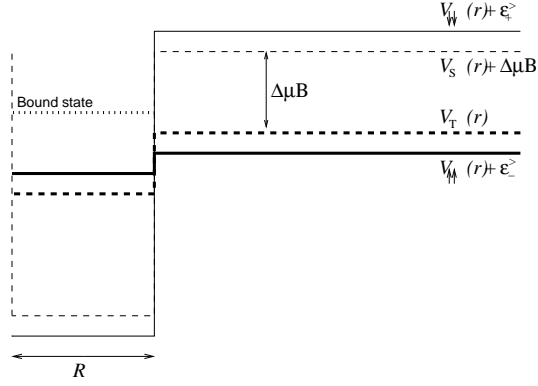}
\caption{\label{fig:fres_sqw}
    Feshbach resonance in a two-channel system with square-well interaction potentials.
   The triplet potential $V_{\rm T} (r)$ is indicated by the thick
   dashed line. The singlet potential that contains the bound
   state responsible for the Feshbach resonance is indicated by
   the thin dashed line. Due to the Zeeman interaction with the magnetic field, the energy
   difference between the singlet and triplet is equal to $\Delta \mu
   B$. The interactions in the open and closed hyperfine channels
   are indicated by $V_{\uparrow\uparrow} (r)$ and
   $V_{\downarrow\downarrow} (r)$, respectively.
   }
\end{center}
\end{figure}

As a specific example, we use for both interaction potentials again square
well potentials,
\begin{equation}
\label{eq:squarewellmultichan} V_{{\rm T},{\rm S}}(r) = \left\{
         \begin{array}{ll}
     -V_{{\rm T},{\rm S}} & \mbox{if $r<R$} \\
     0 & \mbox{if $r>R$}
     \end{array}
         \right.,
\end{equation}
where $V_{{\rm T},{\rm S}}>0$. For convenience we have taken the
range the same for both potentials. Furthermore, we assume that
the potentials are such that $V_{{\rm T}} < V_{\rm S}$ and that
$V_{{\rm S}}$ is just deep enough such that it contains exactly
one bound state. Finally, we assume that $0<V_{\rm hf} \ll V_{{\rm
T}},V_{{\rm S}},\Delta \mu B$. The potentials are shown in
Fig.~\ref{fig:fres_sqw}.

To discuss the scattering properties of the atoms, we have to
diagonalize the hamiltonian for $r>R$, in order to determine the
incoming channels, which are superpositions of the triplet and
singlet states $| {\rm T} \rangle$ and $| {\rm S} \rangle$. Since
the kinetic energy operator is diagonal in the internal space of
the atoms, we have to find the eigenvalues of the hamiltonian
\begin{eqnarray}
\label{eq:outsideham}
  H^> = \left(
    \begin{array}{cc}
     0 & V_{\rm hf}\\
      V_{\rm hf} & \Delta \mu B
    \end{array}
  \right)~.
\end{eqnarray}
These are given by
\begin{equation}
\label{eq:lambdabig}
  \epsilon_{\pm}^> = \frac{\Delta \mu B}{2} \pm \frac{1}{2}\sqrt{(\Delta \mu
  B)^2+(2V_{\rm hf})^2}.
\end{equation}
The hamiltonian $H^>$ is diagonalized by the matrix
\begin{eqnarray}
  Q(\theta) = \left(
    \begin{array}{cc}
      \cos \theta& \sin \theta\\
      -\sin \theta & \cos \theta
    \end{array}
  \right)~,
\end{eqnarray}
according to
\begin{eqnarray}
  Q(\theta^>) H^> Q^{-1} (\theta^>) =
   \left(
    \begin{array}{cc}
      \epsilon_{-}^>& 0\\
     0 & \epsilon_{+}^>
    \end{array}
  \right)~,
\end{eqnarray}
which determines $\tan \theta^>= -2 V_{\rm hf}/\Delta \mu B$. We
define now the hyperfine states $| \uparrow \uparrow \rangle$ and
$| \downarrow \downarrow \rangle$ according to
\begin{eqnarray}
\label{eq:defupdown}
  \left(
    \begin{array}{c}
     | \uparrow \uparrow\rangle \\
     | \downarrow \downarrow \rangle
    \end{array}
  \right)
  = Q(\theta^>)
  \left(
    \begin{array}{c}
     | {\rm T} \rangle \\
     | {\rm S} \rangle
    \end{array}
  \right) ~,
\end{eqnarray}
which asymptotically represent the scattering channels. In this
basis the Schr\"odinger equation for all $\br$ reads
\begin{eqnarray}
\label{eq:hffeshbachham}
  \left(
    \begin{array}{cc}
      -\frac{\hbar^2 {\bf \nabla}^2}{m}+V_{\uparrow\uparrow}(\br) -E&
      V_{\uparrow\downarrow} (\br) \\
      V_{\uparrow\downarrow} (\br) & -\frac{\hbar^2 {\bf \nabla}^2}{m}+
      \epsilon_{+}^>-\epsilon_{-}^>
      +V_{\downarrow\downarrow} (\br)-E
    \end{array}
  \right) \nonumber \\
  \times \left(
   \begin{array}{c}
     \psi_{\uparrow\uparrow} (\br)\\
     \psi_{\downarrow\downarrow} (\br)
     \end{array}
  \right)=0~,
\end{eqnarray}
where the energy $E$ is measured with respect to $\epsilon_{-}^>$
and we have defined the potentials according to
\begin{equation}
\label{eq:potmatrix}
\left(
    \begin{array}{cc}
      V_{\uparrow\uparrow}(\br) &
      V_{\uparrow\downarrow} (\br) \\
      V_{\uparrow\downarrow} (\br) & V_{\downarrow\downarrow} (\br)
    \end{array}
  \right)
  =Q (\theta^>)
    \left(
    \begin{array}{cc}
      V_{{\rm T}}(\br) &
     0  \\
      0 & V_{{\rm S}} (\br)
    \end{array}
  \right)
  Q^{-1} (\theta^>)~.
\end{equation}
Since all these potentials vanish for $r > R$ we can study
scattering of atoms in the states $| \uparrow \uparrow \rangle$
and $| \downarrow \downarrow \rangle$.  Because the hyperfine
interaction $V_{\rm hf}$ is small we have that $\epsilon_{+}^>
\simeq \Delta \mu B$ and $\epsilon_{-}^> \simeq 0$. Moreover, for
the experiments with magnetically-trapped gases we always have
that $\Delta \mu B \gg k_{\rm B} T$ where $k_{\rm B}$ is
Boltzmann's constant and is $T$ the temperature. This means that
in a realistic atomic gas, in which the states $| \uparrow
\uparrow\rangle$ and $| \downarrow \downarrow\rangle$ are
available, there are in equilibrium almost no atoms that scatter
via the latter state. Because of this, the effects of the
interactions of the atoms will be determined by the scattering
amplitude in the state $| \uparrow \uparrow \rangle$. If two atoms
scatter in this channel with energy $E \simeq k_B T \ll \Delta \mu
B$ they cannot come out in the other channel because of energy
conservation. Therefore, the indices $ \uparrow \uparrow$ refers
to an open channel, whereas $\downarrow \downarrow$ is associated
with a closed channel. The situation is further clarified in
Fig.~\ref{fig:fres_sqw}.

To calculate the $s$-wave scattering length in the open channel we
have to solve the Schr\"odinger equation. In the region $r>R$ the
solution is of the from
\begin{eqnarray}
\label{eq:hfsoloutside}
 \left(
   \begin{array}{c}
     u_{\uparrow \uparrow}^> (r)\\
     u_{\downarrow \downarrow}^> (r)
     \end{array}
  \right)
  =\left(
   \begin{array}{c}
     C e^{ikr} + D e^{-ikr}\\
     F e^{-\kappa r}
     \end{array}
  \right)~,
\end{eqnarray}
where $\kappa=
\sqrt{m(\epsilon_{+}^>-\epsilon_{-}^>)/\hbar^2-k^2}$ and, because
we have used the same notation as in Eq.~(\ref{eq:usquarewell}),
the $s$-wave phase shift is again determined by
Eq.~(\ref{eq:phaseshiftcd}). In the region $r<R$ the solutions are
of the form
\begin{eqnarray}
\label{eq:hfsolinside}
 \left(
   \begin{array}{c}
     u^<_{\uparrow \uparrow} (r)\\
     u_{\downarrow \downarrow}^< (r)
     \end{array}
  \right)
  =\left(
   \begin{array}{c}
     A \left( e^{ik_{\uparrow \uparrow}^< r} -  e^{-ik_{\uparrow \uparrow}^< r} \right)\\
     B \left( e^{ik_{\downarrow \downarrow}^< r} -  e^{-ik_{\downarrow\downarrow}^< r} \right)
     \end{array}
  \right)~,
\end{eqnarray}
where
\begin{eqnarray}
k_{\uparrow\uparrow}^< =
\sqrt{m(\epsilon_{-}^>-\epsilon_{-}^<)/\hbar^2+k^2}~;
\nonumber \\
k_{\downarrow\downarrow}^< =
\sqrt{m(\epsilon_{-}^>-\epsilon_{+}^<)/\hbar^2+k^2}~,
\end{eqnarray}
and
\begin{equation}
 \epsilon_{\pm}^< = \frac{\Delta \mu B-V_{\rm T}-V_{\rm S}}{2} \mp \frac{1}{2}
 \sqrt{(V_{\rm S}-V_{\rm T}-\Delta \mu
  B)^2+(2V_{\rm hf})^2}.
\end{equation}
are the eigenvalues of the matrix
\begin{eqnarray}
\label{eq:insideham}
  H^< = \left(
    \begin{array}{cc}
     -V_{\rm T} & V_{\rm hf}\\
      V_{\rm hf} & \Delta \mu B - V_{\rm S}
    \end{array}
  \right)~.
\end{eqnarray}

\begin{figure}
\begin{center}
\includegraphics{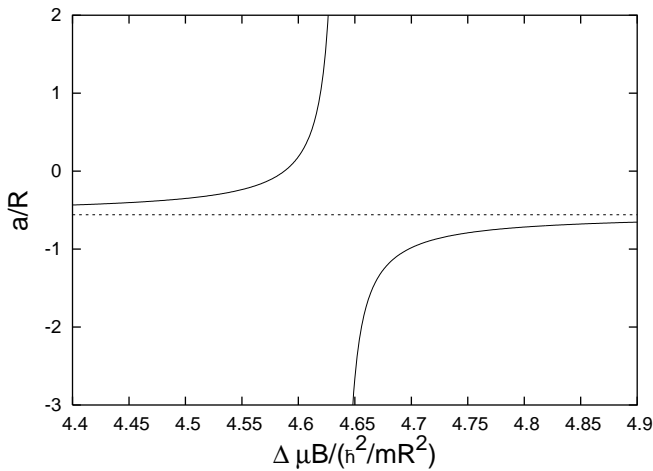}
\caption{\label{fig:asqw_feshbach}
   Scattering length for two coupled square-well potentials as a function
  of $\Delta \mu B$. The depth of the triplet and singlet channel potentials
  is $V_{\rm T} =\hbar^2/m R^2$ and $V_{\rm S} =10\hbar^2/m R^2$,
  respectively. The hyperfine coupling is $V_{\rm hf} =0.1
  \hbar^2/mR^2$. The dotted line shows the background scattering length
  $a_{\rm bg}$.
   }
\end{center}
\end{figure}

In order to determine the phase shift we have to join the solution
for $r<R$ and $r>R$ smoothly. This is done most easily by
transforming to the singlet-triplet basis $\left\{ |{\rm
T}\rangle,|{\rm S}\rangle\right\}$ since this basis is independent
of $r$. Demanding the solution to be continuously differentiable
leads to the equations
\begin{eqnarray}
\label{eq:joinsmoothfeshbach}
  Q^{-1} (\theta^<) \left(
   \begin{array}{c}
     u_{\uparrow\uparrow}^< (R)\\
     u_{\downarrow\downarrow}^< (R)
     \end{array}
  \right) &=&
  Q^{-1} (\theta^>)
  \left(
   \begin{array}{c}
     u_{\uparrow\uparrow}^> (R)\\
     u_{\downarrow\downarrow}^> (R)
     \end{array}
  \right)~; \nonumber \\
  \frac{\partial}{\partial r}\left. Q^{-1} (\theta^<) \left(
   \begin{array}{c}
     u_{\uparrow\uparrow}^< (r)\\
     u_{\downarrow\downarrow}^< (r)
     \end{array}
  \right) \right|_{r=R}&=&
  \frac{\partial}{\partial r}\left.Q^{-1} (\theta^>)
  \left(
   \begin{array}{c}
     u_{\uparrow\uparrow}^> (r)\\
     u_{\downarrow\downarrow}^> (r)
     \end{array}
  \right)\right|_{r=R}~,
\end{eqnarray}
where $\tan \theta^<=2V_{\rm hf}/(V_{\rm S}-V_{\rm T}-\Delta \mu
B)$. These four equations determine the coefficients $A,B,C,D$ and
$F$ up to a normalization factor, and therefore also the phase
shift and the scattering length. Although it is possible to find
an analytical expression for the scattering length as a function
of the magnetic field, the resulting expression is rather
formidable and is omitted here. The result for the scattering
length is shown in Fig.~\ref{fig:asqw_feshbach}, for $V_{\rm S}=10
\hbar^2/m R^2$, $V_{\rm T} =\hbar^2/mR^2$ and $V_{\rm
hf}=0.1\hbar^2/m R^2$, as a function of $\Delta \mu B$. The
resonant behaviour is due to the bound state of the singlet
potential $V_{{\rm S}} (r)$. Indeed, solving the equation for the
binding energy in Eq.~(\ref{eq:boundstatee}) with $V_0=-V_{\rm S}$
we find that $|E_{\rm m}|\simeq 4.62 \hbar^2/m R^2$, which is
approximately the position of the resonance in
Fig.~\ref{fig:asqw_feshbach}. The difference is due to the fact
that the hyperfine interaction leads to a shift in the position of
the resonance with respect to $E_{\rm m}$.

The magnetic-field dependence of the scattering length near a
Feshbach resonance is characterized experimentally by a width
$\Delta B$ and position $B_0$ according to
\begin{equation}
\label{eq:ascatofb}
  a(B) = a_{\rm bg} \left( 1-\frac{\Delta B}{B-B_0} \right).
\end{equation}
This explicitly shows that the scattering length, and therefore
the magnitude of the effective interatomic interaction, may be
altered to any value by tuning the magnetic field. The
off-resonant background scattering length is denoted by $a_{\rm
bg}$ and is, in our example, approximately equal to the scattering
length of the triplet potential $V_{\rm T} (r)$. Using the
expression for the scattering length of a square well in
Eq.~(\ref{eq:asqw}) for $\gamma=1$, we find that $a_{\rm bg}
\simeq -0.56 R$. Furthermore, we have for our example that the
position of the resonance is given by $B_0 \simeq 4.64 \hbar^2/m
\Delta \mu R^2$ and that the width is equal to  $\Delta B \simeq
-0.05 \hbar^2/m \Delta \mu R^2$.

\begin{figure}
\begin{center}
\includegraphics{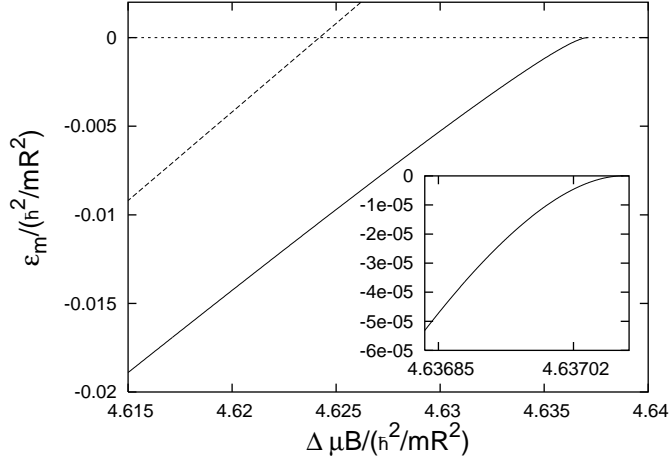}
\caption{\label{fig:esqw_feshbach}
  Bound-state energy of the molecular state near a Feshbach resonance for
 two coupled square-well interaction potentials. The solid line and the
 inset show the result for $V_{\rm hf}=0.1\hbar^2/mR^2$. The dashed line
 corresponds to $V_{\rm hf}=0$. The other parameters are the same as in
 Fig.~\ref{fig:asqw_feshbach}.
   }
\end{center}
\end{figure}

Next, we calculate the energy of the molecular state for the
coupled-channel case which is found by solving
Eq.~(\ref{eq:hffeshbachham}) for negative energy. In particular,
we are interested in its dependence on the magnetic field. In the
absence of the hyperfine coupling between the open and closed
channel we simply have that $\epsilon_{\rm m} (B) = E_{\rm m} +
\Delta \mu B$. Here, $E_{\rm m}$ is the energy of the bound state
responsible for the Feshbach resonance, that is determined by
solving the single-channel Sch\"odinger equation for the singlet
potential. This bound-state energy as a function of the magnetic
field is shown in Fig.~\ref{fig:esqw_feshbach} by the dashed line.
A nonzero hyperfine coupling drastically changes this result. For
our example the bound-state energy is easily calculated. The
result is shown by the solid line in Fig.~\ref{fig:esqw_feshbach}
for the same parameters as before. Clearly, close to the resonance
the dependence of the bound-state energy on the magnetic field is
no longer linear, as the inset of Fig~\ref{fig:esqw_feshbach}
shows. Instead, it turns out to be quadratic. Moreover, the
magnetic field $B_0$ where the bound-state energy is equal to zero
is shifted with respected to the case where $V_{\rm hf}=0$. It is
at this shifted magnetic field that the resonance is observed
experimentally. Moreover, for magnetic fields larger than $B_0$
there no longer exists a bound state and the molecule now decays
into two free atoms due to the hyperfine coupling, because its
energy is above the two-atom continuum threshold.

Close to resonance the energy of the molecular state turns out to be
related to the scattering length by
\begin{equation}
\label{eq:ebres}
  \epsilon_{\rm m}(B)  = - \frac{\hbar^2}{m [a(B)]^2}~,
\end{equation}
as in the single-channel case. As we will see in the next
sections, the reason for this is that close to resonance the
effective two-body T-matrix again has a pole at the energy in
Eq.~(\ref{eq:ebres}). This important result will be proven
analytically in Section~\ref{sec:normalstate}. First, we derive a
description of the Feshbach resonance in terms of coupled atomic
and molecular quantum fields.

  \section{Many-body theory for Feshbach-resonant interactions}
\label{sec:many-body}

In this section we derive the effective quantum field theory that
offers a description of Feshbach-resonant interactions in terms of
an atom-molecule hamiltonian. We start from a microscopic atomic
hamiltonian that involves atoms with two internal states, i.e., we
consider a situation with an open and a closed channel that are
coupled by the exchange interaction. The first step is to
introduce a quantum field that describes the bound state in the
closed channel, which is responsible for the Feshbach resonance.
This is achieved using functional techniques by a so-called
Hubbard-Stratonovich transformation and is described in detail in
Section~\ref{subsec:bareatommolecule}. This section is somewhat
technical and may be omitted in a first reading of this paper. The
most important result is a bare atom-molecule quantum field theory
that is presented in Sec.~\ref{subsec:atommoleham}. In
Section~\ref{subsec:ladders} we subsequently dress the coupling
constants of this bare atom-molecule theory with ladder diagrams,
to arrive at the desired effective quantum field theory that
includes all two-atom physics exactly. The Heisenberg equations of
motion of this effective field theory are presented in
Section~\ref{subsec:eft}.

\subsection{Bare atom-molecule theory}
\label{subsec:bareatommolecule} 
\begin{figure}
\begin{center}
\includegraphics{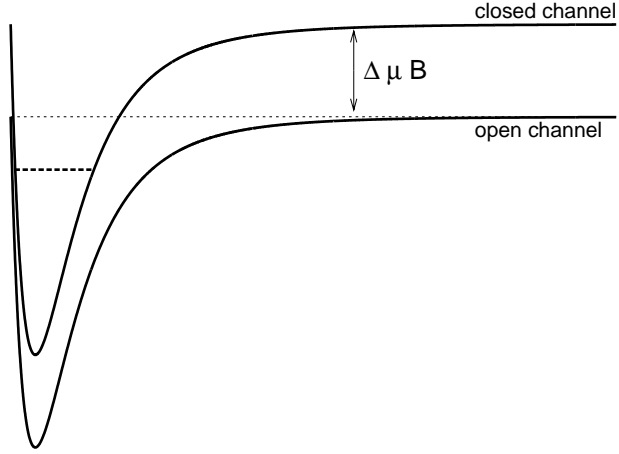}
\caption{\label{fig:fres}
  Illustration of a Feshbach resonance. The upper potential curve
   corresponds to the closed-channel interaction potential $V_{\downarrow
   \downarrow} (\bx-\bx')$ that contains the bound state responsible for the Feshbach
   resonance, indicated by the dashed line. The lower potential curve
   corresponds to the open-channel interaction potential
   $V_{\uparrow\uparrow} (\bx-\bx')$.
   }
\end{center}
\end{figure}

Without loss of generality we can
consider the simplest situation in which a Feshbach resonance
arises, i.e., we consider a homogeneous gas of identical atoms in
a box of volume $V$. These atoms have two internal states, denoted
by $|\!\uparrow \rangle$ and $|\!\downarrow \rangle$, that are
described by the fields $\phiup (\bx,\tau)$ and $\phidown
(\bx,\tau)$, respectively. The atoms in these two states interact
via the potentials $V_{\uparrow \uparrow} ({\bf x}-{\bf x}')$ and
$V_{\downarrow \downarrow} ({\bf x}-{\bf x}')$, respectively. The
state $| \downarrow \rangle$ has an energy $\Delta \mu B/2$ with
respect to the state $|\!\uparrow \rangle$ due to the Zeeman
interaction with the magnetic field $B$. The coupling between the
two states, which from the point of view of atomic physics is due
to the difference in singlet and triplet interactions, is denoted
by $V_{\uparrow \downarrow} (\bx-\bx')$. Putting everything
together we write the grand-canonical partition function for the
gas as a path integral given by
\begin{eqnarray}
\label{eq:zgrcan}
  {\mathcal Z}_{\rm gr} = \int d[\phidup] d[\phiup]
                               d[\phiddown] d[\phidown]
                   \exp \left\{-\frac{1}{\hbar}
                      S[\phidup,\phiup,
                    \phiddown,\phidown]
                     \right\}~.
\end{eqnarray}
Since we are dealing with bosons, the integration is over all
fields that are periodic on the imaginary-time axis ranging from
zero to $\hbar \beta$, with $\hbar$ Planck's constant and
$\beta=1/k_{\rm B} T$ the inverse thermal energy. The Euclidian
action is given by
\begin{eqnarray}
\label{eq:atomatomaction}
   &&S [\phidup,\phiup,\phiddown,\phidown]
   = \nonumber \\ && \ \ \ \ \ \ \int_0^{\hbar \beta} \! d\tau
       \left\{ \int\!d\bx \left[
         \phidup (\bx,\tau) \hbar \frac{\partial}{\partial \tau} \phiup
     (\bx,\tau) + \phiddown (\bx,\tau) \hbar \frac{\partial}{\partial \tau}
     \phidown (\bx,\tau) \right] \right. \nonumber \\
     && \ \ \ \ \ \ \left. +
     H[\phidup,\phiup,\phiddown,\phidown]
        \rule{0mm}{6mm} \right\}~,
\end{eqnarray}
with the grand-canonical hamiltonian functional given by
\begin{eqnarray}
\label{eq:hamiltonian}
   H[\phidup,\phiup,\phiddown,\phidown]
    &=& \int d {\bf x}
        \phidup (\bx,\tau) \left[
     -\frac{\hbar^2 {\bf
      \nabla}^2}{2m}-\mu \right. \nonumber \\
      && \left. \ \ + \half \int d \bx' \phidup (\bx',\tau)
      V_{\uparrow \uparrow} (\bx-\bx') \phiup (\bx',\tau)
    \right] \phiup (\bx,\tau) \nonumber \\
      &+& \int d {\bf x}
        \phiddown (\bx,\tau) \left[
     -\frac{\hbar^2 {\bf
      \nabla}^2}{2m} +\frac{\Delta \mu B}{2} -\mu
                    \right. \nonumber \\
      && \left. \ \ + \half \int d \bx' \phiddown (\bx',\tau)
      V_{\downarrow \downarrow} (\bx-\bx') \phidown (\bx',\tau)
    \right] \phidown (\bx,\tau) \nonumber \\
      &+& \half \int d \bx \int d \bx' \left[ \phidup (\bx,\tau) \phidup
      (\bx',\tau) \right. \nonumber \\
      && \left. \ \
        \times V_{\uparrow \downarrow} (\bx-\bx') \phidown (\bx',\tau) \phidown
     (\bx,\tau)
         + {\rm
     c.c.} \right],
\end{eqnarray}
where $\mu$ is the chemical potential of the atoms. Note that this
hamiltonian functional is the grand-canonical version of the
hamiltonian in Eq.~(\ref{eq:hffeshbachham}). The indices
$\uparrow$ and $\downarrow$ now refer again to single-particle
states, and the two-particle hyperfine states are denoted by
$|\uparrow\uparrow\rangle$ and $|\downarrow\downarrow\rangle$,
respectively. The closed-channel potential is assumed again to
contain the bound state responsible for the Feshbach resonance, as
illustrated in Fig.~\ref{fig:fres}.

As a first step towards the introduction of the molecular field
that describes the center-of-mass motion of this bound state, we
introduce the complex pairing field $\Delta (\bx,\bx',\tau)$ and
rewrite the interaction in the closed channel as a gaussian
functional integral over this field, given by
\begin{eqnarray}
\label{eq:decouple}
  &&\exp \left\{ -\frac{1}{2\hbar} \int d \bx' \phiddown (\bx,\tau) \phiddown (\bx',\tau)
      V_{\downarrow \downarrow} (\bx-\bx') \phidown (\bx',\tau)
   \phidown (\bx,\tau) \right\} \nonumber \\
  && propto \int d[\Delta^*] d[\Delta]
      \exp \left\{ \rule{0mm}{6mm}
      -\frac{1}{2\hbar}\int_0^{\hbar\beta}\!d\tau\int\!d\bx\int\!d\bx'
        \left[\rule{0mm}{6mm}\Delta^*(\bx,\bx',\tau) \phidown (\bx',\tau) \phidown
    (\bx,\tau)
    \right.
      \right. \nonumber \\
   &&  \left. \left. + \phiddown (\bx',\tau) \phiddown
    (\bx,\tau)  \Delta (\bx,\bx',\tau) -
    \Delta^* (\bx,\bx',\tau)  V^{-1}_{\downarrow \downarrow} (\bx-\bx')
    \Delta (\bx,\bx',\tau)
    \rule{0mm}{6mm}\right] \right\}~.
\end{eqnarray}
This step is known as a Hubbard-Stratonovich transformation
\cite{kleinert1978,stoofbook} and  decouples the interaction in
the closed channel. In the BCS-theory of superconductivity this
Hubbard-Stratonovich transformation introduces the order parameter
for the Bose-Einstein condensation of Cooper pairs into the
theory. This order parameter is the macroscopic wave function of
the condensate of Cooper pairs. We shall see below that, in our
case, the role of the Cooper pair is played by the diatomic
molecular state that is responsible for the Feshbach resonance.

The functional integral over the fields $\phiddown(\bx,\tau)$ and
$\phidown(\bx,\tau)$ has now become quadratic and we write this
quadratic part as
\begin{eqnarray}
\label{eq:quadrpart}&& -\frac{\hbar}{2}
  \int_0^{\hbar\beta}\!d\tau
  \int\!d\bx \int_0^{\hbar\beta}\!d\tau'
  \int\!d\bx'
  \left[ \phiddown (\bx,\tau), \phidown(\bx,\tau)\right] \nonumber
  \\ && \ \ \ \ \ \ \ \ \ \ \ \
  \cdot~{\mathbf G}_{\downarrow\downarrow}^{-1}
(\bx,\tau;\bx',\tau') \cdot
  \left[
    \begin{array}{c}
      \phidown (\bx',\tau') \\
      \phiddown (\bx',\tau')
    \end{array} \right]~,
\end{eqnarray}
where the so-called Nambu-space Green's function for the closed
channel obeys the Dyson equation
\begin{equation}
 {\mathbf G}_{\downarrow\downarrow}^{-1}(\bx,\tau;\bx',\tau') =
 {\mathbf G}_{0,\downarrow\downarrow}^{-1}(\bx,\tau;\bx',\tau') -
 {\mathbf \Sigma}_{\downarrow\downarrow} (\bx,\tau;\bx',\tau')~.
\end{equation}
The noninteracting Nambu-space Green's function is given by
\begin{eqnarray}
  {\mathbf G}_{0,\downarrow\downarrow}^{-1}(\bx,\tau;\bx',\tau')
  =\left[
    \begin{array}{ccc}
      G_{0,\downarrow\downarrow}^{-1}(\bx,\tau;\bx',\tau') && 0 \\
       0 && -G_{0,\downarrow\downarrow}^{-1}(\bx',\tau';\bx,\tau)
    \end{array}
   \right]~,
\end{eqnarray}
where
\begin{eqnarray}
  \left[ \hbar \frac{\partial}{\partial \tau} -\frac{\hbar^2 {\bf
      \nabla}^2}{2m} +\frac{\Delta \mu B}{2} -\mu
      \right] G_{0,\downarrow\downarrow} (\bx,\tau;\bx',\tau')
      =-\hbar\delta (\tau-\tau') \delta (\bx-\bx')~,
\end{eqnarray}
is the single-particle noninteracting Green's function. The
self-energy is purely off-diagonal in Nambu space and reads
\begin{eqnarray}
 \hbar {\mathbf \Sigma}_{\downarrow\downarrow} (\bx,\tau;\bx',\tau')
 =   \delta (\tau-\tau') \cdot \left[ \!
   \begin{array}{ccc}
    0 && \kappa (\bx,\bx',\tau) \\
   \kappa^* (\bx,\bx',\tau) && 0
   \end{array}
   \! \right]~,
\end{eqnarray}
where
\begin{equation}
  \kappa (\bx,\bx',\tau) \equiv  \Delta (\bx,\bx',\tau) +
      V_{\uparrow\downarrow} (\bx-\bx') \phiup (\bx,\tau) \phiup
      (\bx',\tau)~.
\end{equation}
Note that a variation of the action with respect to the pairing
field shows that
\begin{equation}
\label{eq:defdelta}
 \langle \Delta (\bx,\bx',\tau) \rangle
   = \langle V_{\downarrow \downarrow} (\bx-\bx') \phidown (\bx) \phidown
   (\bx') \rangle~,
\end{equation}
which relates the auxiliary pairing field to the wave function of
two atoms in the closed channel. Roughly speaking, to introduce
the field that describes a pair of atoms in the closed-channel
bound state we have to consider only contributions from this bound
state to the pairing field. Close to resonance it is this
contribution that dominates. Note that the average of the pairing
field in Eq.~(\ref{eq:defdelta}) indeed shows that the pairing
field is similar to the macroscopic wave function of the
Cooper-pair condensate. However, in this case we are interested in
the phase $\langle \Delta \rangle = 0$ and therefore need to
consider also fluctuations.

Since the integration over the fields $\phiddown(\bx,\tau)$ and
$\phidown(\bx,\tau)$ involves now a gaussian integral, it is
easily performed. This results in an effective action for the
pairing field and the atomic fields that describes the open
channel, given by
\begin{eqnarray}
\label{eq:seff}
  && S^{\rm eff} [\phidup,\phiup,\Delta^*,\Delta] =
      \int_0^{\hbar \beta} \! d\tau \int \! d\bx
       \left\{
         \phidup (\bx,\tau) \hbar \frac{\partial}{\partial \tau} \phiup
     (\bx,\tau) \right. \nonumber \\
    &&  + \left.  \phidup (\bx,\tau) \left[
     -\frac{\hbar^2 {\bf
      \nabla}^2}{2m}-\mu+ \half \int d \bx' \phidup (\bx',\tau)
      V_{\uparrow \uparrow} (\bx-\bx') \phiup (\bx',\tau)
    \right] \phiup (\bx,\tau)\right\} \nonumber \\
    &&  -\frac{1}{2}
   \int_0^{\hbar\beta}\!d\tau\int\!d\bx\int\!d\bx'
        \left[
    \Delta^* (\bx,\bx',\tau)  V^{-1}_{\downarrow \downarrow} (\bx-\bx')
    \Delta (\bx,\bx',\tau)
    \right]   \nonumber \\
    &&
    + \frac{\hbar}{2} {\rm Tr}\left[ \ln (-{\mathbf
    G}^{-1}_{\downarrow\downarrow})\right]~.
\end{eqnarray}
Because we are interested in the bare atom-molecule coupling we
expand the effective action up to quadratic order in the fields
$\Delta^* (\bx,\bx',\tau)$ and $\Delta (\bx,\bx',\tau)$.
Considering higher orders would lead to atom-molecule and
molecule-molecule interaction terms that will be neglected here,
since in our applications we always deal with a small density of
molecules relative to the atomic density.

Hence, we expand the effective action by making use of
\begin{equation}
  {\rm Tr} [\ln (-{\mathbf G}_{\downarrow\downarrow}^{-1})]
  = {\rm Tr} [\ln (-{\mathbf G}^{-1}_{0,\downarrow\downarrow})]
  -\sum_{m=1}^{\infty} \frac{1}{m} {\rm Tr} [ ({\mathbf
  G}_{0,\downarrow\downarrow} {\mathbf \Sigma_{\downarrow\downarrow}})^m]~.
\end{equation}
This results for the part of the effective action that is quadratic in
$\Delta^* (\bx,\bx',\tau)$ and $\Delta (\bx,\bx',\tau)$ in
\begin{eqnarray}
  S [\Delta^*,\Delta] &=& -\frac{1}{2}
  \int_0^{\hbar \beta} \! d \tau \int \! d\bx \int \! d\bx'
  \int_0^{\hbar\beta} \!d \tau' \int \! d {\bf y} \int \! d {\bf y}'
  \nonumber \\ && \times \Delta^* (\bx,\bx',\tau) \hbar G^{-1}_{\Delta} (\bx,\bx',\tau;{\bf y},{\bf
  y'},\tau') \Delta ({\bf y},{\bf y}',\tau')~,
\end{eqnarray}
where the Green's function of the pairing field obeys the equation
\begin{eqnarray}
\label{eq:gfdelta}
 && G_{\Delta} (\bx,\bx',\tau;{\bf y},{\bf
  y'},\tau') = \hbar V_{\downarrow \downarrow} (\bx-\bx') \delta (\bx -
  {\bf y}) \delta (\bx'-{\bf y}') \delta (\tau-\tau') \nonumber \\
  &-& \frac{1}{\hbar} \int_0^{\hbar \beta } d \tau''\int\! d{\bf z}\int\! d{\bf z}'
  \left[ V_{\downarrow \downarrow} (\bx-\bx')
   G_{0,\downarrow\downarrow} (\bx,\tau;{\bf z},\tau'')
    G_{0,\downarrow\downarrow} (\bx',\tau;{\bf z}',\tau'')
   \right.\nonumber \\
   && \left. \times G_{\Delta} ({\bf z},{\bf z'},\tau'';{\bf y},{\bf
  y'},\tau')
  \right]~.
\end{eqnarray}
From this equation we observe that the propagator of the pairing
field is related to the many-body T-matrix in the closed channel.
More precisely, introducing the Fourier transform of the
propagator to relative and center-of-mass momenta and Matsubara
frequencies $\Omega_n = 2 \pi n/\hbar \beta$, denoted by
$G_{\Delta} (\bk,\bk',{\bf K},i\Omega_n)$, we have that
\begin{equation}
  G_{\Delta} (\bk,\bk',{\bf K},i\Omega_n) = \hbar T^{\rm
  MB}_{\downarrow\downarrow} (\bk,\bk',{\bf K},i\hbar\Omega_n-\Delta
  \mu B+2\mu)~,
\end{equation}
where the many-body T-matrix in the closed channel obeys the equation
\begin{eqnarray}
\label{eq:mbtmatrixclosed}
  && T^{\rm MB}_{\downarrow\downarrow}
  (\bk,\bk',{\bf K},z) = V_{\downarrow \downarrow} (\bk-\bk')
   \nonumber \\ && + \frac{1}{V} \sum_{\bk''}
   V_{\downarrow\downarrow} (\bk-\bk'')
    \frac{\left[ 1+ N\left(\epsilon_{{\bf K}/2+\bk''}\!-\!\mu\!+\!\frac{\Delta \mu B}{2} \right)
                  + N\left(\epsilon_{{\bf K}/2-\bk''}\!-\!\mu\!+\!\frac{\Delta \mu B}{2}\right)\right]}
         {z-\epsilon_{{\bf K}/2+\bk''}-\epsilon_{{\bf K}/2-\bk''}}
     \nonumber \\  &&
     T^{\rm MB}_{\downarrow\downarrow}
     (\bk'',\bk',{\bf K},z)~.
\end{eqnarray}
with $N(x)=[e^{\beta x}-1]^{-1}$ the Bose distribution function.
Here, $V_{\downarrow \downarrow} (\bk)=\int d\bx
V_{\downarrow\downarrow} (\bx) e^{i\bk\cdot\bx}$ denotes the
Fourier transform of the atomic interaction potential. This
equation describes the scattering of a pair of atoms from relative
momentum $\bk'$ to relative momentum $\bk$ at energy $z$. Due to
the fact that the scattering takes places in a medium the
many-body T-matrix also depends on the center-of-mass momentum
${\bf K}$, contrary to the two-body T-matrix introduced in the
previous section, which describes scattering in vacuum. The
kinetic energy of a single atom is equal to
$\epsilon_{\bk}=\hbar^2 \bk^2/2m$. The factor that involves the
Bose-Einstein distribution function arises because the probability
of a process where a boson scatters into a state that is already
occupied by $N_1$ bosons is proportional to $1+N_1$. The reverse
process is only proportional to $N_1$. This explains the factor
\begin{eqnarray}
 1+N_1+N_2=(1+N_1)(1+N_2)-N_1 N_2~,
\end{eqnarray}
in the equation for the many-body T-matrix \cite{stoof1996b}.

The many-body T-matrix is discussed in more detail in the next
section when we calculate the renormalization of the interatomic
interactions. For now we only need to realize that, for the
conditions of interest to us, we are always in the situation where
we are allowed to neglect the many-body effects in
Eq.~(\ref{eq:mbtmatrixclosed}) because the Zeeman energy $\Delta
\mu B/2$ strongly suppresses the Bose occupation numbers for atoms
in the closed channel. This is certainly true for the experimental
applications of interest because in the current experiments with
magnetically-trapped ultracold gases the Zeeman splitting of the
magnetic trap is much larger than the thermal energy. This reduces
the many-body T-matrix equation to the Lippmann-Schwinger equation
in Eq.~(\ref{eq:lippmannschwinger1}) for the two-body T-matrix in
the closed channel $T_{\downarrow\downarrow}^{\rm 2B}
(\bk,\bk',z-\epsilon_{\bf K}/2)$, which, in its basis-independent
operator formulation, reads
\begin{equation}
\label{eq:lippmannschwinger2}
 \hat T^{\rm 2B}_{\downarrow\downarrow} (z) =
 \hat V_{\downarrow\downarrow} +
 \hat V_{\downarrow\downarrow} \frac{1}{z-\hat H_0}
 \hat T^{\rm 2B}_{\downarrow\downarrow} (z)~,
\end{equation}
with $\hat H_0 = \hat p^2/m$. As we have seen previously, this
equation is formally solved by
\begin{equation}
 \hat T^{\rm 2B}_{\downarrow\downarrow} (z)
 = \hat V_{\downarrow\downarrow} + \hat V_{\downarrow\downarrow}
 \frac{1}{z-\hat H_{\downarrow\downarrow}} \hat V_{\downarrow\downarrow}~,
\end{equation}
with $\hat H_{\downarrow\downarrow}=\hat H_0 +\hat
V_{\downarrow\downarrow}$. From the previous section we know that
the two-body T-matrix has poles at the bound states of the
closed-channel potential. We assume that we are close to resonance
and hence that one of these bound states dominates. Therefore, we
approximate the two-body T-matrix by
\begin{equation}
\label{eq:t2bapprox}
  \hat T^{\rm 2B}_{\downarrow\downarrow} (z) \simeq
  \hat V_{\downarrow\downarrow} \frac{|\chi_{\rm m} \rangle\langle
  \chi_{\rm m}|}{z-E_{\rm m}}\hat V_{\downarrow\downarrow}~,
\end{equation}
where the properly normalized and symmetrized bound-state wave
function $\chi_{\rm m} (\bx) \equiv \langle \bx|\chi_{\rm
m}\rangle$ obeys the Schr\"odinger equation
\begin{equation}
  \left[ -\frac{\hbar^2 {\bf \nabla}^2}{m}
   + V_{\downarrow\downarrow} (\bx)
  \right] \chi_{\rm m} (\bx) = E_{\rm m} \chi_{\rm m}
  (\bx)~.
\end{equation}
It should be noted that this wave function does not correspond to
the dressed, or true, molecular state which is an eigenstate of
the coupled-channels hamiltonian and determined by
Eq.~(\ref{eq:hffeshbachham}). Rather, it corresponds to the bare
molecular wave function. The coupling $V_{\uparrow\downarrow}
(\bx-\bx')$ of this bare state with the continuum renormalizes it
such that it contains also a component in the open channel.
Moreover, as we have already seen in the previous section, this
coupling also affects the energy of this bound state. Both effects
are important near the resonance and are discussed in detail later
on.

We are now in the position to derive the quadratic action for the
quantum field that describes the bare molecule. To do this, we
consider first the case that the exchange interaction $\hat
V_{\uparrow\downarrow}(\bx-\bx')$ is absent. Within the above
approximations, the two-point function for the pairing field is
given by
\begin{equation}
 \langle \Delta ({\bf k},{\bf K},i \Omega_n)
         \Delta^* ({\bf k}',{\bf K},i \Omega_n) \rangle
     =-2 \hbar \frac{\langle \bk| \hat V_{\downarrow\downarrow}|\chi_{\rm m}\rangle
             \langle \chi_{\rm m} | \hat V_{\downarrow\downarrow}|\bk'\rangle}
      {i\hbar\Omega_n-\epsilon_\bK/2-E_{\rm m} - \Delta \mu B+2 \mu}~.
\end{equation}
We introduce the field $\phim (\bx,\tau)$, that describes the
bound state in the closed channel, i.e, the bare molecule, by
considering configurations of the pairing field such that
\begin{equation}
 \Delta (\bx,\bx',\tau) = \sqrt{2} V_{\downarrow \downarrow} (\bx-\bx')
 \chi_{\rm m} (\bx-\bx') \phim ((\bx+\bx')/2,\tau)~.
\end{equation}
Using this we have that
\begin{equation}
 \langle \phim ({\bf K},\Omega_n) \phimd ({\bf K},\Omega_n) \rangle
  = \frac{\hbar}{-i \hbar \Omega_n+\epsilon_{\bf K}/2+E_{\rm m}+\Delta \mu
  B -2 \mu}~,
\end{equation}
which shows that the quadratic action for the bare molecular field
is, in position representation, given by
\begin{eqnarray}
\label{eq:baremolaction}
  S[\phimd,\phim] &=&
   \int_0^{\hbar\beta}\!d\tau \int d \bx~\phimd (\bx,\tau) \nonumber \\
  && \times \left[  \hbar \frac{\partial}{\partial \tau}
 - \frac{\hbar^2 {\bf
      \nabla}^2}{4m}  + E_{\rm m} + \Delta \mu B - 2 \mu
  \right] \phim (\bx,\tau)~.
\end{eqnarray}
In the absence of the coupling of the bare molecular field to the
atoms, the dispersion relation of the bare molecules is given by
\begin{equation}
\label{eq:dispnonint}
  \hbar \omega_\bk (B)= \epsilon_\bk/2 + E_{\rm m} + \Delta \mu
  B~.
\end{equation}
As expected, the binding energy of the bare molecule is equal to
$\epsilon_{\rm m} (B) = E_{\rm m} + \Delta \mu B$. The momentum
dependence of the dispersion is due to the kinetic energy of the
molecule.

\begin{figure}
\includegraphics{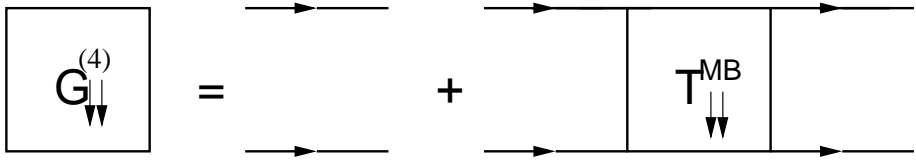}
\caption{\label{fig:tbgreens}  Diagrammatic representation of the
two-particle Green's function in the closed channel. The solid
lines correspond to single-atom propagators. }
\end{figure}

To derive the coupling of this bare molecular field to the fields
$\phidup (\bx,\tau)$ and $\phiup (\bx,\tau)$ it is convenient to
start from the effective action in Eq.~(\ref{eq:seff}) and to
consider again only terms that are quadratic in the self-energy.
Integrating out the pairing fields leads to an interaction term in
the action for the field describing the open channel, given by
\begin{eqnarray}
\label{eq:psiupinteraction}
 \frac{1}{2}
  \int_0^{\hbar \beta} \! d \tau \int \! d\bx \int \! d\bx'
  \int_0^{\hbar\beta} \!d \tau' \int \! d {\bf y} \int \! d {\bf y}'
  \left[ \rule{0mm}{4mm} V_{\uparrow \downarrow} (\bx-\bx')
  \phidup (\bx,\tau) \phidup (\bx',\tau)
 \right. \nonumber \\ \times \left. G^{(4)}_{\downarrow\downarrow}
    (\bx,\bx',\tau;{\bf y},{\bf y}',\tau')
  V_{\uparrow\downarrow} ({\bf y}-{\bf y}')
  \phiup ({\bf y},\tau') \phiup ({\bf y}',\tau') \right],
\end{eqnarray}
where the two-atom four-point Green's function is given
diagrammatically in Fig.~\ref{fig:tbgreens}. For our purposes it
is, for the same reasons as before, sufficient to neglect the
many-body effects on this propagator and to consider again only
the contribution that arises from the bound state in the closed
channel. This gives for the Fourier transform of this Green's
function
\begin{equation}
  G^{(4)}_{\downarrow\downarrow} (\bk,\bk',{\bf K},\Omega_n) \simeq
  \frac{\chi^*_{\rm m} (\bk) \chi_{\rm m} (\bk')}{i \hbar \Omega_n -
  \epsilon_{\bf K}/2 -\Delta \mu B-E_{\rm m}+2\mu}~,
\end{equation}
where $\chi_{\rm m} (\bk)$ is the Fourier transform of the
bound-state wave function. After substitution of this result into
Eq.~(\ref{eq:psiupinteraction}) the resulting interaction term is
decoupled by introducing the field $\phim (\bx,\tau)$ with the
quadratic action given in Eq.~(\ref{eq:baremolaction}). This
procedure automatically shows that the bare atom-molecule coupling
constant is equal to $V_{\uparrow\downarrow} (\bk) \chi_{\rm m}
(\bk)/\sqrt{2}$.

\subsection{Bare atom-molecule hamiltonian}
\label{subsec:atommoleham} In the previous section we have
derived, from a microscopic atomic hamiltonian, a bare
atom-molecule theory for the description of a Feshbach resonance.
It is determined by the action
\begin{eqnarray}
\label{eq:atommolaction}
  && S [\phidup,\phiup,\phimd,\phim]
   =\nonumber \\ && \ \ \ \ \ \ \int_0^{\hbar \beta} \! d\tau
       \left\{ \int\!d\bx \left[
         \phidup (\bx,\tau) \hbar \frac{\partial}{\partial \tau} \phiup
     (\bx,\tau) + \phimd (\bx,\tau) \hbar \frac{\partial}{\partial \tau}
     \phim (\bx,\tau) \right] \right. \nonumber \\
     && \ \ \ \ \ \ + \left. \rule{0mm}{5mm}
     H[\phidup,\phiup,\phim,\phimd]
       \right\}~,
\end{eqnarray}
where the bare or microscopic atom-molecule hamiltonian functional
is given by
\begin{eqnarray}
\label{eq:atommoleham}
   &&H [\phidup,\phiup,\phim,\phimd] = \nonumber \\ && \int d {\bf x}
        \phidup (\bx,\tau) \left[
     -\frac{\hbar^2 {\bf
      \nabla}^2}{2m} - \mu + \half \int d \bx' \phidup (\bx',\tau)
      V_{\uparrow \uparrow} (\bx-\bx') \phiup (\bx',\tau)
    \right] \phiup (\bx,\tau) \nonumber \\
      &&+ \int d {\bf x}
        \phimd (\bx,\tau) \left[
     -\frac{\hbar^2 {\bf
      \nabla}^2}{4m} +\Delta \mu B + E_{\rm m} - 2 \mu
    \right] \phim (\bx,\tau) \nonumber \\
      &&+\int\!d \bx \int\! d \bx'
        \left[ g_{\uparrow\downarrow}(\bx-\bx') \phimd ((\bx+\bx')/2,\tau)
              \phiup (\bx',\tau) \phiup (\bx,\tau) + {\rm
     c.c.} \right]~,
\end{eqnarray}
and the bare atom-molecule coupling is given by
$g_{\uparrow\downarrow}(\bx)=V_{\uparrow\downarrow} (\bx)
\chi_{\rm m} (\bx)/\sqrt{2}$, where $V_{\uparrow\downarrow} (\bx)$
is the coupling between the open and closed atomic collision
channel of the Feshbach problem, that has its origin in the
exchange interaction of the atoms. Note also that the
atom-molecule coupling is proportional to the wave function
$\chi_{\rm m} (\bx)$ for the bound molecular state in the closed
channel responsible for the Feshbach resonance.

Physically, the microscopic hamiltonian in
Eq.~(\ref{eq:atommoleham}) describes bosonic atoms in the open
channel of the Feshbach problem in terms of the fields $\phidup
(\bx,\tau)$ and $\phiup (\bx,\tau) $. These atoms interact via the
interaction potential $V_{\uparrow \uparrow} (\bx-\bx')$. Apart
from this background interaction, two atoms in the gas can also
form a molecular bound state in the closed channel with energy
$E_{\rm m}$ that is detuned by an amount of $\Delta \mu B$ from
the open channel. This bare molecular state is described by the
fields $\phimd (\bx,\tau)$ and $\phim  (\bx,\tau)$. The most
important input in the derivation of Eq.~(\ref{eq:atommoleham}) is
that the energy difference between the various bound states in the
closed channel is much larger than the thermal energy, so that
near resonance only one molecular level is of importance. This
condition is very well satisfied fo almost all the atomic gases of
interest. An exception is $^6$Li, which has two Feshbach
resonances relatively close to each other
\cite{ohara2002b,strecker2003}. The derivation presented in the
previous section is easily generalized to this situation, by
introducing an additional molecular field to account for the
second resonance.

To point out the differences of our approach with work of other
authors a few remarks are in order. First of all, our starting
point was the microscopic two-channel atomic hamiltonian in
Eq.~({\ref{eq:hamiltonian}), from which we derived the microscopic
atom-molecule hamiltonian in Eq.~(\ref{eq:atommoleham}). As we
started with the full interatomic interaction potentials, the
atom-molecule coupling constant and atom-atom interaction have
momentum dependence which cut off the momentum integrals
encountered in perturbation theory. Because of this, no
ultraviolet divergencies are encountered at any order of the
perturbation theory, as we will see in the next section. This
contrasts with the model used by Kokkelmans and Holland
\cite{kokkelmans2002b}, and Mackie {\it et al.} \cite{mackie2002},
who use a phenomenological atom-molecule hamiltonian with
delta-function interactions and therefore need a renormalization
procedure to subtract the ultraviolet divergencies.

In an application of the above microscopic atom-molecule
hamiltonian to realistic atomic gases we have to do perturbation
theory in the interaction $V_{\uparrow \uparrow} (\bx-\bx')$ and
the coupling $g_{\uparrow\downarrow}(\bx-\bx')$. Since the
interatomic interaction is strong, this perturbation theory
requires an infinite number of terms. Progress is made by
realizing that the atomic and molecular densities of interest are
so low that we only need to include two-atom processes. This is
achieved by summing all ladder diagrams as explained in detail in
the next section.

\subsection{Ladder summations} \label{subsec:ladders} From the
bare or microscopic atom-molecule theory derived in the previous
section we now intend to derive an effective quantum field theory
that contains the two-atom physics exactly. This is most
conveniently achieved by renormalization of the coupling
constants. Moreover, the molecules acquire a self-energy. Both
calculations are done within the framework of perturbation theory
to bring out the physics involved most clearly. It is, however,
also possible to achieve the same goal in a nonperturbative manner
by a second Hubbard-Stratonovich transformation.

Because we are dealing with a homogeneous system, it is convenient
to perform the perturbation theory in momentum space. Therefore,
we Fourier transform to momentum space, and expand the atomic and
molecular fields according to
\begin{eqnarray}
 \phiup (\bx,\tau) = \frac{1}{(\hbar \beta V)^{1/2}} \sum_{\bk,n}
 a_{\bk,n} e^{i \bk \cdot \bx - i \omega_n \tau}~,
\end{eqnarray}
and
\begin{eqnarray} \phim (\bx,\tau) = \frac{1}{(\hbar \beta
V)^{1/2}} \sum_{\bk,n}
 b_{\bk,n} e^{i \bk \cdot \bx - i \omega_n \tau}~,
\end{eqnarray}
respectively. The even Matsubara frequencies $\omega_n = 2 \pi
n/\hbar \beta$ account for the periodicity of the fields on the
imaginary-time axis. With this expansion, the grand-canonical
partition function of the gas is written as a functional integral
over the fields $a_{\bk,n}$ and $b_{\bk,n}$ and their complex
conjugates. It is given by
\begin{eqnarray}
\label{eq:zgrmomenta} {\mathcal Z}_{\rm gr} = \int d[a^*] d[a]
                               d[b^*] d[b]  \exp \left\{
                      -\frac{1}{\hbar} S[a^*,a,b^*,b]
                      \right\}~,
\end{eqnarray}
where the action $S[a^*,a,b^*,b]$ is the sum of four terms. The
first two terms describe noninteracting atoms and noninteracting
bare molecules, respectively, and are given by
\begin{equation}
\label{eq:atomactionnonint}
  S_{{\rm a}}[a^*,a] = \sum_{\bk, n}
  \left(- i \hbar \omega_n+\epsilon_\bk - \mu \right) a^*_{\bk,n} a_{\bk,n}~,
\end{equation}
and
\begin{equation}
\label{eq:molactionnonint}
  S_{{\rm m}}[b^*,b] = \sum_{\bk, n}
  \left(- i \hbar \omega_n+\epsilon_\bk/2 +E_{\rm m} + \Delta \mu B- 2\mu
  \right) b^*_{\bk,n} b_{\bk,n}~.
\end{equation}
The atomic interactions are described by the action
\begin{eqnarray}
\label{eq:sint}
  S_{\rm int} [a^*,a] &=&
  \frac{1}{2} \frac{1}{\hbar \beta V}
   \sum_{\stackrel{\bK,\bk,\bk'}{n,m,m'}} V_{\uparrow\uparrow}(\bk-\bk')
    a^*_{\bK/2+\bk,n/2+m} a^*_{\bK/2-\bk,n/2-m}  \nonumber \\
  && \ \ \ \ \ \ \times a_{\bK/2+\bk',n/2+m'} a_{\bK/2-\bk',n/2-m'}~,
\end{eqnarray}
where $V_{\uparrow\uparrow}(\bk)$ is the Fourier transform of the
interatomic interaction potential. This Fourier transform vanishes
for large momenta due to the nonzero range of the interatomic
interaction potential. The last term in the action describes the
process of two atoms forming a molecule and vice versa, and is
given by
\begin{eqnarray}
\label{eq:scoup}
  S_{\rm coup} [a^*,a,b^*,b] &=&
  \frac{1}{(\hbar \beta V)^{1/2}}
   \sum_{\stackrel{\bK,\bk}{n,m}} g_{\uparrow\downarrow}(\bk)
   \nonumber \\
    && \times  \left[ b^*_{\bK,n} a_{\bK/2+\bk,n/2+m} a_{\bK/2-\bk,n/2-m}
   + {\rm c.c.} \rule{0mm}{3mm} \right],
\end{eqnarray}
where $g_{\uparrow\downarrow}(\bk)$ is the Fourier transform of
the bare atom-molecule coupling constant. This coupling constant
also vanishes for large momenta since the bare molecular wave
function has a nonzero extent.

\begin{figure}
\begin{center}
\includegraphics{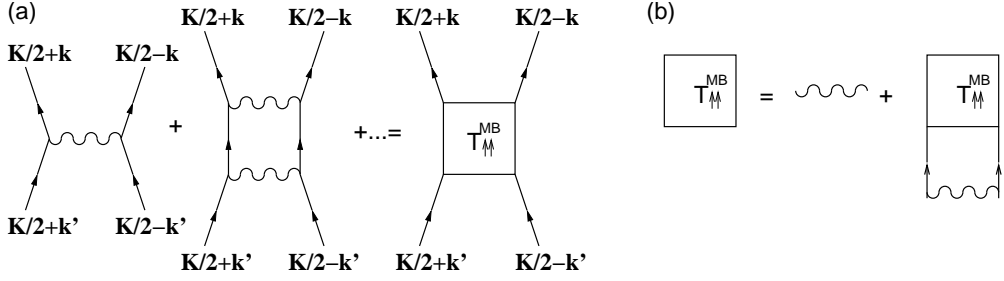}
\end{center}
\caption{\label{fig:tmatrix}  (a) Ladder diagrams that contribute
to the renormalization of the interatomic interaction. (b)
Diagrammatic representation of the Lippmann-Schwinger equation for
the many-body T-matrix. The solid lines correspond to single-atom
propagators. The wiggly lines correspond to the interatomic
interaction $V_{\uparrow\uparrow}$.}
\end{figure}

We first discuss the renormalization of the microscopic atomic
interaction $V_{\uparrow\uparrow}(\bk)$, due to nonresonant
background collisions between the atoms. The first term that
contributes to this renormalization is of second order in the
interaction. It is found by expanding the exponential in the
path-integral expression for the grand-canonical partition
function in Eq.~(\ref{eq:zgrmomenta}). To second order in the
interactions this leads to
\begin{eqnarray}
\label{eq:ptbsecondorder}
   {\mathcal Z}_{\rm gr} &=& \int d[a^*] d[a]
                       \left(
                      1-\frac{1}{\hbar} S_{\rm int}[a^*,a]
                      +\frac{1}{2 \hbar^2} S^2_{\rm int}
                      [a^*,a] + \cdots
                      \right) \nonumber \\
           && \ \ \ \ \ \      \times   \exp \left\{
                      -\frac{1}{\hbar} S_{\rm a}[a^*,a]
                      \right\}.
\end{eqnarray}
After the decoupling of the eight-point function resulting from
the square of the action $S_{\rm int} [a^*,a]$ with the use of
Wick's theorem, it gives rise to various terms in the perturbation
theory which can be depicted by Feynman diagrams
\cite{stoofbook,negeleorlandbook}. As mentioned already, we only
take into account the ladder Feynman diagram. This diagram is
given by the second term of the Born series depicted in
Fig.~\ref{fig:tmatrix}~(a), and corresponds to the expression
\begin{eqnarray}
 - \frac{1}{\hbar \beta V}\sum_{\bk'',m} V_{\uparrow\uparrow}(\bk-\bk'')
  G_{0,{\rm a}} \left(\bK/2+\bk'',i \omega_{n/2+m} \right)
  \nonumber \\
 \times G_{0,{\rm a}} \left(\bK/2-\bk'',i \omega_{n/2-m} \right)
  V_{\uparrow\uparrow}(\bk''-\bk')~,
\end{eqnarray}
where
\begin{equation}
\label{eq:mbpropnonint}
  G_{0,{\rm a}} (\bk, i \omega_n) = \frac{-\hbar}{-i \hbar \omega_n + \epsilon_\bk -
  \mu}~,
\end{equation}
is the noninteracting propagator of the atoms. After performing
the summation over the Matsubara frequencies we find that, to
second order, the renormalization of the interatomic interactions
is given by
\begin{eqnarray}
\label{eq:secondordermb}
 && V_{\uparrow\uparrow}(\bk-\bk') \to V_{\uparrow \uparrow} (\bk-\bk')
   \nonumber \\ && + \frac{1}{V} \sum_{\bk''}
   V_{\uparrow\uparrow} (\bk-\bk'')
    \frac{\left[ 1+ N\left(\epsilon_{{\bf K}/2+\bk''}\!-\!\mu\right)
                  + N\left(\epsilon_{{\bf K}/2-\bk''}\!-\!\mu\right)\right]}
         {i \hbar \omega_n -\epsilon_{{\bf K}/2+\bk''}-\epsilon_{{\bf K}/2-\bk''}+ 2 \mu}
     \nonumber \\  &&
     \times V_{\uparrow\uparrow} (\bk''-\bk')~,
\end{eqnarray}
which is finite due to the use of the true interatomic potential.
In comparing this result with the first two terms of the Born
series for scattering in vacuum in Eq.~(\ref{eq:born2b}), we see
that the only difference between the two-body result and the above
result is the factor involving the Bose distributions. This
so-called statistical factor accounts for the fact that the
scattering takes place in a medium and is understood as follows.
The amplitude for a process where an atom scatters from a state
with occupation number $N_1$ to a state with occupation number
$N_2$ contains a factor $N_1(1+N_2)$. The factor $N_1$ simply
accounts for the number of atoms that can undergo the collision,
and may be understood from a classical viewpoint as well. However,
the additional factor $(1+N_2)$ is a result of the Bose statistics
of the atoms and is therefore called the Bose-enhancement factor.
For fermions this factor would correspond to the Pauli-blocking
factor $(1-N_2)$, reflecting the fact that a fermion is not
allowed to scatter into a state that is already occupied by an
identical fermion. In calculating the Feynman diagram we have to
take into account the forward and backward scattering processes,
which results in the statistical factor in
Eq.~(\ref{eq:secondordermb}).

Continuing the expansion in Eq.~(\ref{eq:ptbsecondorder}) and
taking into account only the ladder diagrams leads to a geometric
series, which is summed by introducing the many-body T-matrix in
the open channel. It is given by
\begin{eqnarray}
\label{eq:mbtmatrixopen}
  && T^{\rm MB}_{\uparrow\uparrow}
  (\bk,\bk',{\bf K},z) = V_{\uparrow \uparrow} (\bk-\bk')
   \nonumber \\ && + \frac{1}{V} \sum_{\bk''}
   V_{\uparrow\uparrow} (\bk-\bk'')
    \frac{\left[ 1+ N\left(\epsilon_{{\bf K}/2+\bk''}\!-\!\mu \right)
                  + N\left(\epsilon_{{\bf K}/2-\bk''}\!-\!\mu \right)\right]}
         {z-\epsilon_{{\bf K}/2+\bk''}-\epsilon_{{\bf K}/2-\bk''}}
     \nonumber \\  &&
   \times T^{\rm MB}_{\uparrow\uparrow}
     (\bk'',\bk',{\bf K},z)~.
\end{eqnarray}
Its diagrammatic representation is given in
Fig.~\ref{fig:tmatrix}~(b). For the moment we neglect the
many-body effects on the scattering atoms and put the
Bose-distribution functions equal to zero. This assumption is
valid at temperatures far below the critical temperature
\cite{bijlsma1996}. This reduces the many-body T-matrix to the
two-body T-matrix $T_{\uparrow\uparrow}^{\rm 2B}
(\bk,\bk',z-\epsilon_\bK/2)$. For the low temperatures of interest
to us here, we are allowed to take the external momenta equal to
zero. For small energies we find, using the result in
Eq.~(\ref{eq:tmatrixcomplexplane}), that the effective interaction
between the atoms reduces to
\begin{eqnarray}
\label{eq:tmatriximagtime}
  && T^{\rm 2B}_{\uparrow\uparrow} ({\bf 0},{\bf 0},i \hbar \omega_n-\epsilon_\bK/2+ 2\mu) =
  \frac{4 \pi a_{\rm bg} \hbar^2}{m}
  \nonumber \\
 && \ \ \ \times \left[ \frac{1}{1-a_{\rm bg}
  \sqrt{\frac{-m\left( i \hbar \omega_n - \epsilon_\bK/2  + 2\mu\right)}{\hbar^2}}
  -\frac{ a_{\rm bg} r_{\rm bg}
  m\left(i \hbar \omega_n - \epsilon_\bK/2+ 2\mu \right)}{2 \hbar^2}}
  \right]~.
\end{eqnarray}
Here $a_{\rm bg}$ and $r_{\rm bg}$ are the scattering length and
the effective range of the open-channel potential $V_{\uparrow
\uparrow} (\bx)$, respectively. Although these could in principle
be calculated with the precise knowledge of this potential, it is
much easier to take them from experiment. For example, the
magnitude of the scattering length can be determined by
thermalization-rate measurements \cite{pethickandsmithbook}. The
effective range is determined by comparing the result of
calculations with experimental data. We will encounter an explicit
example of this in Section~\ref{sec:oscillations}.

\begin{figure}
\begin{center}
\includegraphics{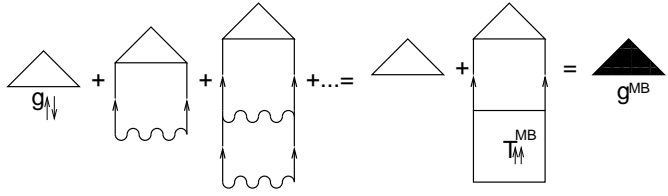}
\end{center}
\caption{\label{fig:gren}  Renormalization of the atom-molecule
coupling constant by interatomic interactions. The solid lines
correspond to single-atom propagators. The wiggly lines
corresponds to the interatomic interaction
$V_{\uparrow\uparrow}$.}
\end{figure}

The next step is the renormalization of the microscopic
atom-molecule coupling constant. Using the same perturbative
techniques as before, we find that the effective atom-molecule
coupling is given in terms of the bare coupling by
\begin{eqnarray}
\label{eq:rengb}
  && g^{\rm MB} (\bk,{\bf K},z) = g_{\uparrow\downarrow}(\bk)
  + \frac{1}{V} \sum_{\bk'}
   T^{\rm MB}_{\uparrow\uparrow} (\bk,\bk',{\bf K},z) \nonumber \\
   &&\ \ \ \times \frac{\left[ 1+ N(\epsilon_{{\bf K}/2+\bk'}-\mu)
                          + N(\epsilon_{{\bf K}/2-\bk'}-\mu)\right]}
                 {z-\epsilon_{{\bf K}/2+\bk'}-\epsilon_{{\bf K}/2-\bk'}}
    g_{\uparrow\downarrow}(\bk')~,
\end{eqnarray}
and is illustrated diagrammatically in Fig~\ref{fig:gren}.
Neglecting again many-body effects, the coupling constant becomes
$g^{\rm 2B}(\bk,z-\epsilon_{\bf K}/2)$ with
\begin{eqnarray}
\label{eq:rengb2b}
  g^{\rm 2B} (\bk,z) = g_{\uparrow\downarrow}(\bk) + \frac{1}{V} \sum_{\bk'}
   T_{\uparrow\uparrow}^{\rm 2B} (\bk,\bk',z) \frac{1}{z-2\epsilon_{\bf k'}}
    g_{\uparrow\downarrow}(\bk')~.
\end{eqnarray}
From the above equation we infer that the energy dependence of
this coupling constant is the same as that of the two-body
T-matrix. This result is easily understood by noting that for a
contact potential $V_{\uparrow\uparrow} (\bk)=V_{\bf 0}$  and we
simply have that $g^{\rm 2B} = g_{\uparrow\downarrow} T^{\rm
2B}_{\uparrow\uparrow}/V_{\bf 0}$. Hence we have for the effective
atom-molecule coupling
\begin{eqnarray}
\label{eq:gedep}
   &&g^{\rm 2B} ({\bf 0},i \hbar \omega_n - \epsilon_\bK/2+ 2\mu )
  = \nonumber \\
  && \ \ \ g  \left[ \frac{1}{1-a_{\rm bg}
  \sqrt{\frac{-m\left( i \hbar \omega_n  - \epsilon_\bK/2+ 2\mu \right)}{\hbar^2}}
  -\frac{ a_{\rm bg} r_{\rm bg}
  m\left(i \hbar \omega_n  - \epsilon_\bK/2+ 2\mu\right)}{2 \hbar^2}}
  \right]~.
\end{eqnarray}
where $g$ is the effective atom-molecule coupling constant at zero
energy. The latter is also taken from experiment. We come back to
this point in Section~\ref{subsec:twoatom} where we discuss the
two-atom properties of our effective many-body theory.

Finally, we have to take into account also the ladder diagrams of
the resonant part of the interaction. This is achieved by
including the self-energy of the molecules. It is in first
instance given by the expression
\begin{eqnarray}
\label{eq:selfenergy}
  \Pi^{\rm MB} ({\bf K},z) &=& \frac{2}{V}
      \sum_{\bk} g_{\uparrow\downarrow}(\bk)
      \frac{\left[ 1+ N(\epsilon_{{\bf K}/2+\bk}-\mu)+N(\epsilon_{{\bf
    K}/2-\bk}-\mu)\right]}{z-\epsilon_{{\bf K}/2+\bk}-\epsilon_{{\bf K}/2-\bk}}
    \nonumber \\ && \times g^{\rm MB} (\bk,{\bf K},z)~,
\end{eqnarray}
and shown diagrammatically in Fig.~\ref{fig:selfenergy}. We
neglect again many-body effects which reduces the self-energy in
Eq.(\ref{eq:selfenergy}) to $ \Pi^{\rm 2B} (z-\epsilon_{\bf K}/2)$
with
\begin{equation}
\label{eq:selfenergytwobody}
 \Pi^{\rm 2B} (z) =  \langle
  \chi_{\rm m} | \hat V_{\uparrow \downarrow} \hat G_{\uparrow\uparrow} (z)
                 \hat V_{\uparrow \downarrow}|
  \chi_{\rm m} \rangle~,
\end{equation}
where the propagator $\hat G_{\uparrow\uparrow}(z)$ is given by
\begin{equation}
  \hat G_{\uparrow\uparrow}(z) = \frac{1}{z-\hat H_{\uparrow\uparrow}}~,
\end{equation}
with the hamiltonian
\begin{equation}
  \hat H_{\uparrow\uparrow} = \frac{\hat {\bf p}^2}{m}
                              + \hat V_{\uparrow \uparrow}
  \equiv \hat H_0 + \hat V_{\uparrow \uparrow}~.
\end{equation}
We insert in Eq.~(\ref{eq:selfenergytwobody}) a complete set of
bound states $|\psi_{\kappa}\rangle$ with energies $E_\kappa$ and
scattering states $|\psi^{(+)}_{\bk} \rangle$ that obey the
equation in Eq.~(\ref{eq:formalsolscatt}). This reduces the
self-energy to
\begin{figure}
\begin{center}
\includegraphics[width=11cm]{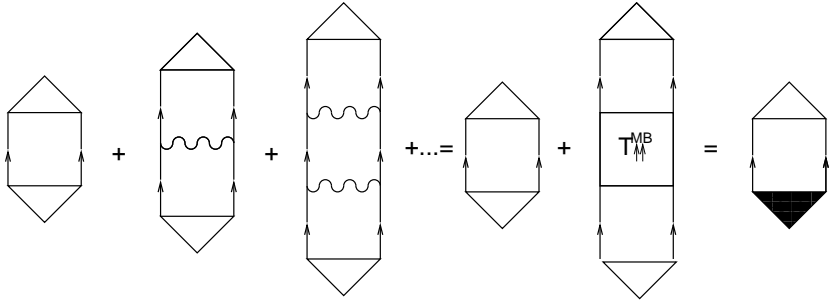}
\end{center}
\caption{\label{fig:selfenergy}  Molecular self-energy. The solid
lines correspond to single-atom propagators. The wiggly lines
corresponds to the interatomic interaction
$V_{\uparrow\uparrow}$.}
\end{figure}
\begin{eqnarray}
\label{eq:selfenergyint}
  \Pi^{\rm 2B}(z) &=&
    \sum_{\kappa} |\langle \chi_{\rm m}|
           \hat V_{\uparrow \downarrow}|\psi_{\kappa}\rangle |^2
    \frac{1}{z-E_{\kappa}} \\ \nonumber
&&  + \int \frac{d \bk}{(2 \pi)^3} |\langle \chi_{\rm m}|
           \hat V_{\uparrow \downarrow}|\psi^{(+)}_{\bk}\rangle |^2
    \frac{1}{z-2 \epsilon_\bk}~,
\end{eqnarray}
where we replaced the sum over the momenta {\bf k} by an integral.
Using Eq.~(\ref{eq:rengb2b}) and the equation for the scattering
states we have that
\begin{equation}
   g^{\rm 2B} (\bk, 2 \epsilon_{\bk}^+)=
  \frac{1}{\sqrt{2}} \langle \chi_{\rm
  m} |\hat V_{\uparrow \downarrow}|\psi^{(+)}_{\bk}\rangle~.
\end{equation}
Neglecting the energy dependence due to the contribution of the
bound states since their binding energies are always large
compared to the thermal energy, we have, using the result for to
the atom-molecule coupling constant in Eq.~(\ref{eq:gedep}), the
intermediate result
\begin{equation}
\label{eq:self}
 \Pi^{\rm 2B} (z) =2 \int \frac{d\bk}{(2 \pi)^3}
  \left|g^{\rm 2B}({\bf 0},2 \epsilon^+_\bk)\right|^2~ \frac{1}{z- 2
  \epsilon_\bk}~.
\end{equation}
The remaining momentum integral yields the final and for our
purposes very important result
\begin{eqnarray}
\label{eq:selfmfinal}
  && \hbar \Sigma^{\rm 2B}_{\rm m} (z) \equiv \Pi^{\rm 2B} (z) - \Pi^{\rm
 2B}(0) \equiv \Pi^{\rm 2B} (z)+\left( \Delta \mu B_0 + E_{\rm m}
 \right) \nonumber \\
 &&=-\frac{g^2 m}{4 \pi^2 \hbar^2} \left\{ \rule{0mm}{8mm}
   - 2 \pi\sqrt{a_{\rm bg}-2r_{\rm bg}}  \sqrt{\frac{-m z}{\hbar^2}} \right.
    \nonumber \\
 && + i \sqrt{a_{\rm bg}} \left[
    \log \left(-\frac{i\sqrt{a_{\rm bg}}r_{\rm bg}}{\sqrt{a_{\rm bg}-2r_{\rm bg}}}\right)
    -\log \left(\frac{i\sqrt{a_{\rm bg}}r_{\rm bg}}{\sqrt{a_{\rm bg}-2r_{\rm bg}}}\right)
    \right]\nonumber \\
    && \times \left. \rule{0mm}{8mm} \frac{m z}{\hbar^2}
    \left[ 3 r_{\rm bg}-2a_{\rm bg}-\frac{a_{\rm bg}r_{\rm bg}^2 m z}{2 \hbar^2}\right]
 \right\} \nonumber \\
 && \times \left\{ \sqrt{a_{\rm bg}\!-\!2r_{\rm bg}}
 \left[ 1+ a_{\rm bg}\left(a_{\rm bg}\!-\!r_{\rm bg}\right)\frac{m z}{\hbar^2}
 +\left(\frac{a_{\rm bg}r_{\rm bg}m z}{2
 \hbar^2}\right)^{2}\right]\right\}^{-1},
\end{eqnarray}
where we have denoted the energy-independent shift $\Pi^{\rm 2B}
(0) $ in such a manner that the position of the resonance in the
magnetic field is precisely at the experimentally observed
magnetic-field value $B_0$. This shift is also shown in the
results of the calculation of the bound-state energy of the
coupled square wells in Fig.~(\ref{fig:esqw_feshbach}).

\subsection{Effective atom-molecule theory} \label{subsec:eft}
Putting the results from the previous section together, we find
that the atom-molecule system is described by the effective action
\begin{eqnarray}
\label{eq:seffatommolkspace}
 && S^{\rm eff} [a^*,a,b^*,b] = \sum_{\bk, n}
  \left(- i \hbar \omega_n+\epsilon_\bk - \mu \right) a^*_{\bk,n}
  a_{\bk,n} \nonumber \\
 &&  \ \ \ + \frac{1}{2} \frac{1}{\hbar \beta V}
   \sum_{\stackrel{\bK,\bk,\bk'}{n,m,m'}} T^{\rm
   2B}_{\rm bg} \left(i \hbar \omega_n - \epsilon_\bK/2 + 2\mu \right)
     \nonumber \\
  && \ \ \ \ \ \ \times
  a^*_{\bK/2+\bk,n/2+m} a^*_{\bK/2-\bk,n/2-m}
  a_{\bK/2+\bk',n/2+m'} a_{\bK/2-\bk',n/2-m'}
   \nonumber \\
 && \ \ \ + \sum_{\bk, n}
  \left[- i \hbar \omega_n+\epsilon_\bk/2 + \delta (B) - 2\mu
  +\hbar \Sigma_{\rm m}^{\rm 2B} (i \hbar \omega_n -\epsilon_\bk/2+2\mu)
  \right] b^*_{\bk,n} b_{\bk,n} \nonumber \\
 && \ \ \ + \frac{1}{(\hbar \beta V)^{1/2}}
   \sum_{\stackrel{\bK,\bk}{n,m}} g^{\rm 2B}\left(
   i \hbar \omega_n - \epsilon_\bK/2 + 2\mu\right)
   \nonumber \\
    && \ \ \ \times  \left[ b^*_{\bK,n} a_{\bK/2+\bk,n/2+m} a_{\bK/2-\bk,n/2-m}
   + {\rm c.c.} \rule{0mm}{3mm} \right]~,
\end{eqnarray}
where $\delta (B) \equiv \Delta \mu (B-B_0)$ is the so-called
detuning. From now on we use the notation $T^{\rm 2B}_{\rm bg} (z)
\equiv T^{\rm 2B}_{\uparrow\uparrow} ({\bf 0},{\bf 0},z)$, and
$g^{\rm 2B} (z) \equiv g^{\rm 2B} ({\bf 0},z)$. Since these
coupling constants are the result of summing all ladder diagrams,
these diagrams should not be taken into account again. In the next
section we discuss how the coupling constants are determined from
experiment.

To consider also the real-time dynamics of the system we derive
the Heisenberg equations of motion for the field operators $\psia
\args$ and $\psim \args$, that annihilate an atom and a molecule
at position $\bx$ and time $t$, respectively. Their hermitian
conjugates are the creation operators. To determine the Heisenberg
equations of motion for these field operators, we first have to
perform an analytic continuation from the Matsubara frequencies to
real frequencies. To ensure that the physical quantities and
equations of motion are causal, this has to be done by a so-called
Wick rotation. This amounts to the replacement of the Matsubara
frequencies by a frequency with an infinitesimally small and
positive imaginary part
\begin{equation}
  i \omega_n \to \omega^+.
\end{equation}
This leads to a subtlety involving the analytic continuation of
the square root of the energy in the various expressions. Due to
the branch cut in the square root we have that
\begin{equation}
   \sqrt{-i \hbar \omega_n} \to \sqrt{-\left( \hbar \omega^+ \right)} =
   -i \sqrt{\hbar \omega}~.
\end{equation}
The last expression on the right-hand side of this equation is
valid for $\hbar \omega$ on the entire real axis.

To obtain the equation of motion in position and time
representation, we have to Fourier transform back from momentum
and frequency space. This amounts to the replacement
\begin{equation}
  \hbar \omega - \epsilon_\bK/2 \to i\hbar \frac{\partial}{\partial
  t}+ \frac{\hbar^2 \nabla^2}{4 m}.
\end{equation}
Note that this combination of time and spatial derivatives is
required due to the Galilean invariance of the theory.

For simplicity we assume that we are so close to resonance that we
are allowed to neglect the energy dependence of the effective
atomic interactions and the effective atom-molecule coupling.
Moreover, for notational convenience we take only the
leading-order energy dependence of the molecular self-energy into
account. Higher orders are straightforwardly included but lead to
somewhat complicated notations in the position and time
representation. The leading-order energy dependence of the
self-energy is, after the Wick rotation to real energies, given by
\begin{equation}
\label{eq:resultselfenergy}
  \hbar \Sigma^{(+)}_{\rm m}  (E)  \simeq - g^2 \frac{m^{3/2}}{2 \pi \hbar^3} i
  \sqrt{E}.
\end{equation}
The additional superscript indicates that we are dealing with the
retarded self-energy, i.e., the self-energy evaluated at the
physically relevant energies $E^+$ so that $\hbar
\Sigma^{(+)}_{\rm m} (E) \equiv \hbar \Sigma^{\rm 2B}_{\rm m}
(E^+)$. Note that for positive energy $E$ this result is in
agreement with the Wigner-threshold law. This law gives the rate
for a state with well-defined positive energy to decay into a
three-dimensional continuum.

Within the above approximations, the Heisenberg equations of
motion for the coupled atom-molecule model read
\begin{eqnarray}
\label{eq:heom}
  && i \hbar \frac{\partial \hat \psi_{\rm a} \args}{\partial t}
    =\left[  -\frac{\hbar^2 {\bf \nabla}^2}{2m}
            + \frac{4\pi a_{\rm bg} \hbar^2}{m} \hat \psi_{\rm a}^{\dagger}
        \args \hat \psi_{\rm a} \args
    \right] \hat \psi_{\rm a} \args \nonumber \\
    && \ \ \ \ \ \ + 2 g \hat \psi_{\rm a}^{\dagger}
          \args \psim \args~, \nonumber \\
   &&i \hbar \frac{\partial \hat \psi_{\rm m} \args}{\partial t}
    =\left[  \rule{0mm}{7mm} -\frac{\hbar^2 {\bf \nabla}^2}{4m}
            +\delta (B(t)) \right. \nonumber \\
   && \ \ \ \ \ \ \left. - g^2 \frac{m^{3/2}}{2 \pi \hbar^3} i
  \sqrt{i \hbar \frac{\partial}{\partial t}
    +\frac{\hbar^2 {\bf \nabla}^2}{4m}}~
    \right] \hat \psi_{\rm m} \args + g \hat \psi_{\rm a}^2
    \args~,
\end{eqnarray}
where we have also allowed for a time-dependent detuning. In the
next section we show that these equations correctly reproduce the
Feshbach-resonant scattering amplitude and the binding energy of
the molecule. Moreover, we apply the effective theory derived in
this section to study many-body effects on this binding energy,
above the critical temperature for Bose-Einstein condensation.

  \section{Normal state} \label{sec:normalstate} In this section we
discuss the properties of the gas in the normal state. In the
first section, we consider the two-atom properties of our
many-body theory. Hereafter, we discuss the equilibrium properties
that follow from our theory. In the last section, we investigate
many-body effects on the energy of the molecular state, above the
critical temperature for Bose-Einstein condensation.

\subsection{Two-atom properties of the many-body theory}
\label{subsec:twoatom} In this section we show that our effective
field theory correctly contains the two-atom physics of a Feshbach
resonance. First, we show that the correct Feshbach-resonant
atomic scattering length is obtained after the elimination of the
molecular field. Second, we calculate the bound-state energy and
show that it has the correct threshold behaviour near the
resonance. To get more insight in the nature of the molecular
state near resonance, we also investigate the molecular density of
states.

\subsubsection{Scattering properties} \label{subsubsec:scattering}
To calculate the effective interatomic scattering length, we have
to eliminate the molecular field from the Heisenberg equations of
motion in Eq.~(\ref{eq:heom}). Since the scattering length is
related to the scattering amplitude at zero energy and zero
momentum, we are allowed to put all the time and spatial
derivatives in the equation of motion for the molecular field
operator equal to zero. This equation is now easily solved, which
leads to
\begin{equation}
\label{eq:psimadiab}
  \psim \args = - \frac{g}{\delta (B)} \psia^2 \args.
\end{equation}
Substitution of this result into the equation for the atomic field
operator leads for the interaction terms to
\begin{eqnarray}
&& \frac{4\pi a_{\rm bg} \hbar^2}{m} \hat \psi_{\rm a}^{\dagger}
        \args \hat \psi_{\rm a} \args
     \hat \psi_{\rm a} \args + 2 g \hat \psi_{\rm a}^{\dagger}
          \args \psim \args = \nonumber \\
&& \ \ \   \left( \frac{4 \pi a_{\rm bg} \hbar^2}{m}
          - \frac{2g^2}{\delta (B)}
          \right) \hat \psi_{\rm a}^{\dagger}
        \args \hat \psi_{\rm a} \args
     \hat \psi_{\rm a} \args~.
\end{eqnarray}
From this result we observe that we have to take the renormalized
atom-molecule coupling constant at zero energy equal to $g=\hbar
\sqrt{2 \pi a_{\rm bg} \Delta B  \Delta \mu /m}$, so that we have
\begin{equation}
  \frac{4 \pi a_{\rm bg} \hbar^2}{m}
          - \frac{2g^2}{\delta (B)} = \frac{4 \pi a(B)
          \hbar^2}{m}~,
\end{equation}
where we recall that the scattering length near a Feshbach
resonance is given by
\begin{equation}
\label{eq:ascatofb2}
  a(B) = a_{\rm bg} \left( 1-\frac{\Delta B}{B-B_0} \right) \equiv a_{\rm bg} + a_{\rm res} (B).
\end{equation}
Since both the width $\Delta B$ and the background scattering
length $a_{\rm bg}$ are known experimentally, the knowledge of the
difference in magnetic moment between the open and the closed
channel $\Delta \mu$ completely determines the renormalized
coupling constant $g$. Since the open and the closed channel
usually correspond to the triplet and singlet potential,
respectively, we always have that $|\Delta \mu | \simeq 2 \mu_{\rm
B}$, with $\mu_{\rm B }$ the Bohr magneton. More precise values of
the difference in magnetic moments are obtained from
coupled-channels calculations using the interatomic interaction
potentials
\cite{tiesinga1993,marte2002,kokkelmans2002b,kempen2002}.

From the above analysis we see that the correct Feshbach-resonant
scattering length of the atoms is contained in our theory exactly.
Next, we show that our effective theory also contains the correct
bound-state energy.

\subsubsection{Bound-state energy} \label{subsubsec:boundstate} The
energy of the molecular state is determined by the poles of the
retarded molecular propagator $G_{\rm m}^{(+)} (\bk,\omega)$. It
is given by
\begin{equation}
\label{eq:gmkw}
  G_{\rm m}^{(+)} (\bk,\omega)=
    \frac{\hbar}{\hbar \omega^+  - \epsilon_{\bk}/2 -\delta (B)
     -\hbar \Sigma_{\rm m}^{(+)} (\hbar\omega -\epsilon_\bk/2)}~.
\end{equation}
For positive detuning $\delta (B)$ there only exists a pole with a
nonzero and negative imaginary part. This is in agreement with the
fact that the molecule decays when its energy is above the
two-atom continuum threshold. The imaginary part of the energy is
related to the lifetime of the molecular state. For negative
detuning the molecular propagator has a real and negative pole
corresponding to the bound-state energy. More precisely, in this
case the poles of the molecular propagator are given by $\hbar
\omega = \epsilon_{\rm m} (B) + \epsilon_\bk/2$, where the
bound-state energy is determined by solving for $E$ in the
equation
\begin{equation}
\label{eq:zeroes}
  E - \delta (B) - \hbar \Sigma_{\rm m}^{(+)} (E) =0.
\end{equation}
In general, this equation cannot be solved analytically but is
easily solved numerically, and in Section~\ref{sec:oscillations}
we discuss its numerical solution for the parameters of $^{85}$Rb.
Close to resonance, however, we are allowed to neglect the
effective range of the interactions. This reduces the retarded
self-energy of the molecules to
\begin{equation}
\label{eq:selfmoleffrzero}
 \hbar \Sigma^{(+)}_{\rm m} (E) \simeq -\frac{g^2 m^{3/2}}{2 \pi \hbar^3}
     \frac{i\sqrt{E}}{1-i|a_{\rm
     bg}|\sqrt{\frac{mE}{\hbar^2}}}~.
\end{equation}
Moreover, the bound-state energy is small in this regime and we
are allowed to neglect the linear terms in the energy with respect
to the square-root terms. This reduces the equation for the
bound-state energy in Eq.~(\ref{eq:zeroes}) to
\begin{equation}
  \frac{g^2 m^{3/2}}{2 \pi \hbar^3}
     \frac{i\sqrt{E}}{1-i|a_{\rm
     bg}|\sqrt{\frac{mE}{\hbar^2}}}=\delta(B)~.
\end{equation}
This equation is easily solved analytically, and yields the result
\begin{equation}
\label{eq:solbse}
  \epsilon_{\rm m} (B) = -\frac{\hbar^2}{m [a(B)]^2}~,
\end{equation}
which analytically proves the numerical result in
Eq.~(\ref{eq:ebres}). This numerical result was obtained for the
specific case of two coupled attractive square wells. The above
analytic proof, which does not depend on the details of the
potential, shows that the result is general.

The same result is found by noting that after the elimination of
the molecular field the effective on-shell T-matrix for the atoms
in the open channel is given by
\begin{equation}
  T^{\rm 2B} (E^+) =
    T^{\rm 2B}_{\rm bg} \left(E^+\right)+
    \frac{2}{\hbar}  |g^{\rm 2B}(E^+)|^2
    G^{(+)}_{\rm m} \left(\sqrt{mE/\hbar^2},E\right)~.
\end{equation}
Close to resonance this expression reduces to
\begin{equation}
\label{eq:tmatrixfeshbachclose}
  T^{\rm 2B} (E) \simeq \frac{4 \pi a_{\rm res} (B) \hbar^2}{m}
  \left[ \frac{1}{1+ i a_{\rm res} (B) \sqrt{\frac{mE}{\hbar^2}}}
  \right]~.
\end{equation}
The pole of this T-matrix, which gives the bound-state energy, is
indeed equal to the result in Eq.~(\ref{eq:solbse}) close to
resonance.

\subsubsection{Molecular density of states}
\label{subsubsec:moldos} The molecular density of states is
obtained by taking the imaginary part of the retarded molecular
propagator \cite{negeleorlandbook}, i.e.,
\begin{equation}
\label{eq:defdos}
  \rho_{\rm m} (\bk,\omega) = -\frac{1}{\pi \hbar}
               {\rm Im} \left[ G_{\rm m}^{(+)} (\bk, \omega) \right]~.
\end{equation}
For simplicity, we discuss here only the situation that we are
close to resonance, and therefore approximate the retarded
molecular self-energy by the square-root term resulting from
Wigner's threshold law as given in
Eq.~(\ref{eq:resultselfenergy}). The extension to situations
further of resonance are straightforward.

For the case of negative detuning, the molecular density of states
is shown by the solid line in Fig.~\ref{fig:dos} and has two
contributions. One arising from the pole at the bound-state energy
and the second from the two-atom continuum. Within the above
approximation, it is given by
\begin{eqnarray}
\label{eq:dos}
 && \rho_{\rm m} (\bk, \omega) =
     Z(B)
     \delta (\hbar \omega -\epsilon_{\bk}/2-\epsilon_{\rm m}(B)) \nonumber \\
    &&+ \frac{1}{\pi} \theta (\hbar \omega - \epsilon_{\bk}/2)
      \nonumber \\
      && \ \ \ \times \frac{(g^2 m^{3/2}/2 \pi \hbar^3) \sqrt{\hbar \omega -
      \epsilon_{\bk}/2}}
      {\left[ \hbar \omega - \epsilon_{\bk}/2 - \delta (B) \right]^2
      +(g^4 m^3/4 \pi^2 \hbar^6)
      (\hbar \omega - \epsilon_{\bk}/2 )}~,
\end{eqnarray}
with $Z(B)$ the so-called wave-function renormalization factor
\begin{eqnarray}
\label{eq:factorz}
  Z(B)&=&\left. \left[1-\frac{\partial \Sigma^{(+)}_{\rm m}(\hbar \omega)}
                      {\partial \omega}\right]^{-1}
                      \right|_{\hbar \omega = \epsilon_{\rm m} (B)}
                      \nonumber \\
                      &\simeq& \left[1 + \frac{g^2 m^{3/2}}{4 \pi \hbar^3
                  \sqrt{|\epsilon_{\rm m}(B)|}}\right]^{-1}~.
\end{eqnarray}
This factor goes to zero as we approach the resonance and it
becomes equal to one far off resonance. Physically, this is
understood as follows. Far off resonance the bound state of the
coupled-channels hamiltonian in Eq.~(\ref{eq:hffeshbachham}),
i.e., the dressed molecule, is almost equal to the bound state of
the closed-channel potential and has zero amplitude in the open
channel. This corresponds to the situation where $Z(B) \simeq 1$.
As the resonance is approached, the dressed molecule contains only
with an amplitude $\sqrt{Z(B)}$ the closed-channel bound state,
i.e., the bare molecule. Accordingly, the contribution of the open
channel becomes larger and gives rise to the threshold behaviour
of the bound-state energy in Eq.~(\ref{eq:solbse}). Of course, the
square of the wave function of the dressed molecule is normalized
to one. This is expressed by the sum rule for the molecular
density of states,
\begin{equation}
\label{eq:sumrule}
  \int d(\hbar \omega) \rho_{\rm m} (\bk,\omega) =1~.
\end{equation}

\begin{figure}
\begin{center}
\includegraphics{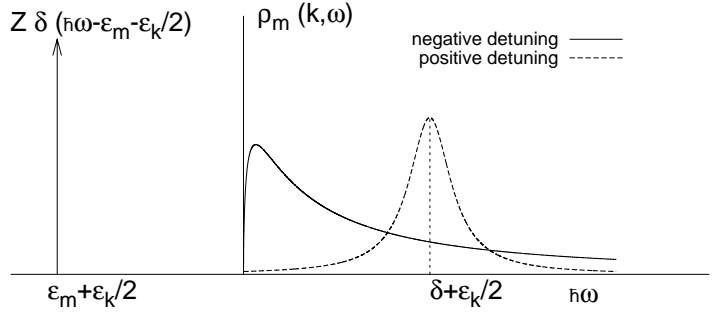}
\caption{\label{fig:dos}
  Molecular density of states. The solid line shows the density of
 states for negative detuning. Since there is a true bound state
 in this case there is a pole in the density of states. For
 positive detuning the density of states is approximately a Lorentzian as shown
 by the dashed line.
   }
\end{center}
\end{figure}

In detail, the dressed molecular state with zero momentum is given
by
\begin{eqnarray}
\label{eq:wavefctmol}
  | \chi_{\rm m}; {\rm dressed} \rangle=
                 \sqrt{Z(B)} \hat b^{\dagger}_{\bf 0} | 0 \rangle
                 + \sum_\bk C_\bk
                          \hat a^{\dagger}_\bk \hat a^{\dagger}_{-\bk} | 0 \rangle
                          ~.
\end{eqnarray}
Here, the second-quantized operator $\hat b^{\dagger}_{\bf 0}$
creates a molecule with zero momentum. It acts on the vacuum state
$|0\rangle$. The bare molecular state is therefore given by $|
\chi_{\rm m} \rangle =   \hat b^{\dagger}_{\bf 0} | 0 \rangle$.
The operator $\hat a^{\dagger}_\bk$ creates an atom with momentum
$\hbar \bk$ and hence the coefficient $C_\bk$ denotes the
amplitude of the dressed molecular state to be in the open channel
of the Feshbach problem.

To gain more insight in the nature of the dressed molecular state
we calculate the coefficients $C_\bk$ in perturbation theory.
Neglecting the off-resonant background interactions and the energy
dependence of the atom-molecule coupling constant, the hamiltonian
appropriate for our purposes is, in terms of the above operators,
given by
\begin{equation}
  \hat H = \hat H_{\rm am} + \hat H_{\rm coup}~,
\end{equation}
with
\begin{equation}
 \hat H_{\rm am}= \sum_{\bk} \epsilon_\bk \hat a^{\dagger}_\bk \hat a_\bk
    + \sum_{\bk} \left[ \frac{\epsilon_\bk}{2} + \delta (B) \right] \hat b^{\dagger}_\bk \hat
    b_\bk~,
\end{equation}
and
\begin{equation}
   \hat H_{\rm coup} = \frac{g}{\sqrt{V}} \sum_{\bK,\bk}  \left[ \hat b^{\dagger}_\bK \hat
    a_{\bK/2+\bk} \hat a_{\bK/2-\bk} + {\rm h.c.} \right]~.
\end{equation}
The zeroth-order state around which we perturb is the bare
molecular state $|\chi_{\rm m} \rangle$ with energy $\delta(B)$.
In first order in $g$ the dressed molecular state is given by
\begin{eqnarray}
 | \chi_{\rm m}; {\rm dressed} \rangle &=&
  \sqrt{Z (B) } \hat b^{\dagger}_{\bf 0} | 0 \rangle
  +\frac{1}{\delta (B)-\hat H_{\rm am}} \hat H_{\rm coup} \hat b^{\dagger}_{\bf 0}| 0 \rangle
   \nonumber \\
 &=& \sqrt{Z (B) } \hat b^{\dagger}_{\bf 0} | 0 \rangle
  +\frac{g}{\sqrt{V}} \sum_\bk
  \frac{1}{\delta (B)-2 \epsilon_\bk} \hat a^{\dagger}_\bk \hat a^{\dagger}_{-\bk} | 0
  \rangle~,
\end{eqnarray}
where $Z(B)=1-\mathcal{O} (g^2)$. This result shows that, to first
order in $g$, the coefficients $C_\bk$ are given by
\begin{equation}
  C_\bk = \frac{g}{\sqrt{V}}
  \frac{1}{\delta (B)-2 \epsilon_\bk}~.
\end{equation}
We now calculate the wave-function renormalization factor $Z(B)$
in a different manner by demanding that the dressed molecular wave
function is properly normalized, i.e.,
\begin{equation}
 \langle \chi_{\rm m} ; {\rm dressed} | \chi_{\rm m}; {\rm dressed}
 \rangle=1~.
\end{equation}
This leads to
\begin{equation}
\label{eq:normdressed}
 1 = Z(B) +  \frac{2 g^2}{V} \sum_\bk \frac{1}{\left[\delta (B) -
 2\epsilon_\bk\right]^2} = Z(B) - \frac{\partial \Sigma_{\rm m}^{(+)} ( \delta(B))}{\partial
 \omega}~.
\end{equation}
The factor of two corresponds to the two contributions arising
from the matrix element $\langle 0 | \hat a_\bk \hat a_{-\bk} \hat
a^{\dagger}_{\bk'} \hat a^{\dagger}_{-\bk'}  | 0 \rangle$. From
this result we find that the wave-function renormalization factor
is given by
\begin{equation}
 Z(B) = 1+\frac{\partial \Sigma_{\rm m}^{(+)} (\delta(B))}{\partial
 \omega} \simeq \left[1-\frac{\partial \Sigma_{\rm m}^{(+)} ( \delta(B))}{\partial
 \omega}\right]^{-1}~,
\end{equation}
in agreement with the result in Eq.~(\ref{eq:factorz}) to second
order in $g$.

Note that the total number of atoms in the dressed molecular state
should be equal to two. The number of atoms is given by
\begin{equation}
\label{eq:natomsdressed}
  N= 2 \sum_\bk \langle \hat b^{\dagger}_\bk \hat b_\bk
  \rangle_{\rm dressed}
              + \sum_\bk \langle \hat a^{\dagger}_\bk \hat a_\bk
              \rangle_{\rm dressed}~,
\end{equation}
where $\langle \cdots \rangle_{\rm dressed} \equiv \langle
\chi_{\rm m};{\rm dressed}| \cdots | \chi_{\rm m}; {\rm dressed}
\rangle$. For the number of atoms with momentum $\hbar \bk$ we
have that
\begin{equation}
\label{eq:natomswithkdressed}
  \langle \hat a^{\dagger}_\bk \hat a_\bk
              \rangle_{\rm dressed} = \frac{4g^2}{V} \frac{1}{\left[\delta (B) - 2 \epsilon_\bk
              \right]^2}~,
\end{equation}
from which, with the use of Eq.~(\ref{eq:normdressed}), we find
that
\begin{equation}
\label{eq:twominusz}
 \sum_\bk \langle \hat a^{\dagger}_\bk \hat a_\bk
              \rangle_{\rm dressed} = 2-2 Z(B)~.
\end{equation}
Using $\langle \hat b^{\dagger}_\bk \hat b_\bk \rangle_{\rm
dressed} = Z(B) \delta_{\bk,{\bf 0}}$ we have indeed that the
total number of atoms $N=2$, as required.

If the magnetic field varies not too rapidly, we are allowed to
make an adiabatic approximation to the Heisenberg equation of
motion for the bare molecular field operator in
Eq.~(\ref{eq:heom}). This amounts to introducing a molecular field
$\psim' \args$ that annihilates a dressed molecule, i.e., a
molecule with internal state given by Eq.~(\ref{eq:wavefctmol}).
This is achieved as follows. In frequency and momentum space the
action for the bare molecular field is given by
\begin{eqnarray}
 S[\phimd,\phim] &=&
 \int \frac{d \omega}{(2 \pi)}
    \sum_\bk
    \phimd (\bk,\omega) \left[
    \hbar \omega - \epsilon_\bk/2 \rule{0mm}{4mm} \right. \nonumber \\
    && \left. \rule{0mm}{4mm} \ \ \ - \delta (B) - \hbar
    \Sigma^{(+)}_{\rm m} (\hbar \omega - \epsilon_\bk/2)
    \right] \phim (\bk,\omega)~.
\end{eqnarray}
Next, we expand this action around the pole of the propagator
$\epsilon_{\rm m} (B)$. To linear order, this yields the result
\begin{equation}
  S[\phimd,\phim] \simeq \int \frac{d \omega}{(2 \pi)}
    \sum_\bk
    \frac{\phimd (\bk,\omega)}{\sqrt{Z(B)}} \left[
    \hbar \omega -\epsilon_\bk/2 - \epsilon_{\rm m} (B) \right]
    \frac{\phim (\bk,\omega)}{\sqrt{Z(B)}}~.
\end{equation}
From this equation we see that the field that describes the
dressed molecule is given by $\phim'=\phim/\sqrt{Z(B)}$. This
leads to the following action for the dressed molecular field in
position and time representation
\begin{eqnarray}
  S[\phim'^{*},\phim']=
  \int dt \int d \bx~\phim'^{*} \args
  \left[ i \hbar \frac{\partial}{\partial t}
  + \frac{\hbar^2 \nabla^2}{4 m} - \epsilon_{\rm m} (B)
  \right]
  \phim' \args~.
\end{eqnarray}
More importantly, the terms that describe the coupling between the
atoms and the molecules are multiplied by a factor $\sqrt{Z(B)}$.
In detail, the coupled Heisenberg equations of motion for the
atomic and dressed molecular field operators are given by
\cite{duine2003b}
\begin{eqnarray}
\label{eq:dressedheom}
   i \hbar \frac{\partial \psia \args}{\partial t}
    &=&\left[  -\frac{\hbar^2 {\bf \nabla}^2}{2m}
            + \frac{4 \pi a_{\rm bg} \hbar^2}{m}\psiad \args
        \psia \args
    \right] \psia \args
    \nonumber \\
    && \ \ \ + 2 g \sqrt{Z(t)} \psiad \args \hat \psi'_{\rm m} \args~, \nonumber \\
  i \hbar \frac{\partial \hat \psi'_{\rm m} \args}{\partial t}
    &=&\left[  -\frac{\hbar^2 {\bf \nabla}^2}{4m}
            +\epsilon_{\rm m} (t)
    \right] \hat \psi'_{\rm m} \args  + g \sqrt{Z(t)} \hat \psi_{\rm a}^2
    \args~,
\end{eqnarray}
where $Z(t) \equiv Z(B(t))$, and $\epsilon_{\rm m} (t) \equiv
\epsilon_{\rm m} (B(t))$. In the derivation of the above coupled
equations we have assumed that we are allowed to make an adiabatic
approximation for the renormalization factor $Z(B)$ and that we
can evaluate it at every time at the magnetic field $B(t)$. In
principle there are retardation effects due to the fact that the
dressed molecular state does not change instantaneously. Following
the above manipulations for time-dependent magnetic field we see
that these effects can be neglected if
\begin{equation}
\label{eq:condition}
  \hbar \left| \frac{\partial \ln Z(t)}{\partial t} \right| \ll | \epsilon_{\rm
m} (t)|~.
\end{equation}
In principle, the Heisenberg equation of motion for the molecular
field operator also contains an imaginary part due to the fact
that the dressed molecule can decay into a pair of atoms with
opposite momenta. The rate for this process will be small,
however, under the condition given in Eq.~(\ref{eq:condition}). We
will come back to this process when we consider its effect on the
coherent atom-molecule oscillations.

For positive detuning the molecular density of states has only a
contribution for positive energy. For large detuning it is in
first approximation given by
\begin{equation}
\label{eq:moldosposdelta}
  \rho_{\rm m} (\bk,\omega) =
    \frac{\hbar \Gamma_{\rm m} (B) /2}{\pi\left[(\hbar \omega - \epsilon_\bk/2 - \delta (B))^2+(\hbar \Gamma_{\rm
    m}(B) /2)^2\right]}~,
\end{equation}
where the lifetime of the molecular state is defined by
\begin{equation}
\Gamma_{\rm m} (B)= \frac{g^2m^{3/2}}{\pi \hbar^4}\sqrt{\delta
(B)}~.
\end{equation}
As expected, the density of states is, in the case of positive
detuning, approximately a Lorentzian centered around the detuning
with a width related to the lifetime of the molecule. It is shown
in Fig.~\ref{fig:dos} by the dashed line.

\subsection{Equilibrium properties} \label{subsec:equilibrium} The
equilibrium properties of the gas are determined by the equation
of state, which relates the total density of the gas to the
chemical potential. This equation can be calculated in two ways,
either by calculating the thermodynamic potential and
differentiating with respect to the chemical potential, or by
directly calculating the expectation value of the operator for the
total density. We discuss both methods, which should, of course,
yield the same result. Nevertheless, to show the equivalence is a
subtle matter.

\begin{figure}
\begin{center}
\includegraphics[width=11cm]{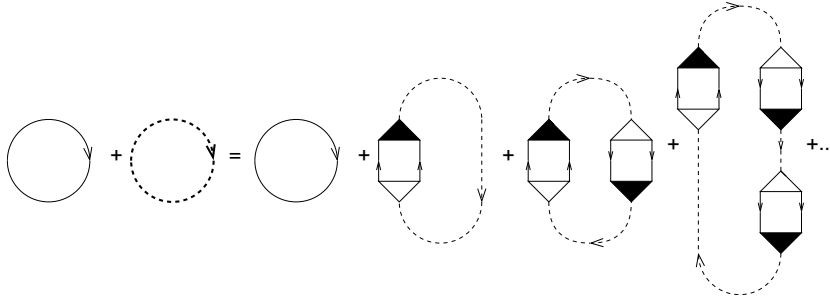}
\caption{\label{fig:freeenergydiag}
  Diagrams contributing to the thermodynamic potential of the gas. The
 noninteracting atomic and molecular propagators are denoted by
 the solid and dashed thin lines, respectively. The full molecular
 propagator is given by the thick dashed line. The bare and
 renormalized atom-molecule coupling constants are denoted by the
 open and filled triangles, respectively.
   }
\end{center}
\end{figure}

First, we calculate thermodynamic potential \cite{2ohashi2002b}.
Within our approximations it is in first instance given by the
expression
\begin{equation}
\label{eq:freeenergyfirst}
  \Omega (\mu,T) = \frac{1}{\beta} {\rm Tr} \left[ \ln \left(G_{0,{\rm a}}^{-1}
  \right)\right]
           +\frac{1}{\beta} {\rm Tr} \left[ \ln \left(G_{\rm m}^{-1} \right)
           \right]~.
\end{equation}
Here, we recall that $G_{0,{\rm a}} (\bk, i \omega_n)$ is the
noninteracting atomic propagator of the atoms in
Eq.~(\ref{eq:mbpropnonint}). The full molecular propagator is
given by
\begin{equation}
\label{eq:fullmolimagtime}
   G_{\rm m} (\bk,i\omega_n)=
    \frac{-\hbar}{-i\hbar \omega_n  + \epsilon_{\bk}/2 +\delta
    (B)-2\mu
     +\hbar \Sigma_{\rm m}^{\rm 2B}(i \hbar\omega_n -\epsilon_\bk/2+2\mu)}~,
\end{equation}
with the molecular self-energy given in Eq.~(\ref{eq:selfmfinal}).
The so-called ring diagrams that contribute to the thermodynamic
potential in our approximation are given in
Fig.~\ref{fig:freeenergydiag}. The full molecular propagator is
denoted by the thick dashed line and the noninteracting molecular
propagator is denoted by the thin dashed line. The noninteracting
atomic propagators are indicated by the thin solid lines. The
total atomic density is calculated by using the thermodynamic
identity $N=-\partial \Omega (\mu,T)/\partial \mu$, which results
in
\begin{eqnarray}
\label{eq:densityfirst}
  n &=& -\frac{1}{\hbar \beta V} \sum_\bk
  \sum_n \left[ \frac{1}{i
  \omega_n-(\epsilon_\bk-\mu)/\hbar}\right]\nonumber \\
  &-& \frac{\partial}{\partial \mu} \frac{1}{\beta V} \sum_\bk
  \sum_n \ln \left[ \beta \left( \rule{0mm}{4mm}
   -i\hbar \omega_n + \epsilon_\bk/2+\delta (B) - 2 \mu \right.
   \right. \nonumber \\
   && \ \ \ \ \ \
   \left. \left. + \hbar \Sigma_{\rm m}^{\rm 2B} (i \hbar\omega_n -\epsilon_\bk/2+2\mu)
  \right)\right]~.
\end{eqnarray}
After performing the summation over the Matsubara frequencies in
this expression, the first term corresponds to the density of an
ideal gas of bosons. The second term in
Eq.~(\ref{eq:densityfirst}) is more complicated and should, in
principle, be dealt with numerically. For negative detuning we can
gain physical insight, however, by expanding the propagator around
its pole at the molecular binding energy $\epsilon_{\rm m} (B)$.
This leads to the approximation
\begin{eqnarray}
\label{eq:poleapproxmolfree}
  && \frac{\partial}{\partial \mu}
  \ln \left[ \beta \left(
   -i\hbar \omega_n + \epsilon_\bk/2+\delta (B) - 2 \mu
   + \hbar \Sigma^{\rm 2B}_{\rm m}(i \hbar\omega_n -\epsilon_\bk/2+2\mu)
  \right)\right] \nonumber \\
  && =\frac{-2\left[1- \left(\hbar \Sigma^{\rm 2B}_{\rm m}\right)'(i \hbar\omega_n -\epsilon_\bk/2+2\mu)
  \right]
  }{
   -i\hbar \omega_n + \epsilon_\bk/2+\delta (B) - 2 \mu
   + \hbar \Sigma^{\rm 2B}_{\rm m}(i \hbar\omega_n -\epsilon_\bk/2+2\mu)
  } \nonumber \\
  && \simeq \frac{2}{-i \hbar \omega_n+ \epsilon_\bk/2+\epsilon_{\rm m} (B) - 2
  \mu}~,
\end{eqnarray}
where we used the expression for the residue of the pole in
Eq.~(\ref{eq:factorz}). With this approximation the sum over the
Matsubara frequencies in Eq.~(\ref{eq:densityfirst}) is performed
easily and leads to the result
\begin{eqnarray}
\label{eq:densityresultpoleapprox}
   n &=& -\frac{1}{\hbar \beta V} \sum_\bk
  \sum_n \left[ \frac{1}{i
  \omega_n-(\epsilon_\bk-\mu)/\hbar} \right . \nonumber \\
  && \ \ \ \ \ \ \left. +
  \frac{2}{i \omega_n-(\epsilon_\bk/2+\epsilon_{\rm m} (B) -2
  \mu)/\hbar}\right] \nonumber \\
  &=& \frac{1}{V} \sum_\bk \left[ N(\epsilon_\bk-\mu) +2  N(\epsilon_\bk/2+\epsilon_{\rm m} (B) - 2 \mu
  ) \right]~.
\end{eqnarray}
This important result shows that in equilibrium in the normal
state and for negative detuning the gas in first approximation
behaves as an ideal-gas mixture of atoms and dressed molecules.
The same result is found if we neglect in the Heisenberg equations
of motion for the atomic and dressed molecular field operators in
Eq.~(\ref{eq:dressedheom}) the interaction terms and calculate the
total density in equilibrium.

Instead of calculating the thermodynamic potential and
differentiating with respect to the chemical potential we can also
calculate the total density directly by using
\begin{equation}
\label{eq:densitysecond}
   n = - G_{\rm a} (\bx,\tau;\bx,\tau^+) - 2 G_{\rm m}
   (\bx,\tau,\bx,\tau^+)~.
\end{equation}
An important difference between directly calculating the density
in this manner and calculating it indirectly from the
thermodynamic potential is that we should use in
Eq.~(\ref{eq:densitysecond}) not the noninteracting atomic
propagator. Instead, we should use an approximation to the atomic
propagator that contains the same self-energy diagrams as the
diagrams shown in Fig.~\ref{fig:freeenergydiag}. Conversely, in
calculating the thermodynamic potential with the use of
Eq.~(\ref{eq:freeenergyfirst}) we should not use the full atomic
propagator. The reason for this is that if we calculate ring
diagrams with this propagator we find diagrams which are already
contained in the ring diagram of the full molecular propagator.
The following explicit example clarifies this further.

\begin{figure}
\begin{center}
\includegraphics[width=10cm]{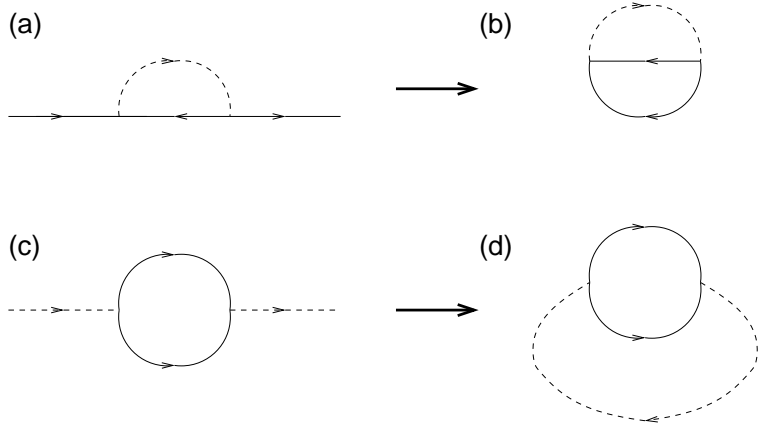}
\caption{\label{fig:densitydiag}
  Examples of approximations for (a) the atomic propagator and (c) the
 molecular propagator. The corresponding ring diagrams that
 contribute to the thermodynamic potential are shown in (b) and (d),
 respectively.
   }
\end{center}
\end{figure}

If we use for the atomic propagator the approximation given
diagrammatically in Fig.~\ref{fig:densitydiag}~(a), the ring
diagram that contributes to the thermodynamic potential is given
in Fig.\ref{fig:densitydiag}~(b). On the other hand, if we use for
the molecular propagator the approximation given in
Fig.~\ref{fig:densitydiag}~(c) the resulting ring diagram, given
in Fig.~\ref{fig:densitydiag}~(d), is exactly the same as
Fig.~\ref{fig:densitydiag}~(b). Clearly, to avoid double counting
problems in the calculation of the thermodynamic potential we
should take only one of these diagrams into account. However, if
we calculate the density directly from the atomic and molecular
propagators we should use both the diagrams given in
Fig.~\ref{fig:densitydiag}~(a)~and~(c).

We now argue that by directly calculating the density, again for
negative detuning, we indeed recover the result in
Eq.~(\ref{eq:densityresultpoleapprox}). We first calculate the
contribution arising from the molecular propagator. It is found to
be equal to
\begin{eqnarray}
\label{eq:contribmols}
 n_{\rm m} &\equiv& -G_{\rm m}
   (\bx,\tau,\bx,\tau^+) = - \frac{1}{\hbar \beta V} \sum_n
   \sum_\bk
   G_{\rm m} (\bk, i \omega_n) \nonumber \\
   &=&\frac{1}{V} \sum_\bk \int d(\hbar \omega) \rho_{\rm m}
   (\bk, \omega) \frac{1}{\hbar \beta} \sum_n \frac{1}{i \omega_n - (\hbar \omega - 2
   \mu)/\hbar} \nonumber  \\
   &=& \int \frac{d\bk}{(2 \pi)^3} \int d(\hbar \omega) \rho_{\rm m}
   (\bk, \omega) N (\hbar \omega - 2\mu)~.
\end{eqnarray}
Taking into account only the pole in the density of states leads
to the result
\begin{equation}
\label{eq:contribmolspole}
  n_{\rm m} = Z(B) \int \frac{d \bk}{(2 \pi)^3} N (\epsilon_\bk/2+\epsilon_{\rm m}
  (B)-2\mu)~.
\end{equation}
At first sight this result seems a factor $Z(B)$ to small to agree
with the result in Eq.~(\ref{eq:densityresultpoleapprox}).
However, we have, in fact, already seen in
Eq.~(\ref{eq:twominusz}) that the contributions from the atoms to
the density results in a term proportional to $2-2 Z(B)$. Taking
this into account, the result from the direct calculation agrees
with the result in Eq.~(\ref{eq:densityresultpoleapprox}) obtained
previously.

\begin{figure}
\begin{center}
\includegraphics[width=2cm]{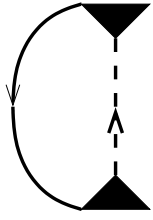}
\caption{\label{fig:atomselfenergy}
  Self-energy of the atoms. The solid and dashed thick lines
 correspond to the full atomic and molecular propagators,
 respectively. The filled triangles correspond to the renormalized
 atom-molecule coupling constant.
   }
\end{center}
\end{figure}

A different way for obtaining the factor $2-2 Z(B)$ in the atomic
density is to include the self-energy diagram shown in
Fig.~\ref{fig:atomselfenergy} in the atomic propagator. The
corresponding mathematical expression is in first instance given
by
\begin{equation}
\label{eq:selfenergyatoms}
  \hbar \Sigma_{\rm a} (\bk,i \omega_n) =- \frac{4 g^2}{V}
  \sum_{{\bf q},n} G_{\rm m} (\bk+{\bf q},i \omega_{n+m}) G_{\rm a}
  (\bk,i\omega_n)~.
\end{equation}
To understand the physics of this expression, we note that if we
neglect the energy and momentum dependence of the molecular
propagator we have that $G_{\rm m} (\bk,i \omega_n) \simeq
-\hbar/\delta(B)$. Within this approximation the self-energy is
given by $8 \pi n_{\rm a} a_{\rm res} (B) \hbar^2/m$, which
corresponds precisely to the Feshbach-resonant part of the
self-consistent Hartree-Fock self-energy of the atoms, as expected
from the diagram in Fig.~\ref{fig:atomselfenergy}.

The full calculation of the expression for the self-energy in
Eq.~(\ref{eq:selfenergyatoms}) is complicated due to the fact that
we have to use the full atomic and molecular propagators, which
makes the calculation self-consistent. To illustrate in
perturbation theory that we are able to reproduce the result in
Eq.~(\ref{eq:twominusz}) let us simply take the noninteracting
atomic and molecular propagators. The self-energy is then given by
\begin{eqnarray}
  \hbar \Sigma_{\rm a} (\bk,i\omega_n) = \frac{4g^2}{V}
   \sum_{\bf q} \frac{N(\epsilon_{\bf q}-\mu)-N(\epsilon_{\bk+{\bf q}}/2+\delta(B)-2\mu)}
   {i \hbar \omega_n - \left( \epsilon_{\bk+{\bf q}}/2-\epsilon_{\bf q}+\delta(B)
   -\mu\right)}~.
\end{eqnarray}
To compare with the two-atom calculation for negative detuning
performed in the previous section, we must take only one other
atom present with momentum $-\hbar \bk$, and no molecules. The
self-energy of the atom with momentum $\hbar \bk$ is then given by
\begin{equation}
 \hbar \Sigma_{\rm a} (\bk, i \omega_n) = \frac{4g^2}{V} \frac{1}{i \hbar
 \omega_n-(\delta(B)-\epsilon_\bk-\mu)}~.
\end{equation}
With this self-energy the retarded propagator of the atoms is
given by
\begin{equation}
  G_{\rm a}^{(+)} (\bk,\omega) = \frac{\hbar}{\hbar \omega^+ -
  \epsilon_\bk-\frac{4g^2}{V}\left[\hbar\omega^+ + \epsilon_\bk - \delta
  (B)\right]^{-1}} ~.
\end{equation}
It has two poles, one close to $\epsilon_\bk$, and one close to
$\delta (B)$. The residue of the latter is given by
\begin{equation}
  Z_\bk \simeq \frac{4g^2}{V} \frac{1}{\left[2\epsilon_\bk-\delta(B)\right]^2}~,
\end{equation}
in agreement with the result in Eq.~(\ref{eq:natomswithkdressed}).
Moreover, we have that
\begin{eqnarray}
  \sum_\bk Z_\bk = 2-2 Z(B).
\end{eqnarray}
Hence, the total density of the atoms is given by
\begin{eqnarray}
  n_{\rm a} \simeq \frac{(2-2 Z(B))}{V} \sum_\bk N(\epsilon_\bk/2+\delta(B)-2\mu)
   +\frac{1}{V} \sum_\bk N(\epsilon_\bk-\mu)~.
\end{eqnarray}
Together with the molecular density from
Eq.~(\ref{eq:contribmolspole}) that becomes
\begin{equation}
  n_{\rm m} \simeq \frac{Z(B)}{V} \sum_\bk N(\epsilon_\bk+\delta (B) -
  2\mu)~,
\end{equation}
the total density $2 n_{\rm m}+n_{\rm a}$ is again equal to the
result in Eq.~(\ref{eq:densityresultpoleapprox}) to lowest order
in the interactions.

\subsection{Applications}
\label{subsec:nsapplications}
\begin{figure}
\begin{center}
\includegraphics{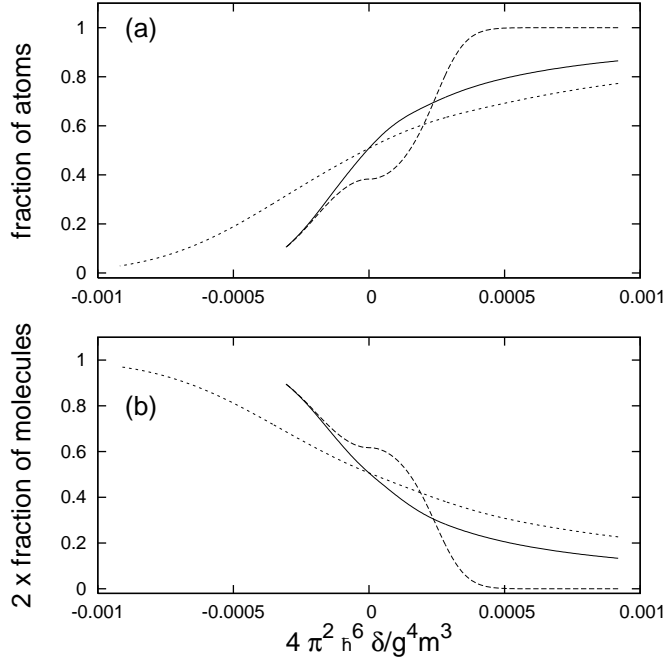}
\caption{\label{fig:nns_delta_t2}
 Fraction of atoms and fraction of atoms in molecules as a function of the
 detuning at a fixed temperature of $2 T_{0}$, for two different total densities. The
 solid line is the exact result that includes all two-atom
 physics, and particular the effects of the nonzero lifetime of
 the molecule at positive detuning,
 for a total atomic density of $n=10^{11}$ cm$^{-3}$. The dashed line shows the result for
 this density if we approximate the gas by an ideal-gas mixture of atoms and dressed molecules.
 The dotted line shows the exact result for a total density
 of $n=10^{12}$ cm$^{-3}$.
   }
\end{center}
\end{figure}
In this section we present results on the properties of the normal
state of the gas. First, we calculate the density of atoms and
molecules as a function of the detuning, at a fixed temperature.
Second, we calculate the density of atoms and molecules, and the
temperature of the gas, as a function of detuning at fixed entropy
and total density. This calculation is of interest because it
gives the outcome of a magnetic-sweep experiment through the
Feshbach resonance in the adiabatic approximation. Finally, we
calculate the critical temperature for Bose-Einstein condensation
as a function of the detuning, at fixed total density.

\subsubsection{Density of atoms and molecules}

As we have seen, the density of the gas is most easily calculated
by means of Eq.~(\ref{eq:densityfirst}). We report all our results
as a function of the detuning in units of the energy $g^4
m^3/4\pi^2\hbar^6$. The temperature is given in units of the
critical temperature for Bose-Einstein condensation of an ideal
gas of atoms with a total density $n$, i.e.,
\begin{equation}
\label{eq:tcritidealgas}
  T_{0} =  \frac{3.31 \hbar^2 n^{2/3}}{m k_{\rm B}}~.
\end{equation}
We compare the exact results, found from numerically performing
the summation over Matsubara frequencies in
Eq.~(\ref{eq:densityfirst}), to the ideal-gas mixture result in
Eq.~(\ref{eq:densityresultpoleapprox}). However, this last
equation was derived for negative detuning where there is a real
pole in the molecular Green's function. We extend this result to
positive detuning by describing the molecular gas at positive
detuning as an ideal gas of molecules with a positive bound-state
energy given by the energy at which the molecular density of
states in Eq.~(\ref{eq:defdos}) has its maximum. It turns out
that, for small detuning, this maximum is at $\hbar^2/ma^2$.
Furthermore, this approximation implies that we ignore the
physical effects of the lifetime of the molecule, that is nonzero
for positive detuning, on the equilibrium properties of the gas.
These lifetime effects are however included in the exact result in
Eq.~(\ref{eq:densityfirst}).

In Fig.~\ref{fig:nns_delta_t2} the results of the calculation of
the density as a function of the detuning are shown, for a
temperature of $T=2T_{0}$. Fig.~\ref{fig:nns_delta_t2}~(a) shows
the fraction of atoms and Fig.~\ref{fig:nns_delta_t2}~(b) shows
the number of atoms in molecules, i.e., twice the fraction of
molecules, as a function of the detuning. The solid and the dotted
lines show the exact result for a total atomic density of
$n=10^{11}$ cm$^{-3}$ and $n=10^{12}$ cm$^{-3}$, respectively. As
expected, for negative detuning most of the atoms in the gas are
bound to molecules. At positive detuning, most of the atoms are
free. Moreover, we find that the width in detuning of this
crossover regime is approximately equal to the temperature. The
solid lines show the ideal-gas mixture result for a density of
$n=10^{11}$ cm$^{-3}$, i.e., the result that does not incorporate
the effects of the nonzero lifetime of the molecules at positive
detuning. For negative detuning, we observe that this ideal-gas
result becomes equal to the exact result. This implies that the
pole approximation in Eq.~(\ref{eq:poleapproxmolfree}) is indeed a
reasonable approximation sufficiently far from resonance. An
important conclusion is therefore that, for sufficiently negative
detuning, we are allowed to treat the gas as an ideal-gas mixture
of atoms and dressed molecules with binding energy $\epsilon_{\rm
m} (B)$. For positive detuning, the ideal-gas result differs
substantially from the exact result. In particular, for relatively
large detuning, the ideal-gas calculation considerably
underestimates the number of molecules. The exact result shows
that there is, even at relatively large detuning, a significant
fraction of molecules in the gas. This is the result of the finite
lifetime of the molecules in this case. Physically, this comes
about because the molecular density of states for positive
detuning has significant spectral weight at low energies. In
equilibrium, this leads to a significant fraction of molecules.
For even larger positive detuning, the ideal-gas result reduces
again to the exact result.

\subsubsection{Adiabatic sweep through the resonance}
\begin{figure}
\begin{center}
\includegraphics{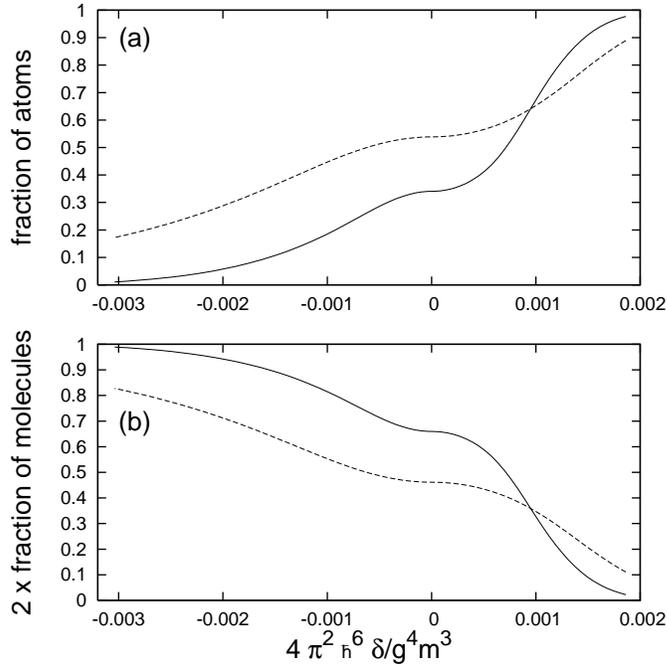}
\caption{\label{fig:n_b_sweep}
 Fraction of atoms and twice the fraction of molecules as a function of the detuning for
 an adiabatic sweep through the resonance. The total atomic density is
 equal to $n=10^{13}$ cm$^{-3}$. The solid lines show the result
 for an initial temperature of $T=2T_0$. The dashed lines show the
 result for $T=4T_0$.
   }
\end{center}
\end{figure}
We now calculate the number of atoms and molecules in the gas
during an adiabatic sweep in the magnetic field, such that the
detuning changes from positive to negative. The condition for
adiabaticity is that the entropy of the gas is constant. The
entropy is given by
\begin{equation}
  S = - \frac{\partial \Omega}{\partial T}~.
\end{equation}
The total number of atoms is, of course, also constant througout
the sweep. As we have seen, for sufficiently large absolute values
of the detuning, the gas is well-described by an ideal-gas
approximation. For simplicity, we will therefore treat the gas
here as an ideal-gas mixture since we are mostly interested in the
final density of atoms and molecules and the final temperature of
the gas after the sweep, for which an ideal-gas treatment is
sufficient \cite{kokkelmans2003}.

In Fig.~\ref{fig:n_b_sweep} the results of the calculation of the
fraction of atoms and twice the fraction of molecules is
presented. The total atomic density is taken equal to $n=10^{13}$
cm$^{-3}$. The solid lines show the result for an initial
temperature of $T=2T_0$, and the dashed lines show the result for
an initial temperature of $T=4T_0$. As we go from positive to
negative detuning, most of the atoms in the gas are converted to
molecules. The range of detuning where the conversion takes place
is proportional to the initial temperature of the gas, as
expected.

\begin{figure}
\begin{center}
\includegraphics{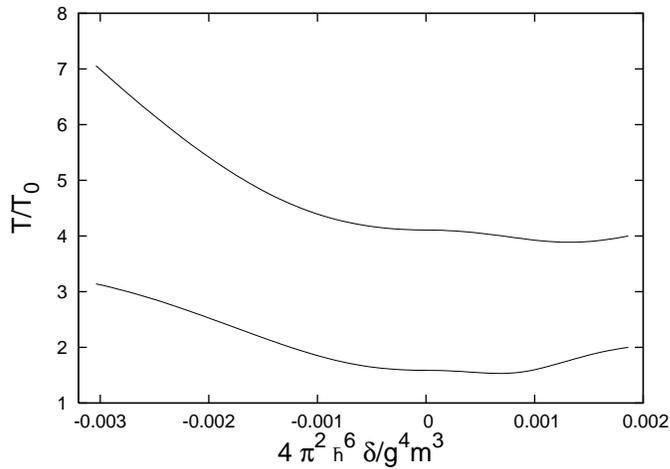}
\caption{\label{fig:t_b_sweep}
 Temperature of the gas as a function of the detuning for a sweep
 through the resonance from positive to negative detuning, for two
 inital temperatures. The total atomic density is equal to $n=10^{13}$
 cm$^{-3}$.
   }
\end{center}
\end{figure}

In Fig.~\ref{fig:t_b_sweep} the temperature is plotted as a
function of the detuning for the two initial temperatures $T=2
T_0$ and $T=4T_0$. The total density is again equal to $n=10^{13}$
cm$^{-3}$. Clearly, the gas is heated as the detuning is changed
from positive to negative. This is easily understood, since
molecules form as the detuning is changed from positive to
negative values, and their binding energy is released as kinetic
energy into the gas.

\subsubsection{Critical temperature}
Finally, we calculate the critical temperature for Bose-Einstein
condensation of the atom-molecule mixture, at a fixed total atomic
density. The results are presented in Fig.~\ref{fig:tcrit}, for a
total density of $n=10^{13}$ cm$^{-3}$. The solid line shows the
exact calculation and the dashed line shows the ideal-gas mixture
result. For positive detuning and far from resonance, we are
essentially dealing with an atomic gas. Hence we have in this
regime that $T_{\rm BEC}=T_0$. For sufficiently negative detuning
we are dealing with a gas of molecules with twice the atomic mass,
and hence we have that $T_{\rm BEC}=2^{-5/3}T_0$. The feature in
the critical temperature at zero detuning turns out to be a
signature of a true thermodynamic phase transition, between a
phase with a single Bose-Einstein condensate of molecules and a
phase containing two Bose-Einstein condensates, one of atoms and
one of molecules, as was first pointed out by Sachdev
\cite{sachdev2003}. This should be contrasted with the situation
of an atomic Fermi gas near a Feshbach resonance, where only a
BCS-BEC crossover exists \cite{nozieres1985}. The calculation of
the full detuning-temperature phase diagram is work in progress
and will be reported in a future publication \cite{romans2003}.

\begin{figure}
\begin{center}
\includegraphics{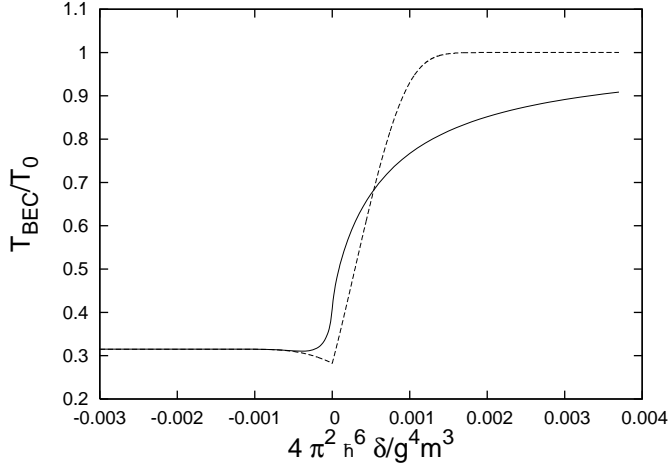}
\caption{\label{fig:tcrit}
 Critical temperature for Bose-Einstein condensation as a function
 of detuning. The total density is equal to $n=10^{13}$ cm$^{-3}$. The solid line shows the result of the
exact calculations. The dashed line shows the result of treating the gas as an ideal-gas mixture.
   }
\end{center}
\end{figure}

\subsection{Many-body effects on the bound-state energy}
\label{subsec:mbeffonbse} In this section we determine the effects
of the atomic gas on the molecular binding energy. The first step
in an examination of these many-body effects is the calculation of
the molecular self-energy given in Eq.~(\ref{eq:selfenergy}). For
simplicity, we neglect the energy dependence of the atom-molecule
coupling constant and the many-body effects on this coupling
constant. After subtraction of the energy-independent shift, the
retarded molecular self-energy that includes many-body effects is
given by the expression
\begin{eqnarray}
\label{eq:mbselfenergy}
  && \hbar \Sigma_{\rm m}^{ (+)} ({\bf K},\omega) = \nonumber \\
  && 2g^2\!
      \int\!\frac{d \bk}{(2 \pi)^3}\!\left\{
      \frac{\left[ 1+ N(\epsilon_{{\bf K}/2+\bk}\!-\!\mu')+N(\epsilon_{{\bf
    K}/2-\bk}\!-\!\mu')\right]}
    {\hbar \omega^+ -\epsilon_\bK/2-2 (\epsilon_\bk-\mu')}\!+\!\frac{1}{2 \epsilon_\bk} \right\}\!~.
\end{eqnarray}
Here, we have treated the atoms in the Hartree-Fock approximation
which effectively implies that the chemical potential is shifted
according to
\begin{equation}
  \mu' = \mu - \frac{8 \pi a(B) \hbar^2n_{\rm a}}{m} \equiv \mu - 2 T^{\rm 2B} n_{\rm a}~,
\end{equation}
where $n_{\rm a}$ is the density of the atoms. In this expression
for the Hartree-Fock self-energy correction to the chemical
potential we have neglected the energy-dependence of the
interactions, which is justified as long as the scattering length
is much smaller than the thermal deBroglie wavelength of the
atoms.

\begin{figure}
\begin{center}
\includegraphics{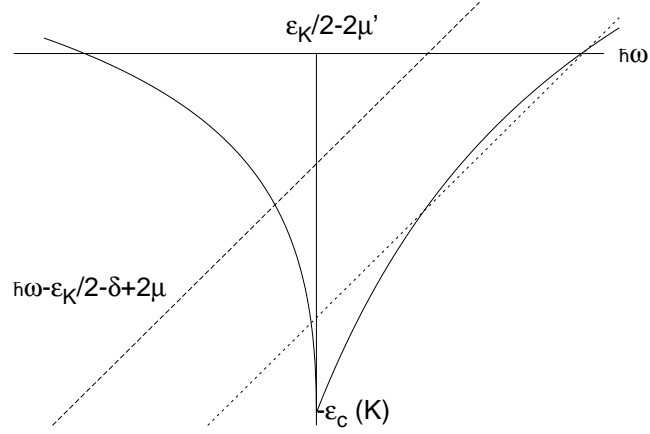}
\caption{\label{fig:graphsol}
  Graphical solution of the equation for the molecular bound-state
 energy. The solid line indicates the real part of the molecular
 self-energy as a function of $\hbar \omega$.
 The dashed and dotted lines indicates the function $\hbar \omega-\epsilon_\bK/2-\delta
 (B)+2\mu$ for different values of the detuning $\delta (B)$. For $\hbar \omega < \epsilon_\bK/2-2\mu'$, the
 value of $\hbar \omega$ at the intersections of the dashed and dotted lines with the solid line
 corresponds to the bound-state energy. For $\hbar \omega >
 \epsilon_\bK/2-2\mu'$ it corresponds to the energy of resonant
 states.
   }
\end{center}
\end{figure}

From now on we restrict ourselves to the regime just above the
critical temperature, where we are able to calculate various
properties analytically. Since the chemical potential approaches
zero from below in this regime, we are allowed to approximate the
Bose distribution function of the atoms by
\begin{equation}
  N(x) \simeq \frac{1}{\beta x}~.
\end{equation}
Within this approximation, the self-energy of the molecules is
given by
\begin{eqnarray}
  && \hbar \Sigma_{\rm m}^{(+)} (\bK,\omega) = \nonumber \\
  && \ \ \ 4 g^2 \int \frac{d \bk}{(2 \pi)^3}
  \frac{1}{\hbar \omega^+ - \epsilon_\bK/2 - 2( \epsilon_\bk -\mu')}
  \frac{1}{\beta (\epsilon_\bK/4+\epsilon_\bk-\mu')}~,
\end{eqnarray}
and we are allowed to also neglect the square-root term that
results from the first and last terms in the integrand in
Eq.~(\ref{eq:mbselfenergy}), and is due to two-atom physics. This
integral is performed analytically. For $\hbar \omega <
\epsilon_\bK/2 - 2 \mu'$ the self-energy is real and given by
\begin{eqnarray}
  \hbar \Sigma_{\rm m}^{(+)} (\bK,\omega) = \frac{2g^2m^{3/2}}{\pi \hbar^3 \beta}
   \left[ \frac{\sqrt{\epsilon_\bK/2-2\mu'-\hbar \omega}
   -\sqrt{\epsilon_\bK/2-2\mu'}}{\hbar \omega}\right]~.
\end{eqnarray}
For $\hbar \omega > \epsilon_\bK/2-2\mu'$ the self-energy contains
an imaginary part and is given by
\begin{eqnarray}
  \hbar \Sigma_{\rm m}^{(+)} (\bK,\omega) =-\frac{2g^2m^{3/2}}{\pi \hbar^3 \beta}
   \left[\frac{\sqrt{\epsilon_\bK/2-2\mu'}
   +i  \sqrt{\hbar \omega - \epsilon_\bK/2+2\mu'}}{{\hbar
   \omega}}\right]~.
\end{eqnarray}

To find the energy of the molecular state we have to solve for
$\hbar \omega$ in the equation
\begin{equation}
 \hbar \omega - \epsilon_\bK/2 - \delta (B) +2 \mu- \hbar
 \Sigma_{\rm m}^{(+)} (\bK,\omega)=0~.
\end{equation}
A great deal of insight is gained by the graphic representation of
this equation which is shown in Fig.~\ref{fig:graphsol}. The solid
line represents the real part of the molecular self-energy as a
function of the energy $\hbar \omega$. The straight dashed and
dotted lines correspond to $\hbar \omega - \epsilon_\bK/2  -
\delta (B)+ 2\mu$, for two different values of $\delta(B)$. From
this figure it is clear that there is a real solution, i.e., a
true bound state, if the detuning is such that
\begin{equation}
\label{eq:detcrit}
  \delta (B) < 4 T^{\rm 2B} n_{\rm a} + \frac{2 g^2 m^{3/2}}{\pi \hbar^3 \beta
  \sqrt{\epsilon_\bK/2-2\mu'}} \equiv 4 T^{\rm 2B} n_{\rm a} + \epsilon_{\rm c} (\bK) \equiv \delta_{\rm max}(\bK)~.
\end{equation}
Note that this also implies that the position of the resonance in
the magnetic field is shifted according to
\begin{equation}
  B_0 \to B_0 + \frac{1}{\Delta \mu} \left( 4 T^{\rm 2B} n_{\rm a} + \frac{2 g^2 m^{3/2}}{\pi \hbar^3 \beta
  \sqrt{-2\mu'}} \right)~,
\end{equation}
due to many-body effects.

\begin{figure}
\begin{center}
\includegraphics{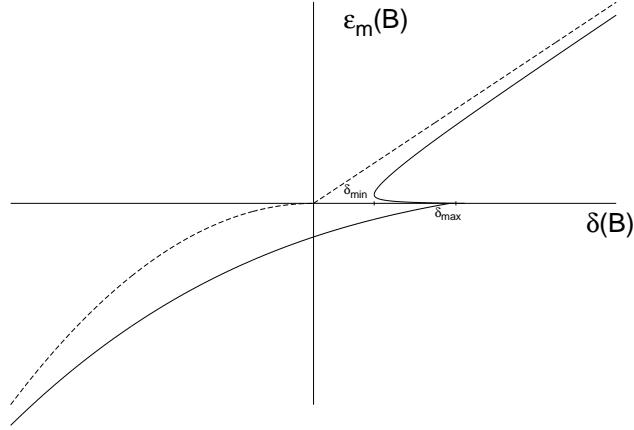}
\caption{\label{fig:emmb}
  Molecular bound-state energy as a function of detuning. The
 dashed line show de molecular bound-state energy in vacuum as a function of
 detuning. The solid line shows the many-body effects
 on the bound-state energy. }
\end{center}
\end{figure}

For a magnetic field such that the detuning is just below the
maximum value $\delta_{\rm max}(\bK)$ given in
Eq.~(\ref{eq:detcrit}), the bound-state energy is given by
\begin{equation}
  \hbar \omega_\bK \simeq
  -2\mu' \left[ 1- \left(\frac{4 T^{\rm 2B} n_{\rm a}-\delta (B)}{\epsilon_{\rm c} ({\bf 0})}
   +1\right)^2 \right] + \frac{\hbar^2 \bK^2}{2m_{\rm eff}}~,
\end{equation}
with an effective mass given by
\begin{equation}
  m_{\rm eff} =
  2m \left[ \frac{3(\delta (B) - 4 T^{\rm 2B} n_{\rm a})}{\epsilon_{\rm c} ({\bf
  0})}
  \left( 1-\frac{2}{3}\frac{(\delta (B) - 4T^{\rm 2B}n_{\rm a})}{\epsilon_{\rm c} ({\bf 0})}
  \right)\right]^{-1}~.
\end{equation}
This effective mass has a minimum value of $4m/3$ at detuning
$\delta (B) = 4 T^{\rm 2B} n_{\rm a} + 3 \epsilon_{\rm c} ({\bf
0})/4$, and diverges for smaller detunings close to $4 T^{\rm 2B}
n_{\rm a}$. In the limit of the detuning $\delta (B) \to -\infty$
we have to recover the two-body bound state with mass $2m$, which
shows that this divergence is due to the approximations we have
adopted. As already discussed, we have in particular neglected the
first and last terms in the integrand in
Eq.~(\ref{eq:mbselfenergy}) that result from two-atom physics.
Nevertheless, the fact that the effective mass is smaller than the
mass of a molecule close to resonance indicates that the molecule
crosses over to a more complex many-body bound state of the
system. Precisely at the shifted resonance at $\delta (B) = 4
T^{\rm 2B} n_{\rm a}+\epsilon_{\rm c} ({\bf 0})$ the effective
mass is again equal to $2m$. Another interesting feature of the
excitation is that for a given detuning it only exist at small
momenta such that Eq.~(\ref{eq:detcrit}) is obeyed.

\begin{figure}
\begin{center}
\includegraphics{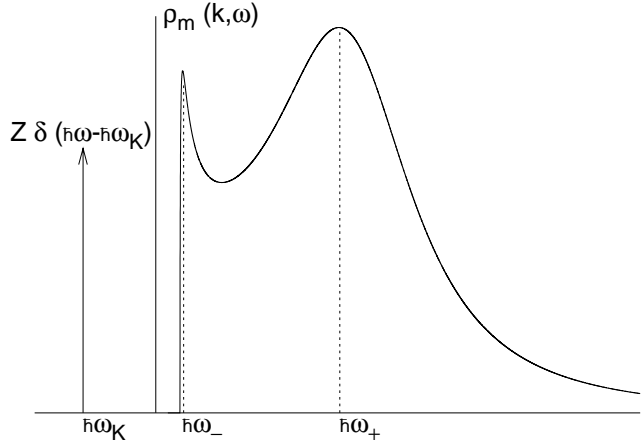}
\caption{\label{fig:dosmb}
  Molecular density of states with many-body effects. Apart from the
 delta function that corresponds to the bound state there are two resonant states,
 indicated by the dashed lines.}
\end{center}
\end{figure}

The intersections at energies $\hbar \omega >
\epsilon_\bK/2-2\mu'$ in Fig.~\ref{fig:graphsol}, as for example
shown by the dotted line, correspond to resonant states since the
self-energy contains an imaginary part at these energies. The
energies of these resonant states is determined by solving for
$\hbar \omega$ in the equation
\begin{equation}
  \hbar \omega - \epsilon_\bK/2 + 2 \mu - \delta (B) +\frac{2g^2m^{3/2}}{\pi \hbar^3 \beta}
   \frac{\sqrt{\epsilon_\bK/2-2\mu'}}{\hbar \omega}=0~.
\end{equation}
For a detuning that obeys the condition in Eq.~(\ref{eq:detcrit})
and such that
\begin{equation}
 \delta (B) > 2 \mu -\epsilon_\bK/2 +
  \sqrt{\frac{8g^2m^{3/2}}{\pi\hbar^3\beta}\sqrt{\epsilon_\bK/2-2\mu'}}
  \equiv \delta_{\rm min}(\bK)~,
\end{equation}
there are two solutions of this equation. They are given by
\begin{eqnarray}
  \hbar \omega_\pm &=& \frac{1}{2}
   \left( \epsilon_\bK/2+\delta(B)-2\mu \right) \nonumber \\
   && \times
   \left(
    1 \pm \sqrt{1-\frac{8g^2m^{3/2}}{\pi\hbar^3\beta}\frac{\sqrt{\epsilon_\bK/2-2\mu'}}
    {(\epsilon_\bK/2+\delta(B)-2\mu)^2}}
   \right)~.
\end{eqnarray}
For large detuning we have that $\hbar \omega_+ \simeq
\epsilon_\bK/2+\delta(B)-2\mu$, from which we see that this
resonant state physically corresponds to the bare molecular state,
which has obtained a finite lifetime due to the interaction with
the atomic continuum. The resonant state at energy $\hbar
\omega_-$ is not present in the two-atom case but arises purely
due to many-body effects. This situation is somewhat similar to
the Kondo-resonant state that arises in a Fermi gas near a
Feshbach resonance \cite{falco2003}.

An illustration of the many-body effects on the molecular
bound-state energy is shown in Fig.~\ref{fig:emmb}. The dashed
line indicates the situation in vacuum. For negative detuning
there is a true molecular state whose energy depends quadratically
on the detuning, as given in Eq.~(\ref{eq:solbse}). For positive
detuning the molecule has a finite lifetime and therefore
corresponds to a resonant state, whose energy is in first
approximation equal to the detuning. Due to many-body effects, the
position of the Feshbach resonance is shifted. Nevertheless, there
is still a molecular state with an energy dependence that is
quadratic on the many-body renormalized detuning. However, for a
detuning larger than $\delta_{\rm min}$ but less than $\delta_{\rm
max}$ this molecular state coexists with two resonant states, one
close to the detuning and one just above the continuum threshold.
The molecular density of states for the latter situation is shown
in Fig.~\ref{fig:dosmb}. The delta function corresponds to the
molecular bound state. The dashed lines indicate the position of
the resonances. For large positive and large negative detuning the
many-body effects are negligible and the result reduces to the
two-atom result.

Finally, we remark that the resonant state at energy $\hbar
\omega_-$, that arises solely due to many-body effects, leads to a
nonzero number of bare molecules, even if the temperature is much
smaller than the detuning. This effect can be measured by directly
measuring the number of bare molecules, as achieved recently by
Chin {\it et al.} \cite{chin2003}. The investigation of the
magnitude and temperature dependence of this effect is intended
for future work.

  \section{Mean-field theories for the Bose-Einstein condensed phase}
\label{sec:meanfieldbec} In the first part of this section we
derive the mean-field theory that results from our effective
quantum field theory. This mean-field theory is appropriate for
the description of the Bose-Einstein condensed phase of the gas.
In the second section we discuss other possible mean-field
theories and discuss the similarities and differences between them
and our mean-field theory.

\subsection{Popov theory} \label{subsec:mft} In this section we
derive the mean-field equations for the atomic and molecular
condensate wave functions. In the first part of this section we
derive the time-independent equations and discuss the excitation
spectrum. In the second part we derive  the time-dependent
mean-field equations.

\subsubsection{Time-independent mean-field equations} The
mean-field equations for the atomic and molecular condensate wave
functions are derived most easily by varying the effective action
in Eq.~(\ref{eq:seffatommolkspace}) with respect to $a^*_{\bk,n}$
and $b^*_{\bk,n}$, respectively. Before doing so, however, we
remark that an important property of this effective action is its
invariance under global $U(1)$ transformations. Namely, any
transformation of the form
\begin{eqnarray}
\label{eq:uone}
  a_{\bk,n} &\to& a_{\bk,n} e^{i \theta}~, \nonumber \\
  b_{\bk,n} &\to& b_{\bk,n} e^{2 i \theta}~,
\end{eqnarray}
with $\theta$ a real parameter, leaves the action unchanged. The
conserved quantity, the so-called Noether charge, associated with
this invariance is the total number of atoms. The appearance of
the atomic and the molecular condensates breaks the $U(1)$
invariance since the wave functions of these condensates have a
certain phase. According to Goldstone's theorem, an exact property
of a system with a broken continuous symmetry is that its
excitation spectrum is gapless \cite{goldstone1961}. Since our
mean-field theory is derived by varying a $U(1)$-invariant action,
this property is automically incorporated in the mean-field
theory.

To derive the time-independent mean-field equations, that describe
the equilibrium values of the atomic and molecular condensate wave
functions, we substitute into the effective action $a_{{\bf 0},0}
\to \phia \sqrt{\beta \hbar V} + a_{{\bf 0},0}$ and $b_{{\bf 0},0}
\to \phim \sqrt{\beta \hbar V} + b_{{\bf 0},0}$. Here, $\phia$ and
$\phim$ correspond to the atomic and molecular condensate wave
functions, respectively. Requiring that the terms linear in
$a_{{\bf 0},0}$ and $b_{{\bf 0},0}$ vanish from the effective
action leads to the equations
\begin{eqnarray}
\label{eq:mfetimeindep}
 \mu \phia\!&\!=\!&\!T^{\rm 2B}_{\rm bg} \left(2\mu-2 \hbar \Sigma^{\rm
  HF}\right) |\phia|^2 \phia + 2\left[g^{\rm 2B} \left( 2\mu-2 \hbar \Sigma^{\rm
  HF} \right)\right]^* \phia^* \phim~, \nonumber \\
2\mu\phim\!&\!=\!&\!\left[ \delta (B) + \hbar \Sigma^{\rm 2B}_{\rm
m} \left( 2\mu-2 \hbar \Sigma^{\rm HF} \right) \right] \phim
 + g^{\rm 2B} \left(2\mu-2 \hbar \Sigma^{\rm HF} \right)
 \phia^2~.
\end{eqnarray}
A crucial ingredient in these equations is the Hartree-Fock
self-energy of the noncondensed atoms. This self-energy is the
mean-field energy felt by the noncondensed atoms due to the
presence of the atomic condensate. Taking into account the
energy-dependence of the interactions, it is determined by the
expression
\begin{eqnarray}
\label{eq:sigmahf}
  \hbar \Sigma^{\rm HF}\!=\!
 2 n_{\rm a}\!\left(
 \frac{2 \left|g^{\rm 2B}\left(\mu-\hbar \Sigma^{\rm
  HF}\right)\right|^2}{\hbar \Sigma^{\rm HF}+\mu
 -\delta (B) - \hbar \Sigma_{\rm m}^{\rm 2B}\left(\mu\!-\!\hbar \Sigma^{\rm HF} \right)
 }\!+\!T^{\rm 2B}_{\rm bg} \left(\mu\!-\!\hbar \Sigma^{\rm
 HF}\right)\right),
\end{eqnarray}
with $n_{\rm a} = |\phia|^2$ the density of the atomic condensate.
Its diagrammatic representation is given in
Fig.~\ref{fig:sigmahf}. The overall factor of two comes from the
constructive interference of the direct and exchange
contributions. Far off resonance we are allowed to neglect the
energy-dependence of the effective atom-atom interactions, and the
Hartree-Fock self-energy of the atoms is given by $8\pi a(B)
\hbar^2 n_{\rm a}/m$, as expected. The Hartree-Fock self-energy is
essential for a correct description of the equilibrium properties
of the system. The physical reason for this is understood as
follows. In the condensed phase the chemical potential is
positive. The energy of a condensate molecule is equal to $2 \mu$,
which is therefore larger than the continuum threshold of two
atoms in vacuum. Without the incorporation of the Hartree-Fock
self-energy, the molecular condensate would therefore always decay
and an equilibrium solution of the mean-field equations would not
exist. However, due to the presence of the atomic condensate the
continuum threshold shifts by an amount $2 \hbar \Sigma^{\rm HF}$,
and the molecular condensate is stable.

\begin{figure}
\begin{center}
\includegraphics[width=5cm]{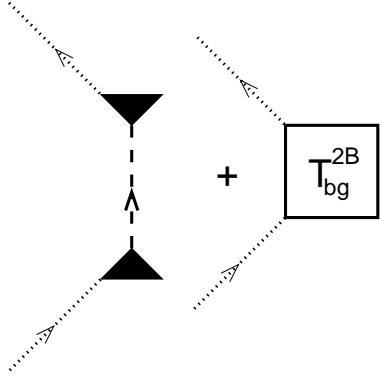}
\caption{\label{fig:sigmahf}
  Hartree-Fock self energy of the atoms.
 The dotted lines correspond to condensate atoms.
 The dashed line corresponds to the full molecular propagator.}
\end{center}\end{figure}

To study the collective excitation spectrum over the ground state
determined by Eq.~(\ref{eq:mfetimeindep}), we consider the
effective action up to second order in the fluctuations, which is
known as the Bogliubov approximation \cite{bogoliubov1947}. To
facilitate the notation we introduce the vector ${\bf u}_{\bk,n}$
by means of
\begin{eqnarray}
 {\bf u}_{\bk,n} \equiv \left(
  \begin{array}{c}
      a_{\bk,n} \\
      a^*_{-\bk,-n} \\
      b_{\bk,n} \\
      b^*_{-\bk,-n}
    \end{array}
 \right)~.
\end{eqnarray}
With this definition, the quadratic part of the effective action
is given by
\begin{equation}
  S_{\rm B}[{\bf u}^{\dagger},{\bf u}] = -\frac{\hbar}{2} \sum_{\bk,n} {\bf u}^{\dagger}_{\bk,n} \cdot
  {\bf G}^{-1}_{\rm B} (\bk,i\omega_n)  \cdot {\bf u}_{\bk,n}~,
\end{equation}
where the Green's function of the fluctuations is determined by
\begin{eqnarray}
  {\bf G}_{\rm B}^{-1}  =
 \left( \begin{array}{cc}
   {\bf G}^{-1}_{\rm a} & {\bf G}^{-1}_{\rm coup} \\
   \left[ {\bf G}^{-1}_{\rm coup} \right]^* & {\bf G}^{-1}_{\rm m}
 \end{array} \right)~.
\end{eqnarray}
The atomic part of this Green's function is found from
\begin{eqnarray}
\label{eq:bogoatoms}
  && -\hbar {\bf G}_{\rm a}^{-1} (\bk,i\omega_n)=
  \left( \begin{array}{cc}
   -\hbar G_{0,{\rm a}}^{-1} (\bk,i\omega_n)& 0\\
  0 & -\hbar G_{0,{\rm a}}^{-1} (\bk,-i\omega_n)
   \end{array} \right) +\nonumber \\
  &&\!\!\!\!\!\!\!\!\!\! \left( \begin{array}{ll}
   2 T^{\rm 2B}_{\rm bg}
  \left( i \hbar \omega_n - \epsilon_\bk/2+2\mu'\right) n_{\rm a}
  &   T^{\rm 2B}_{\rm bg} \left( 2\mu'\right) \phia^2 +2 \left[g^{\rm 2B} (2\mu') \right]^*\phim \\
  T^{\rm 2B}_{\rm bg} \left(2\mu' \right) \left(\phia^*\right)^2+2 g^{\rm 2B} (2 \mu') \phim^*
   &
    2 T^{\rm 2B}_{\rm bg}
  \left( i \hbar \omega_n - \epsilon_\bk/2 +2\mu'\right) n_{\rm a}
 \end{array} \right)
\end{eqnarray}
where $\mu'\equiv \mu-\hbar \Sigma^{\rm HF}$. Note that in the
absence of the coupling to the molecular condensate, this result
reduces to the well-known result for the Green's function that
describes phonon propagation in a weakly-interacting Bose
condensate. We have in this case, however, also explicitly taken
into account the energy dependence of the coupling constants.
Therefore we know that in the limit of vanishing coupling $g^{\rm
2B}$ the propagator in Eq.~(\ref{eq:bogoatoms}) has a pole that
determines the gapless dispersion relation for the phonons. For
energy-independent interactions this so-called Bogoliubov
dispersion is given by
\begin{equation}
\label{eq:bogodisp}
  \hbar \omega_\bk = \sqrt{\epsilon_\bk^2+\frac{8 \pi a_{\rm bg} \hbar^2 n_{\rm
  a}}{m} \epsilon_\bk}~.
\end{equation}

The molecular part of the Green's function ${\bf G}_{\rm B}
(\bk,i\omega_n)$ is determined by
\begin{eqnarray}
\label{eq:bogomols}
  && {\bf G}_{\rm m}^{-1} (\bk,i\omega_n)=
  \left( \begin{array}{cc}
    G_{\rm m}^{-1} (\bk,i\omega_n)& 0\\
  0 &  G_{\rm m}^{-1} (\bk,-i\omega_n)
   \end{array} \right) ~,
\end{eqnarray}
where the single-molecule propagator is given by
\begin{eqnarray}
\label{eq:gfatomshf}
  -\hbar G_{\rm m}^{-1} (\bk,i\omega_n) &=& - i \hbar
  \omega_n+\epsilon_\bk/2+\delta (B) -2\mu \nonumber \\
  && + \hbar \Sigma^{\rm 2B}_{\rm m} \left( i \hbar
  \omega_n-\epsilon_\bk/2+2\mu-2\hbar \Sigma^{\rm HF}\right)~.
\end{eqnarray}
From the previous section we know that the Green's function in
Eq.~(\ref{eq:gfatomshf}) for negative detuning has a pole at the
molecular binding energy. There are now, however, mean-field
effects on this binding energy due to the presence of the atomic
condensate, incorporated by the Hartree-Fock self-energy $\hbar
\Sigma^{\rm HF}$ \cite{duine2003a}. Finally, the Green's function
that describes the coupling between the atomic and molecular
fluctuations is given by
\begin{eqnarray}
\label{eq:bogoatommols}
  && -\hbar {\bf G}_{\rm coup}^{-1} (\bk,i\omega_n)= \nonumber \\
  &&
  \left( \begin{array}{cc}
  2\left[g^{\rm 2B} \left( i \hbar \omega_n\!-\!\epsilon_\bk/2\!+\!2\mu' \right) \right]^* \phia^* & 0\\
  0 & 2g^{\rm 2B} \left( i \hbar \omega_n\!-\!\epsilon_\bk/2\!+\!2\mu'\right) \phia
   \end{array} \right)
\end{eqnarray}

The spectrum of the collective excitations is determined by the
poles of the retarded Green's function for the fluctuations ${\bf
G}_{\rm B} (\bk,\omega^+)$. This implies that we have to solve for
$\hbar \omega$ in the equation
\begin{equation}
\label{eq:collmodesfreq}
  \det {\bf G}_{\rm B}^{-1} (\bk,\omega^+) =0~.
\end{equation}
This is achieved numerically in the next section to determine the
frequency of the Josephson oscillations between the atomic and the
molecular condensate. However, we are already able to infer some
general features of the excitation spectrum of the collective
modes. We have seen that in the absence of the coupling between
the atomic and molecular condensate, we have that one dispersion
is equal to the gapless Bogoliubov dispersion with scattering
length $a_{\rm bg}$. In the presence of the coupling this branch
corresponds again to phonons, but the dispersion is now
approximately equal to the Bogoliubov dispersion for the full
scattering length $a(B)$. There is a second dispersion branch that
for small coupling $g^{\rm 2B}$ lies close to the molecular
binding energy. At nonzero coupling this branch corresponds to
coherent atom-molecule oscillations, i.e., pairs of atoms
oscillating back and forth between the atomic and molecular
condensate. Physically, the difference between the two branches is
understood by realizing that for the phonon modes the phases of
the atomic and the molecular condensate are locked to each other
and oscillate in phase. Since the action is invariant under the
transformations in Eq.~(\ref{eq:uone}) we conclude that the
phonons are indeed gapless, and, in fact, correspond to the
Goldstone mode associated with the breaking of the $U(1)$ symmetry
by the condensates. For the coherent atom-molecule oscillations
the phases of the atomic and molecular condensate oscillate out of
phase and hence the associated dispersion is gapped. As a final
remark we note that we indeed have that
\begin{equation}
 \det {\bf G}_{\rm B}^{-1} ({\bf 0},0) =0~,
\end{equation}
which shows that there is indeed a gapless excitation, in
agreement with Goldstone's theorem.

\subsubsection{Time-dependent mean-field equations}
\label{subsubsec:tdpdtmfe} The time-dependent mean-field equations
are found most easily by taking the expectation value of the
Heisenberg equations of motion in Eq.~(\ref{eq:heom}). For
notational convenience we restrict ourselves to the situation that
we are close to resonance and hence neglect the energy-dependence
of the various couplings. Moreover, we only take into account the
leading-order energy dependence of the molecular self-energy, as
given in Eq.~(\ref{eq:resultselfenergy}). Furthermore, we assume
that we are at such low temperatures that the effects of the
thermal cloud may be neglected. Within these approximations, the
mean-field equations are given by
\begin{eqnarray}
\label{eq:mfe}
   i \hbar \frac{\partial \phia ({\bf x},t)}{\partial t}
    &=& \left[-\frac{\hbar^2 {\bf \nabla}^2}{2m}
        +\frac{4\pi a_{\rm bg} \hbar^2}{m} |\phia({\bf x},t)|^2
    \right]
      \phia ({\bf x},t) \nonumber \\ &&+ 2 g  \phia^* ({\bf x},t)
    \phim ({\bf x},t)~, \nonumber \\
   i \hbar \frac{\partial \phim ({\bf x},t)}{\partial t}
    &=&\left[-\frac{\hbar^2 {\bf \nabla}^2}{4m}+ \delta (B(t))
  \right]  \phim ({\bf x},t) +
  g \phia^2 ({\bf x},t) \nonumber \\ && \hspace*{-0.7in}
  - g^2 \frac{m^{3/2}}{2 \pi \hbar^3} i \sqrt{i \hbar \frac{\partial}{\partial t}
    + \frac{\hbar^2 {\bf \nabla}^2}{4m} -2 \hbar \Sigma^{\rm HF}}
  \phim ({\bf x},t)~.
\end{eqnarray}
Note that, since we use renormalized coupling constants in these
equations, we should not explicitly include also the so-called
anomalous averages because this leads to double-counting of the
interatomic interactions. This is explained in detail in the next
section.

The equilibrium solutions of these mean-field equations are
space-independent and of the form
\begin{eqnarray}
\label{eq:mfeequilsols}
  \phia (\bx,t) = \phia e^{-i\mu t/\hbar}~, \nonumber \\
  \phim (\bx,t) = \phim e^{-2i\mu t/\hbar}~.
\end{eqnarray}
Substitution in Eq.~(\ref{eq:mfe}) reproduces the time-independent
equations for $\phia$ and $\phim$ within the above approximations.
Moreover, by linearizing around these equilibrium solutions we
find again the collective-mode spectrum discussed in the previous
subsection.

We now discuss the solution of the homogeneous version of the
time-dependent mean-field equations in Eq.~(\ref{eq:mfe}). These
equations are given by
\begin{eqnarray}
\label{eq:mfehomo}
 i \hbar \frac{\partial \phim (t)}{\partial t}
    \!&=&\!\left[ \delta (B(t))
    - g^2 \frac{m^{3/2}}{2 \pi \hbar^3} i \sqrt{i \hbar \frac{\partial}{\partial t}
    -2 \hbar \Sigma^{\rm HF}} \right]  \phim (t) + g \phia^2 (t)~,
    \nonumber \\
   i \hbar \frac{\partial \phia (t)}{\partial t}
    \!&=&\!
        \frac{4\pi a_{\rm bg} \hbar^2}{m}  |\phia(t)|^2
      \phia (t) + 2 g  \phia^* (t)
    \phim (t)~.
\end{eqnarray}
Two different situations can occur, that of time-independent
detuning and that of time-dependent detuning. Let us first discuss
the case of time-independent detuning. In this case we are able to
solve the equation for the molecular condensate wave function by
introducing the Fourier transform of the zero-momentum part of the
retarded molecular Green's function. This Fourier transform is,
for the most interesting case of negative detuning, given by
\begin{eqnarray}
  && G^{(+)}_{\rm m} (t-t')  \equiv \int \frac{d \omega}{2 \pi}
  G^{(+)}_{\rm m} ({\bf 0},\omega) e^{-i \omega (t-t')} \nonumber \\
     && =- \frac{i \theta (t-t') g^2 m^{3/2}}{\pi \hbar^2}
      \int_0^{\infty} \frac{d \omega}{2 \pi}
      \frac{\sqrt{\hbar \omega} e^{- i \left(\omega+2\Sigma^{\rm HF}\right) (t-t')}}
      {\left[ \hbar \omega+2\hbar\Sigma^{\rm HF}- \delta (B) \right]^2
      +(g^4 m^3/4 \pi^2 \hbar^6) \hbar \omega
      } \nonumber \\
      &&  \ \ \
       -i\theta (t-t') Z (B)
     \exp \left[ -\frac{i}{\hbar}\epsilon_{\rm m} (B) (t-t')
     \right]~,
\end{eqnarray}
where $\epsilon_{\rm m} (B)$ is the molecular binding energy that
includes also the effects of the Hartree-Fock self-energy. The
molecular condensate wave function is, in terms of this Green's
function, given by
\begin{eqnarray}
  \phim (t) = \frac{g}{\hbar} \int_0^\infty d t' G^{(+)}_{\rm m} (t-t') \phia^2
  (t')+\phim (0) e^{-i \epsilon_{\rm m} (B) t/\hbar}~,
\end{eqnarray}
for $t \geq 0$. This result is substituted in the equation for the
atomic condensate wave function, which can subsequently be solved
numerically.

The second situation we can have is that of a time-dependent
detuning. To take into account the fractional derivative acting on
the molecular wave function in the second equation in
Eq.~(\ref{eq:mfehomo}), we use its definition in frequency space.
Hence we have that
\begin{eqnarray}
  \sqrt{i \hbar \frac{\partial}{\partial t}} \phim (t) &=&
  \sqrt{i \hbar \frac{\partial}{\partial t}} \int_{-\infty}^{\infty} dt'
  \int_{-\infty}^{\infty} \frac{d \omega}{2 \pi} e^{-i \omega
  (t-t')} \phim (t') \nonumber \\
  &\equiv&\int_{-\infty}^{\infty} dt'
  \int_{-\infty}^{\infty} \frac{d \omega}{2 \pi} \sqrt{\hbar \omega} e^{-i \omega
  (t-t')} \phim (t')~.
\end{eqnarray}
This specific definition is referred to in the literature as the
Weyl definition of a fractional derivative \cite{hilferbook}.
Unfortunately, the integral over $\omega$ in the above expression
does not converge. This problem is overcome by considering also
the next-order energy-dependence of the molecular self-energy.
Therefore, we take for the molecular self-energy the expression in
Eq.~(\ref{eq:selfmoleffrzero}), i.e., the molecular self-energy
with the effective range $r_{\rm bg}=0$. The equation for the
molecular mean field is then given by
\begin{eqnarray}
 \left[i \hbar \frac{\partial}{\partial t} -\delta (B(t))
    +  \frac{i\frac{g^2 m^{3/2}}{2 \pi \hbar^3} \sqrt{i \hbar \frac{\partial}{\partial t}
    -2 \hbar \Sigma^{\rm HF}}}{1-i \frac{|a_{\rm bg}| \sqrt{m}}{\hbar} \sqrt{i \hbar \frac{\partial}{\partial t}
    -2 \hbar \Sigma^{\rm HF}}} \right]  \phim (t)= g \phia^2 (t)~.
\end{eqnarray}
The term that involves the fractional derivatives is now rewritten
as
\begin{eqnarray}
&& \frac{i\frac{g^2 m^{3/2}}{2 \pi \hbar^3} \sqrt{i \hbar
\frac{\partial}{\partial t}
    -2 \hbar \Sigma^{\rm HF}} }{1-i \frac{|a_{\rm bg}| \sqrt{m}}{\hbar} \sqrt{i \hbar \frac{\partial}{\partial t}
    -2 \hbar \Sigma^{\rm HF}}} \phim (t) \nonumber \\
    && =
    \int_{-\infty}^{\infty} dt' \int \frac{d \omega}{2 \pi}
    \frac{i\frac{g^2 m^{3/2}}{2 \pi \hbar^3} \sqrt{\hbar \omega
    -2 \hbar \Sigma^{\rm HF}} e^{-i \omega (t-t')}\phim (t') }{1-i \frac{|a_{\rm bg}| \sqrt{m}}{\hbar}
    \sqrt{\hbar \omega
    -2 \hbar \Sigma^{\rm HF}}}~.
\end{eqnarray}
For large $\omega$ the integrand becomes equal to a constant which
gives rise to a delta function $\delta (t-t')$. Taking this into
account, the final result for this term is given by
\begin{eqnarray}
&& \frac{i\frac{g^2 m^{3/2}}{2 \pi \hbar^3} \sqrt{i \hbar
\frac{\partial}{\partial t}
    -2 \hbar \Sigma^{\rm HF}} }
    {1-i \frac{|a_{\rm bg}| \sqrt{m}}{\hbar} \sqrt{i \hbar \frac{\partial}{\partial t}
    -2 \hbar \Sigma^{\rm HF}}} \phim (t) = \nonumber \\
 &&  \ \ \ -\frac{g^2}{2\pi \hbar^2 |a_{\rm bg}| m} \left(
       \phim (t) - i \int_0^{\infty} dx \phim \left(t
       -x \tau \right) \right. \nonumber \\ && \ \ \ \left.
       \times e^{-2ix \Sigma^{\rm HF} \tau}
       \left[ \frac{1}{\sqrt{\pi i x}}-e^{i x} {\rm Erfc} \left( \sqrt{ix} \right) \right]
       \right)~,
\end{eqnarray}
where the characteristic time $\tau \equiv m a_{\rm bg}^2/\hbar$
and the complementary error function is defined by means of
\begin{equation}
  {\rm Erfc} (z) \equiv \frac{2}{\sqrt{\pi}} \int_{z}^{\infty} d w
  e^{-w^2} \equiv 1- {\rm Erf} (z)~.
\end{equation}
This final result shows that the term involving the fractional
derivatives may be dealt with numerically as a term that is
nonlocal in time. In the next section we present results of
numerical solutions of the time-dependent mean-field equations
using the Green's function method.

\subsection{Hartree-Fock-Bogoliubov theory} \label{subsec:hfb} A
completely different approach to arrive at mean-field equations
that describe the Bose-Einstein condensed phase of a system with
Feshbach-resonant interactions has been put forward by Kokkelmans
and Holland \cite{kokkelmans2002b} and Mackie {\it et al.}
\cite{mackie2002}. Their treatments are physically similar but
differ in some technical details. We discuss here the approach of
Kokkelmans and Holland.

Their starting point is the microscopic atom-molecule hamiltonian
in Eq.~(\ref{eq:atommoleham}). The first step is to approximate
the interatomic potential and the atom-molecule coupling as
contact interactions, according to
\begin{eqnarray}
\label{eq:contacts}
    V_{\uparrow \uparrow} (\bx-\bx') &\simeq& V_{\bf 0} \delta
    (\bx-\bx')~, \nonumber \\
    g_{\uparrow\downarrow} (\bx-\bx') &\simeq& g_{\bf 0} \delta
    (\bx-\bx')~.
\end{eqnarray}
Roughly speaking, this approximation is validated by the fact that
the deBroglie wavelength of the atoms and molecules is much larger
than the range of the interactions. However, the use of contact
interactions leads to ultraviolet divergencies in the theory which
have to be regularized by introducing a ultraviolet cut-off
$k_\Lambda$ in momentum space. The unknown microscopic interaction
parameters $V_{\bf 0}$ and $g_{\bf 0}$ are then expressed in terms
of the experimentally known parameters $g$, $\Delta \mu$, and
$a_{\rm bg}$, and the cut-off $k_\Lambda$, in such a way that the
final equations correctly describe the two-atom physics and are
cut-off independent in the limit of a large cut-off. This
renormalization procedure is discussed in detail below.

First we derive the so-called Hartree-Fock-Bogoliubov equations of
motion. Within the above approximation, the hamiltonian for the
system is given by
\begin{eqnarray}
\label{eq:contactham} \hat H &=&  \int d {\bf x}
        \psiad (\bx) \left[
     -\frac{\hbar^2 {\bf
      \nabla}^2}{2m} + \frac{V_{\bf 0}}{2} \psiad (\bx)
       \psia (\bx)
    \right] \psia (\bx) \nonumber \\
      &&+ \int d {\bf x}
        \psim (\bx) \left[
     -\frac{\hbar^2 {\bf
      \nabla}^2}{4m} +\nu (B)
    \right] \psim (\bx) \nonumber \\
      &&+g_{\bf 0} \int\!d \bx
        \left[ \psimd (\bx)
              \psia (\bx) \psia (\bx) + {\rm
     h.c.} \right]~,
\end{eqnarray}
where $\nu (B)$ is a bare and also cut-off dependent detuning for
the molecular state. In this hamiltonian, the Schr\"odinger
operators that annihilate an atom and a molecule are denoted by
$\psia (\bx)$ and $\psim (\bx)$, respectively. Their hermitian
conjugates are the creation operators.

The starting point in the derivation of the
Hartree-Fock-Bogoliubov equations of motion are the equations of
motion for the Heisenberg operators $\psia \args$ and $\psim
\args$, that follow from the hamiltonian in
Eq.~(\ref{eq:contactham}). They are given by
\begin{eqnarray}
 i \hbar \frac{\partial \hat \psi_{\rm a} \args}{\partial t}
    &=&\left[  -\frac{\hbar^2 {\bf \nabla}^2}{2m}
            + V_{\bf 0} \hat \psi_{\rm a}^{\dagger}
        \args \hat \psi_{\rm a} \args
    \right] \hat \psi_{\rm a} \args + 2 g_{\bf 0} \hat \psi_{\rm a}^{\dagger}
          \args \psim \args~, \nonumber \\
   i \hbar \frac{\partial \hat \psi_{\rm m} \args}{\partial t}
    &=&\left[  -\frac{\hbar^2 {\bf \nabla}^2}{4m}
            + \nu (B)
    \right] \hat \psi_{\rm m} \args + g_{\bf 0} \hat \psi_{\rm a}^2
    \args~.
\end{eqnarray}
The next step is to separate out the expectation value of the
Heisenberg operators. These expectation values are constant in
space since we are dealing with a homogeneous system. We write the
Heisenberg operators as a sum of their expectation values and an
operator for the fluctuations according to
\begin{eqnarray}
  \psia \args &=& \langle \psia \args \rangle + \chia \args \equiv
  \phia (t) + \chia \args~, \nonumber \\
  \psim \args &=& \langle \psim \args \rangle + \chim \args \equiv
  \phim (t) + \chim \args~.
\end{eqnarray}
We substitute this result into the Heisenberg equations of motion
and take the expectation values of these equations. These
expectation values are then decoupled in a manner that is similar
to Wick theorem. This is, of course, an approximation in this case
since we are dealing with an interacting system. In detail, we
only take into account the expectation values $\langle \psia
\rangle$, $\langle \psim \rangle$, $\langle \chia \chia \rangle$,
and $\langle \chiad \chia \rangle$. This leads to four coupled
equations of motion for these expectation values. We define the
so-called normal and anomalous expectation values according to
\begin{eqnarray}
  G_{\rm N} (\br,t ) &\equiv& \langle \chiad (\bx,t) \chia (\bx',t)
  \rangle~, \nonumber \\
  G_{\rm A} (\br,t) &\equiv& \langle \chia \args \chia (\bx',t)
  \rangle~,
\end{eqnarray}
which only depend on the difference $\br = \bx-\bx'$ due to
translational invariance of the system. Note that the normal
average yields the density of non-condensed atoms according to $n'
(t) = G_{\rm N} ({\bf 0},t)$. Including the normal average does
not alter the conclusions of the following discussion. Therefore,
we assume from now on that we are at such low temperatures that
there is essentially no thermal cloud present, and therefore take
$G_{\rm N} (\br,t) =0 $.

The Hartree-Fock-Bogoliubov equations of motion are given by
\begin{eqnarray}
\label{eq:hfbeqs}
  i \hbar \frac{\partial \phia (t)}{\partial t} &=&
    V_{\bf 0} |\phia (t)|^2 \phia (t) + \left[ V_{\bf 0} G_{\rm A} ({\bf 0},t) +
    2g_{\bf 0} \phim (t) \right] \phiad
    (t)~, \nonumber \\
  i \hbar \frac{\partial \phim (t)}{\partial t} &=&
  \nu (B) \phim (t) + g_{\bf 0} \left[ \phia^2 (t) +G_{\rm A} ({\bf 0},t)
  \right]~, \nonumber \\
   i \hbar \frac{\partial}{\partial t} G_{\rm A} (\br,t)
   &=& \left[ -\frac{\hbar^2 {\bf \nabla}^2}{m} + 4V_{\bf 0} |\phia (t)|^2
   \right]G_{\rm A} (\br,t) \nonumber \\ &&  +
   \left[ V_{\bf 0} \phia^2(t)+V_{\bf 0} G_{\rm A} ({\bf 0},t) +
   2 g_{\bf 0} \phim (t) \right] \delta (\br)~.
\end{eqnarray}
Note that, as they stand, these equations cannot be derived by
varying a $U(1)$-invariant action. However, we have seen that this
$U(1)$ invariance is an exact property of the theory. This problem
is overcome by realizing that the anomalous average $G_{\rm A}$ is
in fact proportional to the atomic condensate wave function, since
it is zero in the normal phase of the gas. More precisely, we have
that $G_{\rm A} \propto \phia^2$ which renders the equations for
the atomic and molecular condensate wave function
$U(1)$-invariant. Moreover, elimination of the anomalous average
for the Hartree-Fock-Bogoliubov equations of motion in
Eq.~(\ref{eq:hfbeqs}) leads to renormalization of the bare
couplings $V_{\bf 0}$ and $g_{\bf 0}$. We have already seen in
Section~\ref{subsec:bareatommolecule} that introducing a pairing
field into the theory leads to a summation of the ladder Feynman
diagrams. We expect something similar to occur in this case
\cite{proukakis1996,proukakis1998}.

To study how this renormalization works in detail we study the
equilibrium solutions of the Hartree-Fock-Bogliubov equations.
Therefore, we substitute
\begin{eqnarray}
  \phia (t) &=& \phia e^{-i\mu t/\hbar}~, \nonumber \\
  \phim (t) &=& \phim e^{-2i  \mu t/\hbar}~, \nonumber \\
  G_{\rm A} (\br,t) &=& G_{\rm A} (\br) e^{-2 i \mu t/\hbar}~,
\end{eqnarray}
from which we find the time-independent Hartree-Fock-Bogoliubov
equations
\begin{eqnarray}
\label{eq:hfbeqstimeindep}
  \mu \phia  &=&
    V_{\bf 0} |\phia |^2 \phia + \left[ V_{\bf 0} G_{\rm A} ({\bf 0}) +
    2g_{\bf 0} \phim  \right] \phiad
    ~, \nonumber \\
 2 \mu \phim  &=&
  \nu (B) \phim  + g_{\bf 0} \left[ \phia^2  +G_{\rm A} ({\bf 0})
  \right]~, \nonumber \\
  2 \mu G_{\rm A} (\br)
   &=& \left[ -\frac{\hbar^2 {\bf \nabla}^2}{m} + 4V_{\bf 0} |\phia |^2
   \right]G_{\rm A} (\br) \nonumber \\ &&  +
   \left[ V_{\bf 0} \phia^2 +V_{\bf 0} G_{\rm A} ({\bf 0}) +
   2 g_{\bf 0} \phim  \right] \delta (\br)~.
\end{eqnarray}
The equation for the anomalous average $G_{\rm A} (\br)$ is solved
by Fourier transformation. This gives the result
\begin{equation}
\label{eq:anomalousresult}
  G_{\rm A} ({\bf 0}) = \left ( \frac{
  \frac{V_{\bf 0}}{V} \sum_{|\bk | < k_{\Lambda}}
   \frac{1}{2 \mu^+-2\epsilon_\bk-4V_{\bf 0} |\phia|^2}
  }
  {1-\frac{V_{\bf 0}}{V} \sum_{|\bk | < k_{\Lambda}}
   \frac{1}{2 \mu^+-2\epsilon_\bk-4V_{\bf 0} |\phia|^2}} \right)
   \left( \phia^2 +2 \frac{g_{\bf 0}}{V_{\bf 0}} \phim \right)~,
\end{equation}
which explicitly shows that the anomalous average is proportional
to the atomic condensate wave function. Note also that we have to
regularize this expression by using the ultraviolet cut-off
$k_\Lambda$, since it would be ultraviolet divergent otherwise.
Converting the sum over momenta to an integral, we find the final
result for the anomalous average
\begin{equation}
\label{eq:anomalousresultfinal}
  G_{\rm A} ({\bf 0}) = \left ( \frac{
  \frac{V_{\bf 0} m^{3/2}}{2 \pi \hbar^3} i \sqrt{2 \mu-4V_{\bf 0} |\phia|^2}
  - \frac{V_{\bf 0} m k_{\Lambda}}{2 \pi^2 \hbar^2}}
  {1-\frac{V_{\bf 0} m^{3/2}}{2 \pi \hbar^3} i \sqrt{2 \mu-4V_{\bf 0} |\phia|^2}
  + \frac{V_{\bf 0} m k_{\Lambda}}{2 \pi^2 \hbar^2}} \right)
   \left( \phia^2 +2 \frac{g_{\bf 0}}{V_{\bf 0}} \phim \right)~.
\end{equation}
Substitution of this result into the equations of motion for the
atomic and molecular condensate wave functions gives in first
instance
\begin{eqnarray}
\label{eq:hfbren}
  \mu \phia  &=&
    V_{\rm r} |\phia |^2 \phia + 2g_{\rm r} \phiad \phim
    ~, \nonumber \\
 2 \mu \phim  &=&
  \nu_{\rm r} (B) \phim  + g_{\rm r} \phia^2~,
\end{eqnarray}
where the renormalized interaction and atom-molecule coupling are
given by
\begin{eqnarray}
  V_{\rm r} &=&  \frac{
  \frac{V^2_{\bf 0} m^{3/2}}{2 \pi \hbar^3} i \sqrt{2 \mu-4V_{\bf 0} |\phia|^2}
  - \frac{V_{\bf 0} m k_{\Lambda}}{2 \pi^2 \hbar^2}}
  {1-\frac{V_{\bf 0} m^{3/2}}{2 \pi \hbar^3} i \sqrt{2 \mu-4V_{\bf 0} |\phia|^2}
  + \frac{V_{\bf 0} m k_{\Lambda}}{2 \pi^2 \hbar^2}} + V_{\bf 0}~,
  \nonumber \\
  g_{\rm r} &=& \frac{
  \frac{g_{\bf 0} V_{\bf 0} m^{3/2}}{2 \pi \hbar^3} i \sqrt{2 \mu-4V_{\bf 0} |\phia|^2}
  - \frac{V_{\bf 0} m k_{\Lambda}}{2 \pi^2 \hbar^2}}
  {1-\frac{V_{\bf 0} m^{3/2}}{2 \pi \hbar^3} i \sqrt{2 \mu-4V_{\bf 0} |\phia|^2}
  + \frac{V_{\bf 0} m k_{\Lambda}}{2 \pi^2 \hbar^2}} + g_{\bf 0}~,
\end{eqnarray}
and the renormalized detuning is given by
\begin{equation}
  \nu_{\rm r} (B) = 2 \frac{g_{\bf 0}^2}{V_{\bf 0}} \left( \frac{
  \frac{V_{\bf 0} m^{3/2}}{2 \pi \hbar^3} i \sqrt{2 \mu-4V_{\bf 0} |\phia|^2}
  - \frac{V_{\bf 0} m k_{\Lambda}}{2 \pi^2 \hbar^2}}
  {1-\frac{V_{\bf 0} m^{3/2}}{2 \pi \hbar^3} i \sqrt{2 \mu-4V_{\bf 0} |\phia|^2}
  + \frac{V_{\bf 0} m k_{\Lambda}}{2 \pi^2 \hbar^2}} \right) + \nu (B)~.
\end{equation}
Finally, we have to express these renormalized quantities in terms
of the experimentally known parameters $a_{\rm bg}$, $g$, and
$\delta (B)$. Moreover, this has to be performed in a manner that
does not depend on the cut-off in the limit $k_\Lambda \to
\infty$.

The renormalization procedure used by Kokkelmans and Holland is
given by
\begin{eqnarray}
\label{eq:renprockh}
  V_{\bf 0} &=& \frac{\frac{4 \pi a_{\rm bg} \hbar^2}{m}}{1 - \frac{m k_\Lambda}{2 \pi^2 \hbar^2}
  \frac{4 \pi a_{\rm bg} \hbar^2}{m}}~, \nonumber \\
  g_{\bf 0} &=& \frac{g}{1 - \frac{m k_\Lambda}{2 \pi^2 \hbar^2}
  \frac{4 \pi a_{\rm bg} \hbar^2}{m}}~, \nonumber \\
  \nu (B) &=& \delta (B) + \frac{m k_{\Lambda} g_{\bf 0} g}{4 \pi^2
  \hbar^2}~.
\end{eqnarray}
Eliminating the microscopic parameters $V_{\bf 0}$, $g_{\bf 0}$,
and $\nu (B)$ in favor of $a_{\rm bg}$, $g$, and $\delta (B)$
finally yields the renormalized mean-field equations for the
atomic and molecular wave functions
\begin{eqnarray}
\label{eq:hfbfinal}
  \mu \phia  &=&
    \frac{4 \pi a_{\rm bg}\hbar^2/m}{ 1+ia_{\rm bg}\sqrt{\frac{m }{\hbar^2}(2\mu - 4V_{\bf 0}
    |\phia|^2)}}
    |\phia |^2 \phia + \frac{2g}{1+ia_{\rm bg}\sqrt{\frac{m }{\hbar^2}(2\mu - 4V_{\bf 0}
    |\phia|^2)}} \phiad \phim
    ~, \nonumber \\
 2 \mu \phim  &=&
 \left[ \delta (B) - i\frac{g^2 m^{3/2}}{2 \pi \hbar^3}
  \frac{\sqrt{2 \mu - 4 V_{\bf 0}|\phia|^2}}{ 1-i |a_{\rm bg}|\sqrt{\frac{m}{\hbar^2}(2\mu - 4V_{\bf 0}
    |\phia|^2)}}
 \right]\phim  \nonumber \\ &&
  \ \ \ + \frac{g}{1+ia_{\rm bg}\sqrt{\frac{m }{\hbar^2}(2\mu - 4V_{\bf 0}
    |\phia|^2)}} \phia^2~,
\end{eqnarray}
where we have retained the term $4V_0 |\phia|^2$ in the energy
arguments of the coupling constants. In the limit $k_\Lambda \to
\infty$ this term vanishes and the above renormalized equations no
longer depend on the microscopic parameters and the cut-off.

The above equations are very similar to the mean-field equations
of our effective field theory in Eq.~(\ref{eq:mfetimeindep}), if
we neglect the effective range of the interactions in the
couplings and the self-energy of the molecules in the latter
equations. There is, however, another and much more important
difference between the two mean-field theories. In the mean-field
theory that we have derived from our effective quantum field
theory we have included the Hartree-Fock self-energy that is due
to the mean-field interactions of the condensate on the thermal
atoms. This Hartree-Fock self-energy is crucial for a correct
description of the equilibrium properties of the system. In the
Hartree-Fock-Bogoliubov equations the Hartree-Fock self-energy is
replaced by the energy $4 V_{\bf 0} |\phia|^2$, which corresponds
to the mean-field energy resulting from the unrenormalized
interaction. The fact that the interaction between the condensed
and noncondensed atoms is not renormalized is a well-known problem
of the Hartree-Fock-Bogoliubov theory \cite{bijlsma1997}. Note
also that for a nonzero effective range $r_{\rm bg}$ the two-atom
physics is not incorporated exactly, and this will lead to a
discrepancy with experiment as shown in the following section.
Although the renormalization of the interactions between
condensate atoms is, for $r_{\rm bg}=0$, correctly achieved, the
interactions between condensate atoms and thermal atoms is not
correctly incorporated. In the limit where the cut-off $k_\Lambda$
goes to infinity this mean-field energy actually vanishes and we
conclude from our previous discussion that the
Hartree-Fock-Bogoliubov equations in
Eq.~(\ref{eq:hfbeqstimeindep}) have no equilibrium solution. As a
result also a linear-response analysis, similar to the one carried
out in Section~\ref{sec:oscillations}, is not possible with this
approach. Moreover, the above renormalization procedure relies on
the presence of the anomalous average $G_{\rm A} (\br)$ which
makes the theory inapplicable above the critical temperature for
Bose-Einstein condensation. Hence also a description of the
thermal cloud of a Bose-Einstein condensed gas cannot be obtained
in this manner. Note also that the above result explicitly shows
that the inclusion of the pairing field $G_{\rm A} (\br)$ indeed
leads to the summation of the ladder diagrams. This is the reason
why it is exact not to include anomalous averages in our
mean-field equations. Their effect is already incorporated by
using properly renomalized coupling constants.

Finally, we make some remarks about the theory put forward by
K\"ohler {\it et al.} \cite{kohler2002}. These authors do not
explicitly include the molecular field responsible for the
Feshbach resonance into their theory, but instead use a separable
pseudopotential for the interaction between the atoms that, when
inserted in the Lippmann-Schwinger equation, reproduces the
energy-dependent T-matrix. Subsequently, they use the
single-channel version of the above-described
Hartree-Fock-Bogoliubov theory to arrive at their mean-field
equations. The theory of K\"ohler {\it et al.} is derived from our
effective atom-molecule approach by neglecting the effect of the
molecular condensate on the atoms. The molecular field can then be
integrated out, which leads to an energy-dependent T-matrix for
the atoms. We have seen in Eq.~(\ref{eq:tmatrixfeshbachclose})
that close to resonance the energy-dependence of this T-matrix is
equivalent to the energy-dependence of the T-matrix in the
single-channel case. Close to resonance, therefore, the mean-field
theory of K\"ohler {\it et al.} incorporates the correct two-atom
physics. However, their approach cannot fully recover all the
properties of the molecules, which have been integrated out of the
problem. This can for instance be seen from the fact that the
theory contains only the ratio $g^2/\Delta \mu$ instead of the
independent quantities $g$ and $\Delta \mu$, seperately. Their
theory also does not incorporate the mean-field shift on the
noncondensed atoms due to the atomic condensate, as we have seen
explicitly above. The latter feature again disables a
linear-response analysis of the beautiful experiments we are going
to discuss next.

  \section{Coherent atom-molecule oscillations}
\label{sec:oscillations} In this section we discuss the
experimental observation of atom-molecule coherence in a
Bose-Einstein condensate \cite{donley2002,claussen2003}, and its
theoretical description in terms of the mean-field theory derived
in the previous section. In the first section we discuss the
experimental results. In the next section we calculate the
magnetic-field dependence of the frequency of the coherent
atom-molecule oscillations in linear-response theory. In the final
section we present the results of calculations that go beyond this
linear approximation.

\subsection{Experiments}
\label{subsec:expts}
\begin{figure}
\begin{center}
\includegraphics{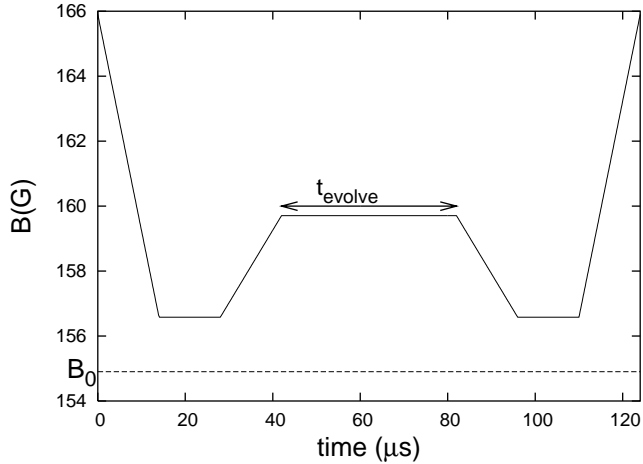}
\end{center}
\caption{\label{fig:dpulse} Typical magnetic-field pulse sequence
as used in the experiments of Donley {\it et al.}
\cite{donley2002} and Claussen {\it et al.} \cite{claussen2003}. }
\end{figure}
In the experiments of Donley {\it et al.} \cite{donley2002} and
Claussen {\it et al.} \cite{claussen2003}, performed both in
Wieman's group at JILA, one makes use of the Feshbach resonance at
$B_0 =155.041 (18)$ G(auss) in the $|f=2;m_f\!=\!-2\rangle$
hyperfine state of $^{85}$Rb. The width of this resonance is equal
to $\Delta B = 11.0(4)$ G and the off-resonant background
scattering length is given by $a_{\rm bg} = -443 a_0$, with $a_0$
the Bohr radius. The difference in the magnetic moment between the
open channel and the closed channel is given by $\Delta \mu =
-2.23 \mu_{\rm B}$, with $\mu_{\rm B}$ the Bohr magneton
\cite{kokkelmans2002b}.

In both experiments, one starts from a stable and essentially pure
condensate of about $N_{\rm c} = 10000$ atoms at a magnetic field
such that the effective scattering length is close to zero. This
implies that, since the condensate is in the noninteracting limit,
its density profile is determined by the harmonic-oscillator
groundstate wave function. The harmonic external trapping
potential is axially symmetric, with trapping frequencies $\nu_r =
17.4$ Hz and $\nu_z=6.8$ Hz in the radial and axial direction,
respectively.

Starting from this situation, one quickly ramps the magnetic field
to a value $B_{\rm hold}$ close to the resonant value and keeps it
there for a short time $t_{\rm hold}$ before ramping to a value
$B_{\rm evolve}$. The magnetic field is kept at this last value
for a time $t_{\rm evolve}$ before performing a similar pulse to
go back to the initial situation. The duration of all four
magnetic-field ramps is given by $t_{\rm ramp}$. A typical pulse
is illustrated in Fig.~\ref{fig:dpulse}. Both the ramp time
$t_{\rm ramp}$ and the hold time $t_{\rm hold}$ are kept fixed at
values of $10\!-\!15$ $\mu$s. The time $t_{\rm evolve}$ between
the pulses is variable.

Such a double-pulse experiment is generally called a Ramsey
experiment. Its significance is most easily understood from a
simple system of two coupled harmonic oscillators. Consider
therefore the hamiltonian
\begin{eqnarray}
\label{eq:twolevelham}
  \hat H =\half \left( \hat a^{\dagger}~\hat b^{\dagger} \right)
    \cdot \left(
     \begin{array}{cc}
     \delta (t) & \Delta \\
      \Delta& -\delta (t)
     \end{array}
    \right) \cdot
    \left(
      \begin{array}{c}
      \hat a \\
      \hat b
      \end{array}
    \right)~,
\end{eqnarray}
where $\hat a^{\dagger}$ and $\hat b^{\dagger}$ create a quantum
in the oscillators $a$ and $b$, respectively, and $\Delta$ denotes
the coupling between the two oscillators.

We consider first the situation that the detuning $\delta (t)$ is
time independent. The exact solution is found easily by
diagonalizing the hamiltonian. We assume that initially there are
only quanta in oscillator $a$ and none in $b$, so that we have
that $\langle \hat b^{\dagger} \hat b \rangle (0) = 0$. The number
of quanta in oscillator $a$ as a function of time is then given by
\begin{equation}
\label{eq:aat}
  \langle \hat a^{\dagger} \hat a \rangle (t) =
  \left[ 1-\frac{\Delta^2}{(\hbar \varpi)^2} \sin^2 \left( \varpi t/2 \right) \right]
  \langle \hat a^{\dagger} \hat a \rangle (0)~,
\end{equation}
with the frequency $\varpi$ given by
\begin{equation}
\label{eq:tloscfreq}
  \hbar \varpi = \sqrt{\delta^2+\Delta^2}~.
\end{equation}
We see that the number of quanta in the oscillator $a$ oscillates
in time with frequency $\varpi$. Such oscillations are called Rabi
oscillations. Note that the number of quanta in oscillator $b$ is
determined by
\begin{equation}
 \langle \hat b^{\dagger} \hat b \rangle (t)  = -\frac{\Delta^2}{(\hbar \varpi)^2} \sin^2 \left( \varpi t/2 \right)
  \langle \hat a^{\dagger} \hat a \rangle (0)~,
\end{equation}
so that the total number of quanta is indeed conserved.

Suppose now that we start from the situation with all quanta in
the oscillator $a$ and none in $b$ and that the detuning is such
that $\delta (t) \gg \Delta$. Then we have from Eq.~(\ref{eq:aat})
that $\langle \hat a^{\dagger} \hat a \rangle (t) \simeq \langle
\hat a^{\dagger} \hat a \rangle (0)$ and $ \langle \hat
b^{\dagger} \hat b \rangle (t) \simeq 0$. Starting from this
situation, we change the detuning instantaneously to a value
$\delta (t) \simeq 0$ and keep it at this value for a time $t_{\rm
hold}$. During this hold time quanta in oscillator $a$ will go to
oscillator $b$. Moreover, if $t_{\rm hold}$ is such that
\begin{equation}
\label{eq:piovertwo}
  t_{\rm hold} \simeq \frac{\pi}{2} \frac{\hbar}{\Delta}~,
\end{equation}
on average half of the quanta in oscillator $a$ will go to
oscillator $b$. Such a pulse is called a $\pi/2$-pulse. The
defining property of a $\pi/2$-pulse is that it creates a
superposition of the oscillators $a$ and $b$, such that the
probabilities to be in oscillators $a$ and $b$ are equal, and
therefore equal to $1/2$. This is indicated by the average
$\langle \hat a^{\dagger} \hat b \rangle (t)$. At $t=0$ this
average is equal to zero because there is no superposition at that
time. We can show that after the above $\pi/2$-pulse the average
$\langle \hat a^{\dagger} \hat b \rangle (t)$ reaches its maximum
value. In detail, the state after the $\pi/2$-pulse is equal to
\begin{equation}
  \frac{1}{\sqrt{N!}} \left[ \frac{\hat a^{\dagger} + \hat b^{\dagger}
  }{\sqrt{2}}\right]^N|0\rangle~,
\end{equation}
where the ground state is denoted by $|0\rangle$, and $N=\langle
\hat a^{\dagger} \hat a \rangle (0)$~.

We can now imagine the following experiment. Starting from the
situation $\delta (t) \gg \Delta$, we perform a $\pi/2$-pulse.
Then jump to a certain value $\delta_{\rm evolve}$ for a time
$t_{\rm evolve }$, and after this perform another $\pi/2$-pulse
and jump back to the initial situation. The number of quanta in
the oscillator $a$, a measurable quantity, then oscillates as a
function of $t_{\rm evolve}$ with the oscillation frequency
determined by Eq.~(\ref{eq:tloscfreq}) evaluated at the detuning
$\delta_{\rm evolve}$. The second $\pi/2$-pulse enhances the
contrast of the measurement thus providing a method of measuring
the frequency $\varpi$ as a function of the detuning with high
precision.

\begin{figure}
\begin{center}
\includegraphics{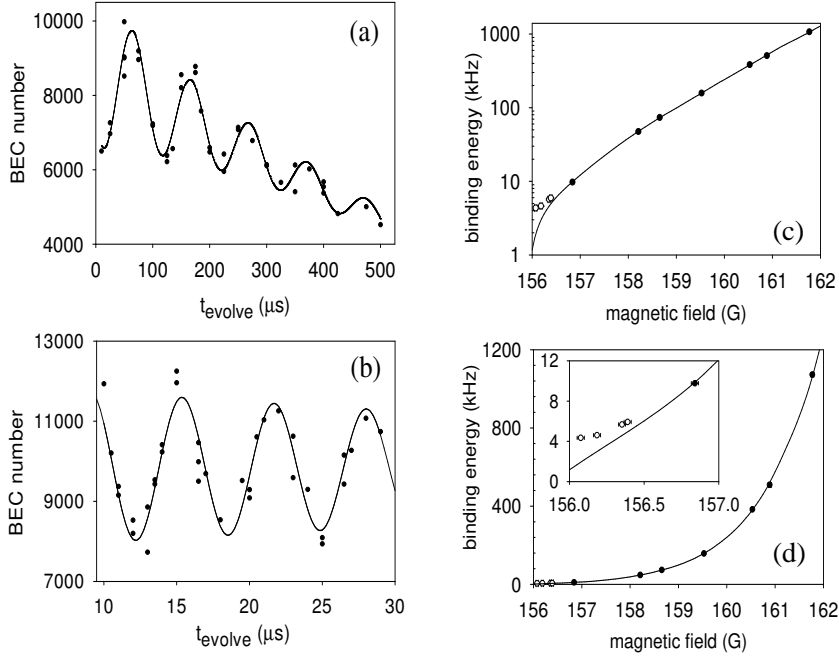}
\end{center}
\caption{\label{fig:exptresults} Experimental observation of
coherent atom-molecule oscillations. The figures are taken from
Ref.~\cite{claussen2003}. Figures (a) and (b) show the number of
atoms in the atomic condensate as a function of the time between
the two pulses in the magnetic field. The solid line indicates the
fit in Eq.~(\ref{eq:exptfit}). For (a) we have that $B_{\rm
evolve} = 156.840(25)$ G. The frequency and damping rates are
respectively given by $\nu_{\rm e}=2\pi\times0.58(12)$ kHz,
$\alpha=7.9(4)$ atom/$\mu$s, and $\beta=2\pi\times0.58(12)$ kHz.
For (b) the magnetic field $B_{\rm evolve}=159.527(19)$ G and
$\nu_{\rm e}=157.8(17)$ kHz. The damping is negligible for the
time that is used to determine the frequency. Note that the
frequency has increased for the magnetic field further from
resonance. Figures (c) and (d) show the observed frequency of the
coherent atom-molecule oscillations as a function of the magnetic
field. The solid line is the result for the molecular binding
energy found from a two-body coupled-channels calculation using
the experimental results for the frequency to accurately determine
the interatomic potential \cite{claussen2003}. Only the black
points were included in the fit. The inset shows that, close to
resonance, the observed frequency deviates from the two-body
result.}
\end{figure}

This is basically the idea of the Ramsey experiments performed by
Donley {\it et al.} \cite{donley2002} and Claussen {\it et al.}
\cite{claussen2003}. Roughly speaking, the atomic condensate
corresponds to oscillator $a$ and the molecular condensate to
oscillator $b$. Therefore, after performing the double-pulse
sequence in the magnetic field one makes a light-absorption image
of the atomic density from which one extracts the number of
condensate and noncondensed atoms. Since this imaging technique is
sensitive to a specific absorption line of the atoms it does not
measure the number of molecules.

From the above discussion we expect to observe oscillations in the
number of condensate atoms. Moreover, if the situation is such
that the detuning between the pulses is relatively large the
effect of the coupling can be neglected and the frequency of the
observed oscillations corresponds to the energy difference between
the atoms and the molecules, i.e., the molecular binding energy.
This is indeed what is observed, thereby providing compelling
evidence for the existence of coherence between atoms and
molecules.

In Fig.~\ref{fig:exptresults} the experimental results of Claussen
{\it et al.} \cite{claussen2003} are presented.
Fig.~\ref{fig:exptresults} (a) and (b) show the number of atoms in
the atomic Bose-Einstein condensate as a function of $t_{\rm
evolve}$ after a double-pulse sequence. Clearly, there is an
oscillation in the number of atoms in both cases. In
Fig.~\ref{fig:exptresults}~(a) the magnetic field between the
pulses is $B_{\rm evolve} = 156.840(25)$ G. In
Fig.~\ref{fig:exptresults}~(b) we have $B_{\rm evolve} =
159.527(19)$ G which is further from resonance. This explains also
the increase in frequency from (a) to (b) since further from
resonance the molecular binding energy is larger.

What is also observed is that there is a damping of the
oscillations and an overall loss of condensate atoms.
Experimentally, the number of atoms in the condensate is fit to
the formula
\begin{equation}
\label{eq:exptfit}
 N_{\rm c} (t) = N_{\rm average} - \alpha t + A \exp (-\beta t)
 \sin (\omega_{\rm e}t+\phi)~,
\end{equation}
where $N_{\rm average}$ is the average number of condensate atoms,
$A$ and $\phi$ are the oscillation amplitude and phase,
respectively, and $\beta$ is the damping rate of the oscillations.
The overall atom loss is characterized by a rate constant
$\alpha$. The experimentally observed frequency is equal to
$\omega_{\rm e} = 2 \pi \sqrt{\nu_{\rm e}^2-[\beta/2\pi]^2}$. By
defining the frequency of the coherent atom-molecule oscillation
in this way one compensates for the effects of the damping on the
frequency. For the results in Fig.~\ref{fig:exptresults}~(a) we
have that $\beta =2 \pi \times 0.58(12)$ kHz and $\alpha=7.9(4)$
atom/$\mu$s. The frequency is equal to $\nu_{\rm e}=9.77(12)$ kHz.
For Fig.~\ref{fig:exptresults}~(b) the frequency is equal to
$\nu_{\rm e} = 157.8(17)$ kHz. The damping and loss rate are
negligible for the short time used to determine the frequency. It
is found experimentally that both the damping rate and the loss
rate increase as $B_{\rm evolve}$ approaches the resonant value.

In Fig.~\ref{fig:exptresults}~(c)~and~(d) the results for the
frequency as a function of $B_{\rm evolve}$ are presented. The
solid line shows the result of a two-body coupled-channels
calculation of the molecular binding energy \cite{claussen2003}.
The parameters of the interatomic potentials are fit to the
experimental results for the frequency. Clearly, the frequency of
the coherent atom-molecule oscillations agrees very well with the
molecular binding energy in vacuum over a large range of the
magnetic field. Moreover, in the magnetic-field range $B_{\rm
evolve} \simeq 157-159$ G the frequency of the oscillations is
well described by the formula $|\epsilon_{\rm m} (B)|
=\hbar^2/ma^2(B)$ for the binding energy, derived in
Section~\ref{subsubsec:boundstate}. Close to resonance, however,
the measured frequency deviates from the two-body result. The
deviating experimental points are shown by open circles and are
not taken into account in the determination of the interatomic
potential. This deviation is due to many-body effects
\cite{duine2003c}.

Although some of the physics of these coherent atom-molecule
oscillations can roughly be understood by a simple two-level
picture, it is worth noting that the physics of a Feshbach
resonance is much richer. First of all, during Rabi oscillations
in a simple two-level system {\it one} quantum in a state
oscillates to the other state. In the case of a Feshbach resonance
{\it pairs} of atoms oscillate back and forth between the
dressed-molecular condensate and the atomic condensate. Therefore,
the hamiltonian is not quadratic in the annihilation and creation
operators and the physics is more complicated. In particular the
dressed molecule may decay into two noncondensed atoms instead of
forming two condensate atoms. This process is discuss in detail
below. Second, the observed atom-molecule oscillations are
oscillations between an atomic condensate and a dressed molecular
condensate. The fact that one of the levels is a dressed molecule
implies that by changing the magnetic field not only the detuning
is altered, but also the internal state of the molecule itself.

This is seen most easily by considering the linearized version of
the time-dependent mean-field equation in Eq.~(\ref{eq:mfehomo}).
Writing $\phia (t) = \phia e^{-i \mu t/\hbar} + \delta \phia (t)$
and $\phim (t) = \phim e^{-2i\mu t/\hbar} + \delta \phim (t)$, we
have that
\begin{eqnarray}
\label{eq:mfehomolinearized}
 i \hbar \frac{\partial \delta \phim (t)}{\partial t}
    \!&=&\!\left[ \delta (B)
    - g^2 \frac{m^{3/2}}{2 \pi \hbar^3} i \sqrt{i \hbar \frac{\partial}{\partial t}
    -2 \hbar \Sigma^{\rm HF}} \right] \delta \phim (t) + 2 g \phia  \delta \phia (t)~,
    \nonumber \\
   i \hbar \frac{\partial \delta \phia (t)}{\partial t}
    \!&=&\!
        2 g  \phia^*
    \delta \phim (t)~,
\end{eqnarray}
where we neglected the off-resonant part of the interatomic
interactions. This is justified sufficiently close to resonance,
where we are also allowed to neglect the energy-dependence of the
atom-molecule coupling constant.

Consider first the situation that the fractional derivative is
absent in the linearized mean-field equations in
Eq.~(\ref{eq:mfehomolinearized}), i.e., we are dealing with the
model of Drummond {\it et. al.} \cite{drummond1998}, and
Timmermans {\it et al.} \cite{timmermans1999a,timmermans1999b}.
These coupled equations describe exactly the same Rabi
oscillations as the coupled harmonic oscillators in
Eq.~(\ref{eq:twolevelham}), with the coupling equal to $\Delta =
|4 g \phia|$. In the context of particle-number oscillations
between condensates, Rabi oscillations are referred to as
Josephson oscillations and the associated frequency is called the
Josephson frequency. The Josephson frequency in the absence of the
fractional derivative term in Eq.~(\ref{eq:mfehomolinearized}) is
given by
\begin{equation}
\hbar \omega^{\rm bare}_{\rm J} = \sqrt {\delta^2 (B) + 16 g^2
n_{\rm a}}~,
\end{equation}
which reduces to $\hbar \omega^{\rm bare}_{\rm J} \simeq |\delta
(B)|$ sufficiently far off resonance where the coupling may be
neglected. This result does not agree with the experimental result
because, by neglecting the fractional derivative, which
corresponds to the molecular self-energy, we are describing
Josephson oscillations between an atomic condensate and a
condensate of bare molecules instead of dressed molecules.
Furthermore, using the result in Eq.~(\ref{eq:aat}) we have that
the amplitude of these oscillations is given by
\begin{equation}
\label{eq:amplbare}
  A^{\rm bare}_{\rm J} = \frac{16 g^2 n_{\rm a}}{[\delta (B)]^2}~.
\end{equation}

In first approximation we take the dressing of the molecules into
account as follows. If we are in the magnetic-field range where
the Josephson frequency deviates not too much from the molecular
binding energy, we are allowed to expand the propagator of the
molecules around the pole at the bound-state energy. As we have
seen in Section~\ref{subsubsec:moldos} this corresponds to
introducing the dressed molecular field and leads to the
Heisenberg equations of motion in Eq.~(\ref{eq:dressedheom}). The
linearized mean-field equations that describe the Josephson
oscillations of a atomic and a dressed-molecular condensate are
therefore given by
\begin{eqnarray}
\label{eq:mfedressedlinearized}
 i \hbar \frac{\partial \delta \phim (t)}{\partial t}
    \!&=&\! \epsilon_{\rm m}  (B)
   \delta \phim (t) + 2 g \sqrt{Z(B)} \phia  \delta \phia (t)~,
    \nonumber \\
   i \hbar \frac{\partial \delta \phia (t)}{\partial t}
    \!&=&\!
        2 g  \sqrt{Z(B)} \phia^*
    \delta \phim (t)~,
\end{eqnarray}
and lead to the Josephson frequency
\begin{eqnarray}
\label{eq:josephdress}
  \hbar \omega_{\rm J} = \sqrt{\epsilon_{\rm m}^2 (B)+16 g^2 Z(B) n_{\rm
  a}}~,
\end{eqnarray}
which reduces to $\hbar \omega_{\rm J} \simeq |\epsilon_{\rm m}
(B)|$ in the situation where the coupling is much smaller than the
binding energy. This result agrees with the experimental fact that
the measured frequency is, sufficiently far from resonance, equal
to the molecular binding energy. Moreover, the initial deviation
from the two-body result in the measured frequency is
approximately described by the equation for the Josephson
frequency in Eq.~(\ref{eq:josephdress}). The amplitude of the
oscillations is in this case given by
\begin{equation}
\label{eq:ampldressed}
  A_{\rm J} = \frac{16 g^2 Z(B) n_{\rm a}}{[\epsilon_{\rm m}
  (B)]^2}~,
\end{equation}
which close to resonance is much larger than the result in
Eq.~(\ref{eq:amplbare}).

To get more quantitative understanding of the magnetic-field
dependence of the Josephson frequency over the entire
experimentally investigated range of magnetic field we calculate
this frequency in a linear-response approximation, including the
energy-dependence of the atom-molecule coupling and the atom-atom
interactions.

Before doing so, we make some remarks about the origin of the
damping of the coherent atom-molecule oscillations and the overall
loss of atoms that is observed in the experiments. One
contribution to the damping is expected to be due to rogue
dissociation \cite{mackie2002}. Physically, this process
corresponds to a pair of condensate atoms forming a dressed
condensate molecule that then breaks up into two noncondensed
atoms instead of oscillating back to the atomic condensate. This
process is incorporated into our theory by the imaginary part of
the molecular self-energy. As explained in
Section~\ref{subsubsec:moldos} in the derivation of the Heisenberg
equations of motion in Eq.~(\ref{eq:dressedheom}), that involve
the dressed molecules, we have neglected such a process. It is,
however, incorporated in the full solution of the mean-field
equation in Eq.~(\ref{eq:mfehomo}). In the last section of the
section we present the results of numerical solutions of these
equations.

The overall loss of atoms from the atomic condensate is also
partially due to the rogue-dissociation process. The experimental
fact that a significant thermal component is formed during the
double-pulse sequence supports this idea. Apart from this process,
it may also be that conventional loss processes, such as dipolar
decay and three-body recombination play a role. Although such
processes are expected to become more important near a Feshbach
resonance, they are, however, not included in our simulations
since there is no detailed knowledge about the precise
magnetic-field dependence near the resonance. In principle,
however, these loss processes could be straightforwardly included
in our calculations, by adding the appropriate imaginary terms to
the mean-field equations. Another possible mechanism is the loss
of atoms due to elastic collisions, the so-called quantum
evaporation process \cite{duine2003b}. This process is also not
included in our present calculations.

\subsection{Josephson frequency} \label{subsec:josephson}
\begin{figure}
\begin{center}
\includegraphics{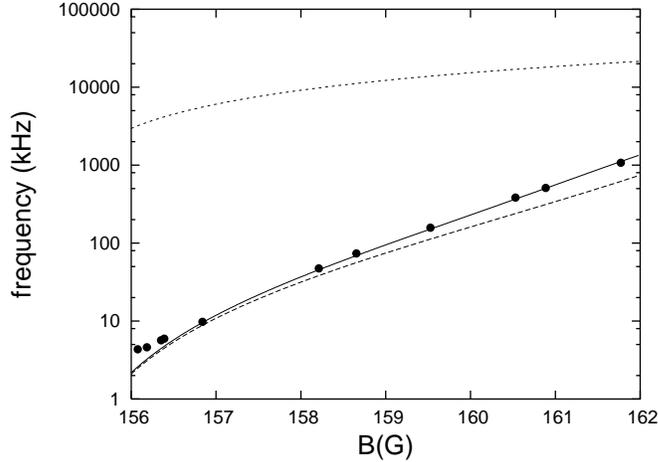}
\end{center}
\caption{\label{fig:bindingvac} Molecular binding energy in
vacuum. The solid line shows the result of a calculation with
$r_{\rm bg}=185 a_0$. The dashed line shows
$|\epsilon(B)|=\hbar^2/ma^2$. The experimental points are taken
from \cite{claussen2003}. The dotted line shows the detuning
$|\delta (B)|$. }
\end{figure}
With the mean-field theory derived in the previous sections we now
calculate the magnetic-field and density dependence of the
Josephson frequency of the coherent atom-molecule oscillations, in
a linear approximation. The only parameter that has not been
determined yet is the effective range of the interatomic
interactions $r_{\rm bg}$. All other parameters are known for
$^{85}$Rb.

The effective range is determined by calculating the molecular
binding energy in vacuum and comparing the result with the
experimental data. We have seen that far off resonance the
Josephson frequency is essentially equal to the molecular binding
energy. Since the effect of a nonzero effective range only plays a
role for large energies, and thus is important far off resonance,
this comparison uniquely determines the effective range. As
explained in detail in Section~\ref{subsubsec:boundstate}, the
molecular binding energy is determined by solving for $E$ in the
equation
\begin{equation}
\label{eq:zeroesagain}
  E - \delta (B) - \hbar \Sigma_{\rm m}^{(+)} (E) =0.
\end{equation}
For $^{85}$Rb the background scattering length is negative and the
effective range turns out to be positive. The retarded molecular
self-energy is therefore given by
\begin{eqnarray}
\label{eq:selfmretrb85} && \hbar \Sigma^{(+)}_{\rm m} (E)
   = \nonumber \\
   && -\frac{g^2 m}{2 \pi \hbar^2 \sqrt{1-2\frac{r_{\rm bg}}{a_{\rm bg}}}}
   \left[ \frac{i \sqrt{\left(1-2\frac{r_{\rm bg}}{a_{\rm bg}}\right)
   \frac{m E}{\hbar^2}}-
    \frac{r_{\rm bg} m E}{2 \hbar^2}}
   {1+ia_{\rm bg}\sqrt{\left(1-2\frac{r_{\rm bg}}{a_{\rm bg}}\right)\frac{m E}{\hbar^2}}
   - \frac{r_{\rm bg} a_{\rm bg} m E}{2 \hbar^2}}
   \right]~.
\end{eqnarray}

In Fig.~\ref{fig:bindingvac} the result of the numerical solution
of Eq.~(\ref{eq:zeroesagain}) is shown for $r_{\rm bg}=185 a_0$.
Also shown in this figure are the experimental data points.
Clearly, far off resonance there is good agreement between our
results and the experimental data points. Therefore, we use this
value for the effective range from now on in all our calculations.
The absolute value of the detuning is shown by the dotted line,
and deviates significantly from the binding energy. The dashed
line in Fig.~\ref{fig:bindingvac} indicates the formula
$|\epsilon_{\rm m}|=\hbar^2/ma^2$. As we have derived in
Section~\ref{subsubsec:boundstate} this formula should accurately
describe the magnetic-field dependence of the binding energy close
to resonance. Clearly, the solid line that indicates the result
that includes the nonzero effective range becomes closer to the
dashed line as we approach resonance. However, there is a
significant range of magnetic field where we need to include the
effective range in our calculations. Closer to the resonance, the
experimental points start to deviate from the two-atom binding
energy. This deviation is taken into account by considering
many-body effects. Note, therefore, that the expected oscillation
frequency $\hbar^2/ma^2$ never leads to a quantitative agreement
with experiment.

\begin{figure}
\begin{center}
\includegraphics{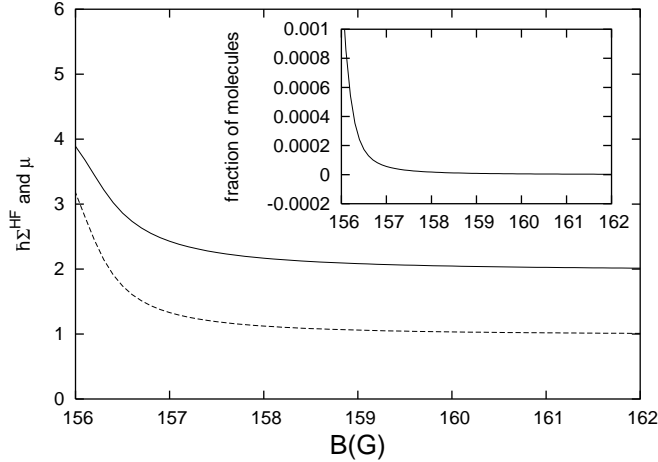}
\end{center}
\caption{\label{fig:equil} Hartree-Fock self-energy (solid line)
and chemical potential (dashed line) as a function of the magnetic
field for an atomic condensate density of $n_{\rm a} = 2 \times
10^{12}$ cm$^{-3}$. Both quantities are shown in units of $4 \pi
a(B) \hbar^2 n_{\rm a}/m$. Far off resonance, where the energy
dependence of the interactions can be safely neglected we have
that $\hbar \Sigma^{\rm HF} = 8 \pi a(B) \hbar^2 n_{\rm a}/m $ and
$\mu=4 \pi a(B) \hbar^2 n_{\rm a}/m$, as expected. The inset shows
the fraction of bare molecules as a function of the magnetic
field.}
\end{figure}

As mentioned previously, we calculate the many-body effects on the
frequency of the coherent atom-molecule oscillations in linear
approximation. Therefore, we first need to determine the
equilibrium around which to linearize. In detail, the equilibrium
values of the atomic and molecular condensate wave functions are
determined by solving the time-independent mean-field equations in
Eq.~(\ref{eq:mfetimeindep}) together with the equation for the
Hartree-Fock self-energy in Eq.~(\ref{eq:sigmahf}) at a fixed
chemical potential $\mu$. To compare with the experimental results
it is more convenient to solve these equations at a fixed
condensate density. The chemical potential is then determined from
these equations.

In Fig.~\ref{fig:equil} we show the result of this calculation for
an atomic condensate density of $n_{\rm a} = 2 \times 10^{12}$
cm$^{-3}$. The solid line shows the Hartree-Fock self-energy
$\hbar \Sigma^{\rm HF}$ and the dashed line the chemical potential
as a function of the magnetic field, both in units of the energy
$4 \pi a(B) \hbar^2 n_{\rm a}/m$. Note that far off resonance,
where the energy dependence of the interaction may be neglected,
we have that $\mu = 4 \pi a(B) \hbar^2 n_{\rm a}/m$ and $\hbar
\Sigma^{\rm HF} = 2\mu$. This is the expected result. The inset of
Fig.~\ref{fig:equil} shows the fraction of bare molecules
$|\phi_{\rm m}|^2/n_{\rm a}$. Note that this fraction is always
very small. This justifies neglecting the atom-molecule and
molecule-molecule interactions since from this figure we see that
the mean-field energies associated with these interactions are at
least three orders of magnitude smaller. {\it A posteriori} this
observation justifies neglecting the effect of the presence of the
molecular condensate on the atoms in the approach of K\"ohler {\it
et al.} \cite{kohler2002}.

\begin{figure}
\begin{center}
\includegraphics[width=7cm]{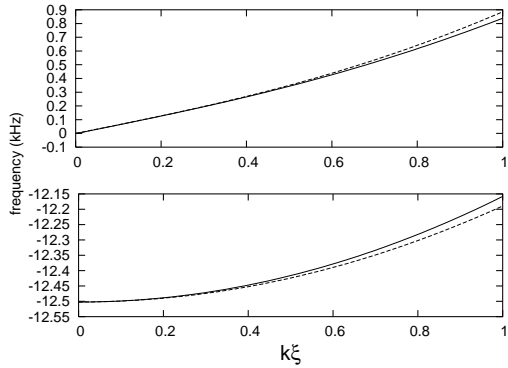}
\end{center}
\caption{\label{fig:dispersion} The dispersion relation for the
collective modes of an atom-molecule system for a condensate
density of $n_{\rm a}=2 \times 10^{12}$ cm$^{-3}$ at a magnetic
field of $B=157$ G. The momentum is measured in units of the
inverse coherence length $\xi^{-1}=\sqrt{16\pi a(B)n_{\rm a}}$.
The upper branch corresponds to the gapless dispersion for
phonons. The solid line is the result of the full calculation, the
dashed line shows the Bogoliubov dispersion for the scattering
length $a(B)$. The lower branch corresponds to the coherent
atom-molecule oscillations. The solid line is the result of the
full calculation whereas the dashed line shows the result with the
same zero-momentum part, but with the momentum dependence
determined by $\hbar^2 \bk^2/4m$. }
\end{figure}

Since the coherent atom-molecule oscillations are a collective
mode where the amplitude of the atomic and molecular condensate
wave functions oscillate out-of-phase, we study the collective
modes of the system. As explained in detail in the previous
section, the frequencies of the collective modes are determined by
Eq.~(\ref{eq:collmodesfreq}). This equation is solved numerically
and yields a dispersion relation with two branches.

The result of this calculation is shown in
Fig.~\ref{fig:dispersion} for an atomic condensate density of
$n_{\rm a}=2 \times 10^{12}$ cm$^{-3}$ and a magnetic field of
$B=157$ G. The momentum is indicated in units of the inverse
coherence length $\xi^{-1}=\sqrt{16\pi a(B)n_{\rm a}}$. The upper
branch corresponds to the gapless phonon excitations. For small
momenta this branch has a linear momentum dependence. The upper
dashed line indicates the Bogoliubov dispersion in
Eq.~(\ref{eq:bogodisp}) evaluated at the scattering length $a(B)$.
For small momentum the solid and the dashed line are almost
identical. For larger momenta the numerically exact result is
smaller, due to the energy-dependence of the interactions that
effectively reduce the scattering length.

\begin{figure}
\begin{center}
\includegraphics{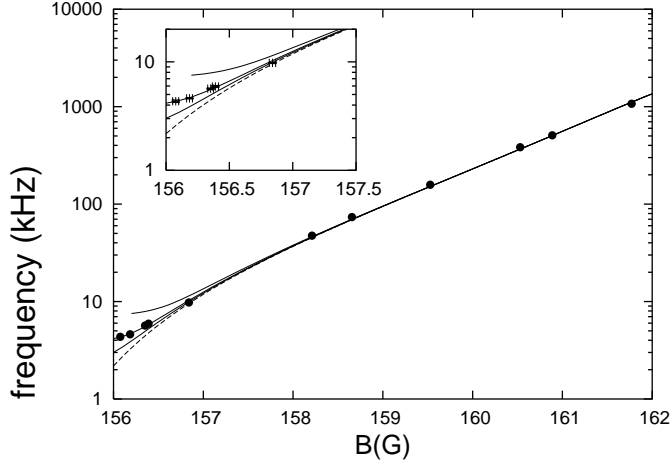}
\caption{\label{fig:joseph_b} Josephson frequency of coherent
atom-molecule oscillations for various values of the condendate
density. The solid lines are the results of calculations for
nonzero condensate density. The different lines correspond from
top to bottom to the decreasing condensate densities $n_{\rm a}=5
\times 10^{12}$ cm$^{-3}$, $n_{\rm a}=2 \times 10^{12}$ cm$^{-3}$,
and $n_{\rm a}=10^{12}$ cm$^{-3}$. The dashed line corresponds to
the molecular binding energy in vacuum, i.e., $n_{\rm a}=0$. The
experimental data points, taken from Ref.~\cite{claussen2003}, are
also shown. }
\end{center}
\end{figure}

The lower branch corresponds to the coherent atom-molecule
oscillations and is gapped. The solid line indicates the result of
the full calculations. For small momenta it is well described by
\begin{equation}
\label{eq:approxdisp}
  \hbar \omega_\bk \simeq - \hbar \omega_{\rm J} +
  \epsilon_\bk/2~,
\end{equation}
where $\omega_{\rm J}$ is the Josephson frequency. The dispersion
resulting from this last equation is shown in the lower part
Fig.~\ref{fig:dispersion} by the dashed line. This momentum
dependence is to be expected since sufficiently far from resonance
the atom-molecule oscillations reduce to a two-body excitation.
The fact that the dispersion is negative is due to the fact that
we are linearizing around a metastable situation with more atoms
than molecules. Although this is the experimentally relevant
situation, the true equilibrium situation for negative detuning
corresponds to almost all atoms in the molecular state
\cite{timmermans1999b}.

In Fig.~\ref{fig:joseph_b} we present the results for the
Josephson frequency as a function of the magnetic field, for
different values of the condensate density. The solid lines in
this figure show, from top to bottom, the results for an
decreasing nonzero condensate density. The respective condensate
densities are given by $n_{\rm a}=5 \times 10^{12}$ cm$^{-3}$,
$n_{\rm a}=2 \times 10^{12}$ cm$^{-3}$, and $n_{\rm a}=10^{12}$
cm$^{-3}$. The dashed line shows the molecular binding energy in
vacuum. The Josephson frequency reduces to the molecular binding
energy for all values of the condensate density, in agreement with
previous remarks. Nevertheless, sufficiently close to resonance
there is a deviation from the two-body result due to many-body
effects. This deviation becomes larger with increasing
condensate density.

In order to confront our results with the experimental data we
have to realize that the experiments are performed in a magnetic
trap. Taking only the ground states $\phi_{\rm a} ({\bf x})$ and
$\phi_{\rm m} ({\bf x})$ into account for both the atomic and the
molecular condensates, respectively, this implies effectively that
the atom-molecule coupling $g$ is reduced by an overlap integral.
Hence we define the effective homogeneous condensate density by
means of $n_{\rm a} = N_{\rm a} \left[ \int d {\bf x} \phi_{\rm
a}^2 ({\bf x} ) \phi_{\rm m} (\bf x) \right]^2=16 \sqrt{2} N_{\rm
a} m^{3/2} \nu_r \sqrt{\nu_z}/(125 \pi^3 \hbar^{3/2})$, where
$N_{\rm a}$ denotes the number of condensed atoms and $\nu_r$ and
$\nu_z$ the radial and axial trapping frequencies, respectively.
For the experiments of Claussen {\it et al.} we have that $N_{\rm
a} \simeq 8000$ during the oscillations close to resonance as seen
from Fig.~\ref{fig:exptresults}, which results in an effective
density of $n_{\rm a}\simeq 2 \times 10^{12}$ cm$^{-3}$. This
agrees also with the effective homogeneous density quoted by
Claussen {\it et al.} \cite{claussen2003}. The solid curve in
Fig.~\ref{fig:joseph_b} clearly shows an excellent agreement with
the experimentally observed frequency for this density.

It is important to note that there are two hidden assumptions in
the above comparison. First, we have used that the dressed
molecules are trapped in the same external potential as the atoms.
This is not obvious because the bare molecular state involved in
the Feshbach resonance is high-field seeking and therefore not
trapped. However, Eq.~(\ref{eq:wavefctmol}) shows that near
resonance almost all the amplitude of the dressed molecule is in
the low-field seeking open channel and its magnetic moment is
therefore almost equal to twice the atomic magnetic moment.
Second, we have determined the frequency of the coherent
atom-molecule oscillations in equilibrium. In contrast, the
observed oscillations in the number of condensate atoms is clearly
a nonequilibrium phenomenon. This is, however, expected not to
play an important role because the Ramsey-pulse sequence is
performed on such a fast time scale that the response of the
condensate wave function can be neglected. By variationally
solving the Gross-Pitaevskii equation for the atomic condensate
wave function, we have explicitly checked that after a typical
pulse sequence its width is only a few percent larger than the
harmonic oscillator ground state.
\begin{figure}
\begin{center}
\includegraphics{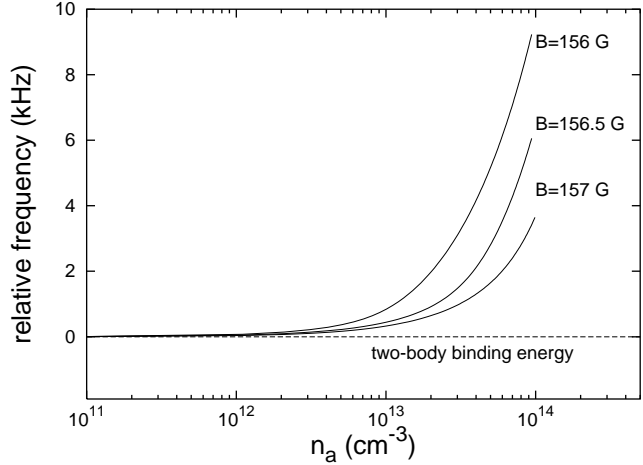}
\end{center}
\caption{\label{fig:joseph_na} Josephson frequency of coherent
atom-molecule oscillations as a function of the condensate
density, for fixed magnetic field. We have subtracted the
molecular binding energy.}
\end{figure}

Finally, we calculate the Josephson frequency as a function of the
condensate density. The results of this calculation are presented
in Fig.~\ref{fig:joseph_na}, for various values of the magnetic
field which is kept fixed in these calculations. In the
presentation of the results we have subtracted the molecular
binding energy to bring out the many-body effects  more clearly.
As expected, the difference between the Josephson frequency and
the molecular binding energy increases with increasing condensate
density. Moreover, for values of the magnetic field closer to
resonance the difference is also larger.

The above calculations in the linear approximation give already a
great deal of insight in the coherent atom-molecule oscillations,
and, in particular, in their many-body aspects. In the next
section we aim at achieving also insight in the nonlinear dynamics
and damping resulting from the time-dependent mean-field equations
for the double-pulse experiments. In particular, we also discuss
the rogue-dissociation process. The nonlinear effects in these
experiments has first been discussed by Kokkelmans and Holland
\cite{kokkelmans2002b}, Mackie {\it et al.} \cite{mackie2002}, and
K\"ohler {\it et al.} \cite{kohler2002}, on the basis of their
mean-field approaches summarized in Section~\ref{subsec:hfb}.

\subsection{Beyond linear response} \label{subsec:beyondlinearresp}
\begin{figure}
\begin{center}
\includegraphics{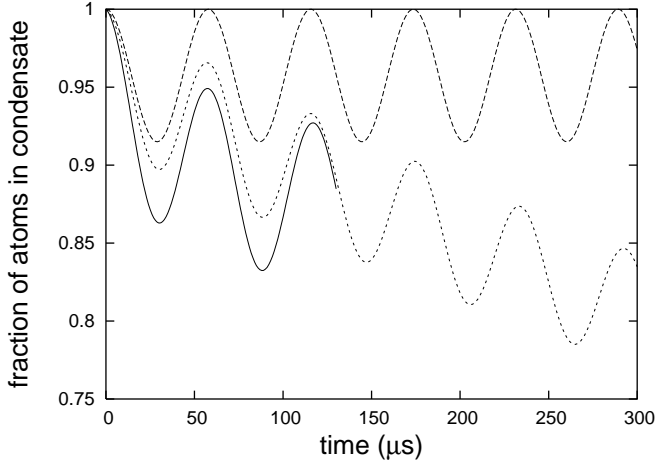}
\end{center}
\caption{\label{fig:na_t} Fraction of atoms in the atomic
condensate. The solid line shows the result of the inclusion of
the rogue-dissociation process into the calculations. The dashed
line shows the result of a calculation without this process. The
dotted line shows the result for a calculation that includes the
estimate in Eq.~(\ref{eq:gpestimate}). We have taken the
parameters $B_{\rm init}=162$ G, $B_{\rm evolve}=158$ G, and
$n_{\rm a}= 2 \times 10^{12}$ cm$^{-3}$.}
\end{figure}
In this section we discuss the numerical solution of the
time-dependent mean-field equations using the methods described in
Sec.~\ref{subsubsec:tdpdtmfe}. We focus here on the situation
where the detuning is only changed instantaneously, so that we are
allowed to use the Green's function method discussed in this
section. After the elimination of the molecular condensate wave
function from the mean-field equations, the effective equation for
the atomic condensate wave function is then given by
\begin{eqnarray}
\label{eq:effeqnatoms}
    && i \hbar \frac{\partial \phia (t)}{\partial t}
    =
        \frac{4\pi a_{\rm bg} \hbar^2}{m}  |\phia(t)|^2
      \phia (t) \nonumber \\
    && + 2 g \phia^* (t) \phim (0) e^{-i \epsilon_{\rm m} (B) t/\hbar} - \frac{2ig^2\phia^* (t)}{\hbar}
      \int_0^t dt' \left\{ Z(B)
    e^{-\frac{i}{\hbar} \epsilon_{\rm m} (B) (t-t')}
     \phia^2(t') \rule{0mm}{7mm} \right. \nonumber \\
      && \left.  + \frac{g^2 m^{3/2}}{\pi \hbar^2}
      \int_0^{\infty} \frac{d \omega}{2 \pi}
      \frac{\sqrt{\hbar \omega} e^{- i \left(\omega+2\Sigma^{\rm HF}\right) (t-t')}\phia^2 (t') }
      {\left[ \hbar \omega\!+\!2 \hbar \Sigma^{\rm HF}\!-\!\delta (B) \right]^2
      +(g^4 m^3/4 \pi^2 \hbar^6) \hbar \omega
      } \right\}~.
\end{eqnarray}
In this equation, the term that involves the integral over
frequencies describes the fact that a pair of condensate atoms
that forms a molecule can decay into a pair of noncondensed atoms
with opposite momenta, i.e., the rogue-dissociation process. In
the absence of this term the equation effectively takes into
account the dressing of molecules in an adiabatic manner, and
describes Josephson oscillations between a condensate of atoms and
dressed molecules.

As we have discussed in the previous section, the above equation
is only applicable to the situation of a sudden change in magnetic
field. Therefore, we perform the following calculation. For a
given magnetic field $B_{\rm init}$ and atomic condensate density
we calculate the equilibrium values of the molecular wave
functions and the Hartree-Fock self-energy, using the
time-independent mean-field equations in
Eq.~(\ref{eq:mfetimeindep}) and Eq.~(\ref{eq:sigmahf}). Then we
change the magnetic field instantaneously to the value $B_{\rm
evolve}$ and keep it at this value. In Fig.~\ref{fig:na_t} the
results of the calculations for this situation are shown, with
$B_{\rm init}=162$~G and $B_{\rm evolve}=158$~G. The atomic
condensate density is taken equal to $n_{\rm a}=2\times10^{12}$
cm$^{-3}$. The dashed line shows the result for a calculation
without the rogue-dissociation process and shows oscillations
where a fraction of the atoms is converted into molecules and
oscillates back and forth between the atomic and dressed molecular
condensate. Since there is no decay mechanism, all of the atoms
come back into the atomic condensate at times equal to a multiple
of the oscillation period. The solid line shows the result of a
calculation that includes the rogue-dissociation process. Clearly,
the number of condensate atoms oscillates in this case as well.
However, not all of the atoms come back into the atomic condensate
and there is a decay of the number of atoms in the atomic
condensate. This is precisely due to the above-mentioned
rogue-dissociation process.

Although the preliminary calculations presented in this section
are limited to the case of a step in the magnetic field, they
nevertheless present some insight in the effects of the
rogue-dissociation process on the coherent atom-molecule
oscillations in a Ramsey experiment. In future work we intend to
study also the case of time-dependent magnetic fields, by an exact
numerical treatment of the fractional derivative in our
time-dependent mean-field equations. In particular, we are
interested in the magnetic-field dependence of the damping that is
caused by the rogue-dissociation process.

We can estimate this dependence as follows. The Green's function
associated with the rogue-dissociation process,
\begin{eqnarray}
\label{eq:grogues}
   && G^{(+)}_{\rm rog} (t-t')
     =  - \frac{i \theta (t-t') g^2 m^{3/2}}{\pi \hbar^2}
     \nonumber \\
     && \times \int_0^{\infty} \frac{d \omega}{2 \pi}
      \frac{\sqrt{\hbar \omega} e^{- i \left(\omega+2\Sigma^{\rm HF}\right) (t-t')}}
      {\left[ \hbar \omega+2\hbar\Sigma^{\rm HF}- \delta (B) \right]^2
      +(g^4 m^3/4 \pi^2 \hbar^6) \hbar \omega
      }~,
\end{eqnarray}
is sharply peaked in time. Hence we approximate this Green's
function by
\begin{equation}
\label{eq:grogueapprox}
  G^{(+)}_{\rm rog} (t-t') \simeq \tau (B) G^{(+)}_{\rm rog} (0) \delta
  (t-t')~,
\end{equation}
with the timescale $\tau (B)$ given by
\begin{equation}
  \tau (B) = \int_{-\infty}^{t_{\rm c}} dt~G_{\rm rog}^{(+)} (t)~,
\end{equation}
with $t_{\rm c}$ a positive cut-off that is determined such that
the result for $\tau (B)$ depends only very weakly on $t_{\rm c}$.
The Green's function evaluated at zero time equals $G^{(+)}_{\rm
rog} (0)= 1-Z(B)$, a result which follows from the sum rule for
the molecular density of states in Eq.~(\ref{eq:sumrule}). This
gives the contribution
\begin{equation}
\label{eq:gpestimate}
 \simeq \frac{-2 i [1-Z(B)] g^2 \tau (B)}{\hbar} |\phia (t)|^2 \phia
  (t)~,
\end{equation}
to the right-hand side of Eq.~(\ref{eq:effeqnatoms}). The rate
equation for the atomic density that follows from this term is
given by
\begin{equation}
  \frac{dn_{\rm a}}{dt} \simeq- \frac{4 [1-Z(B)] g^2 \tau(B)}{\hbar^2} n_{\rm
  a}^2 (t)~,
\end{equation}
which after linearization leads to the following equation for the
number of condensate atoms
\begin{equation}
  \frac{d \delta N_{\rm a} (t)}{dt} \simeq - \beta \delta N_{\rm a}
  (t)~,
\end{equation}
with the rate $\beta$ given by
\begin{equation}
\label{eq:betaest}
  \beta \simeq \frac{8 [1-Z(B)] g^2 \tau (B) n_{\rm a}}{\hbar^2}~.
\end{equation}

We observe from this equation that the loss rate of atoms from the
atomic condensate due to the rogue-dissociation process increases
as the magnetic field approaches its resonant value. This is
indeed what is observed experimentally \cite{claussen2003}. Far
off resonance the loss rate vanishes since the wave function
renormalization factor $Z(B) \to 1$ in this limit. For the
parameters of Fig.~\ref{fig:exptresults}~(a) at the effective
homogeneous density $n_{\rm a} = 2 \times 10^{12}$ cm$^{-3}$, we
have that $\tau (B) \simeq 1.28 \times 10^{-9}$ s, which leads to
$\beta \simeq 0.45$ kHz. The dotted line in Fig.~\ref{fig:na_t}
shows the result of a calculation that includes the term in
Eq.~(\ref{eq:gpestimate}). The exact result, shown by the solid
line, and this approximate result show the same overall damping
rate. This justifies the approximation for the Green's function in
Eq.~(\ref{eq:grogueapprox}). The result for the damping rate
$\beta$ is about a factor of eight smaller than the experimental
result.

\begin{figure}
\begin{center}
\includegraphics{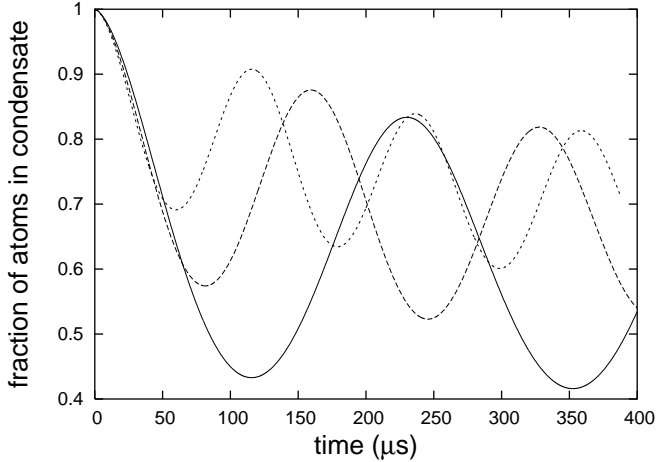}
\caption{\label{fig:gfmethod} Fraction of atoms in the atomic
condensate after a step in the magnetic field. The solid line
corresponds to $B_{\rm evolve} = 156.1$ G. The dashed and dotted
line correspond to a magnetic field of $B_{\rm evolve} = 156.5$ G
and $B_{\rm evolve}=156.9$ G, respectively. The initial magnetic
field is $B_{\rm init}=162$ G and the density of the atomic
condensate is $n_{\rm a}=2 \times 10^{12}$ cm$^{-3}$.}
\end{center}
\end{figure}

To further investigate the magnetic-field dependence of the
damping of the coherent atom-molecule oscillations, we have
calculated the numerical solution of the effective equation of
motion for the atomic condensate wave function for a step in the
magnetic field, for three different final magnetic fields. The
results of these calculations are shown in
Fig.~\ref{fig:gfmethod}. The solid, dashed, and dotted lines
corresponds to a magnetic field of $B_{\rm evolve}=156.1$ G,
$B_{\rm evolve}=156.5$ G, and $B_{\rm evolve}=156.9$ G,
respectively. The initial equilibrium corresponds to an atomic
condensate density of $n_{\rm a}=2\times10^{12}$ cm$^{-3}$ at a
magnetic field of $B_{\rm init} = 162$ G. Note the increase in the
frequency with increasing magnetic field.

The magnetic-field dependence of the frequency and damping of the
coherent atom-molecule oscillations is found from these numerical
results by fitting with the equation in Eq.~(\ref{eq:exptfit}).
The results are presented in Fig.~\ref{fig:damping_b}. The solid
line corresponds to the Josephson frequency of the coherent
atom-molecule oscillations that was found by means of the
linear-response calculation of the previous section. The deviation
for large magnetic fields is understood because we have, in our
numerical solution of the effective mean-field equation, not taken
into account the higher-order energy-dependences of the molecular
self-energy that are fully taken into account in the
linear-response theory. The inset shows the damping as a function
of the magnetic field. Note the increase of the damping as the
magnetic field approaches its resonant value. This is expected
from the estimate in Eq.~(\ref{eq:betaest}).

\begin{figure}
\begin{center}
\includegraphics{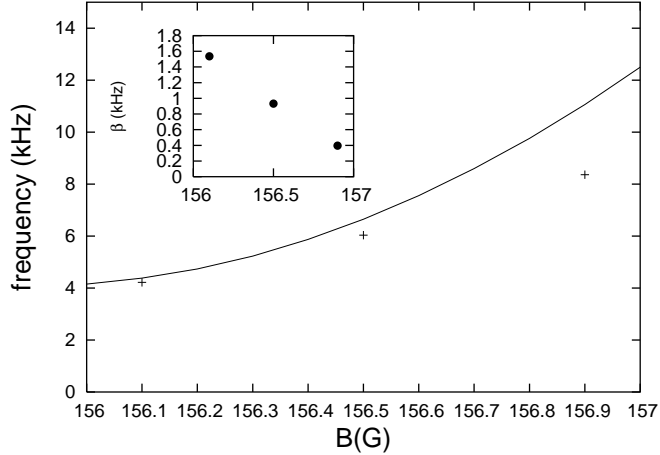}
\caption{\label{fig:damping_b} Frequency and damping as a function
of the magnetic field. The solid line corresponds to the frequency
found by means of linear-response theory.}
\end{center}
\end{figure}

The above analysis indicates that the rogue-dissociation process
gives possibly a contribution to the experimentally observed
damping of the coherent atom-molecule oscillations. Presumably,
however, also other mechanisms contribute to the observed damping.
In particular, we mention here the quantum evaporation process,
that was shown to be important in the single-pulse experiments
\cite{duine2003b}. The detailed investigation of the damping of
the coherent atom-molecule oscillation is a subject for further
study.

  \section{Conclusions and outlook} \label{sec:concl} In this review
paper we have presented the derivation of an effective quantum
field theory suitable for the description of a Bose gas near a
Feshbach resonance, since it incorporates the two-atom physics
exactly. We have presented several applications of this theory,
both above and below the critical temperature for Bose-Einstein
condensation. In the last part of this paper we have studied in
detail the magnetic-field dependence of the frequency of the
coherent atom-molecule oscillations and have obtained excellent
agreement with the experimental results. In particular, we have
been able to quantitatively explain the many-body effects on this
frequency by making use of a linear-response approximation to our
mean-field equations. Although we have already presented some
numerical solutions of the mean-field equations that improve on
this approximation, a great deal of work still has to be done. The
numerical solution of these equations for the situation of
time-dependent detuning is rather involved. Nevertheless, work in
this direction is in progress and will be reported in a future
publication.

As already mentioned, we have also discussed the properties of the
gas above the critical temperature. This discussion was mainly
concerned with the equilibrium properties of the gas and we
studied the many-body effects on the bound-state energy of the
molecular state. An important conclusion of this study is that,
for certain values of the parameters, there exists a many-body
induced resonant state with a relatively small energy. In future
work we intend to study the effects of the appearance of this
resonant state in the molecular density of states on the
properties of the gas. In particular we expect that due to this
effect the number of molecules in the gas will be large even at
relatively large detuning, which can not be explained on the basis
of two-atom physics.

Furthermore, to study the normal state also in an
out-of-equilibrium situation, we should derive a quantum kinetic
theory that describes the evolution of the local occupation
numbers of the atoms and molecules. Moreover, the description of
the Bose-Einstein condensed phase of the gas at nonzero
temperatures requires a modification of the mean-field equations
such that they include the effects of the thermal clouds of atoms
and molecules, and we need equations for the evolution of the
local occupation numbers of the latter. The extension of the
theory presented in this paper to these situations can be derived
in a unifying manner by using a functional formulation of the
Schwinger-Keldysh nonequilibrium theory \cite{stoof1999}, and is
especially important in view of the ongoing effort to produce
ultracold molecules by means of a sweep in the magnetic field
through the Feshbach resonance \cite{regal2003b}.

The theory presented in this paper is generalized to a gas of
fermionic atoms in a straightforward manner
\cite{falco2003,duine2003a}. One modification is that to have
$s$-wave scattering between fermionic atoms we have to have a
mixture of atoms with two hyperfine states, since the Pauli
principle forbids $s$-wave scattering between identical fermions.
Furthermore, the properties of the dressed molecular state is
altered due to the presence of the Fermi sphere. A molecule with
zero momentum only decays if its energy is above twice the Fermi
energy. If the molecular state lies below twice the Fermi energy,
the equilibrium situation is a Bose-Einstein condensate of
molecules. If we start from this situation and increase the
detuning, the Bose-Einstein condensate of molecules crosses over
to a Bose-Einstein condensate of Cooper pairs, i.e., a BCS-BEC
crossover occurs \cite{nozieres1985,ohashi2003}. In view of the
ongoing experiments with atomic Fermi gases near a Feshbach
resonance
\cite{regal2003b,strecker2003,xu2003,jochim2003,greiner2003,zwierlein2003},
it is particularly interesting to study the effects of
nonadiabticity on the crossover from a Bose-Einstein condensate of
molecules to a degenerate Fermi gas. In particular, the atomic
distribution function after such a sweep, and its dependence on
the duration of the sweep, is of great interest, since this will
determine whether or not a BCS-state will form after
equilibration. Determination of the atomic distribution function
requires, in first instance, knowledge of the solution of the
mean-field equation for the molecular condensate for
time-dependent detuning. Work in this direction is in progress. We
also intend to study the equilibrium properties of this crossover,
and in particular the behaviour of the critical temperature, in
detail in future work.

Clearly, Feshbach resonances present an exciting opportunity for
the experimental and theoretical study of the many-body properties
of atomic and molecular Bose and Fermi gases. There is little
doubt that these Feshbach resonances will find many new
applications in the years to come.

  \ack It is a great pleasure to thank Frieda van Belle, Eric
Cornell, Neil Claussen, Gianmaria Falco, Behnam Farid, Randy
Hulet, Niels de Keijzer, Mathijs Romans, Subir Sachdev, Kareljan
Schoutens, Peter van der Straten, Bart Vlaar, and Carl Wieman for
their contributions to this review paper. Furthermore, we would
like to acknowledge the hospitality of the European Centre for
Theoretical Studies in Nuclear Physics and Related Areas (ECT*)
during the Summer Program on Bose-Einstein condensation. This work
is supported by the Stichting voor Fundamenteel Onderzoek der
Materie (FOM) and by the Nederlandse Organisatie voor
Wetenschappelijk Onderzoek (NWO).

\end{document}